\documentclass[twocolumn,showpacs,preprintnumbers,amsmath,amssymb]{revtex4}
\usepackage{epsfig}
\usepackage{dcolumn}
\usepackage{bm}

\newcommand{\remove}[1]{}
\newcommand{\dd}{\mathrm{d}}

\def\be{\begin{equation}}
\def\ee{\end{equation}}

\newcommand{\beq}{\begin{equation}}
\newcommand{\eeq}{\end{equation}}
\newcommand{\beqa}{\begin{eqnarray}}
\newcommand{\eeqa}{\end{eqnarray}}

\renewcommand{\pl}{\partial}
\newcommand{\lag}{\langle}
\newcommand{\rag}{\rangle}

\newcommand{\vv}{{\bf v}}

\newcommand{\vx}{{\bf x}}
\newcommand{\vk}{{\bf k}}
\newcommand{\vq}{{\bf q}}
\renewcommand{\vr}{{\bf r}}
\newcommand{\vs}{{\bf s}}

\newcommand{\tdelta}{{\tilde{\delta}}}

\newcommand{\tg}{{\tilde{g}}}

\newcommand{\tpsi}{{\tilde{\psi}}}

\newcommand{\cG}{{\cal G}}

\newcommand{\cM}{{\cal M}}
\newcommand{\cO}{{\cal O}}

\newcommand{\rhob}{\overline{\rho}}
\newcommand{\Om}{\Omega_{\rm m}}

\newcommand{\bea}{\begin{array}}
\newcommand{\ea}{\end{array}}

\newcommand{\MPl}{M_{\rm Pl}}

\begin{document}

\title{K-mouflage Cosmology: Formation of Large-Scale Structures}

\author{Philippe Brax}
\affiliation{Institut de Physique Th\'eorique,\\
CEA, IPhT, F-91191 Gif-sur-Yvette, C\'edex, France\\
CNRS, URA 2306, F-91191 Gif-sur-Yvette, C\'edex, France}
\author{Patrick Valageas}
\affiliation{Institut de Physique Th\'eorique,\\
CEA, IPhT, F-91191 Gif-sur-Yvette, C\'edex, France\\
CNRS, URA 2306, F-91191 Gif-sur-Yvette, C\'edex, France}
\vspace{.2 cm}

\date{\today}
\vspace{.2 cm}

\begin{abstract}
 We study structure formation in K-mouflage cosmology whose main feature is the absence of screening effect on quasilinear scales. We show that the growth of structure at the linear level is affected by both a new time dependent Newton constant and a friction term which depend on the background evolution. These combine with the modified background evolution to change the growth rate by up to ten percent since $z\sim 2$. At the one loop level, we find that the nonlinearities of the K-mouflage models are mostly due to the matter dynamics and that the scalar perturbations can be treated at tree level.  We also study the spherical collapse in K-mouflage models and show that the critical density contrast deviates from its $\Lambda$-CDM value and that, as a result, the halo mass function is modified for large masses by an order one factor. Finally we consider the deviation of the matter spectrum from $\Lambda$-CDM on nonlinear scales where a halo model is utilized. We find that the discrepancy peaks around $1\ h{\rm Mpc}^{-1}$ with a relative difference which can reach fifty percent. Importantly, these features are still true at larger redshifts, contrary to models of the chameleon-$f(R)$ and Galileon types.

\end{abstract}

 \vskip2pc

\maketitle

\section{Introduction}
\label{Introduction}

Scalar fields could be crucial  in explaining the recent acceleration of the expansion of the Universe \cite{Copeland:2006wr}. They could also modify gravity as described by  General Relativity (GR) \cite{Khoury:2010xi}. Such scalar fields with low masses could affect the growth of structures on very large scales in the Universe. On the other hand,  in the Solar System or the laboratory, modifications of  General Relativity are tightly constrained \cite{Will:2004nx}. The compatibility between the two behaviors on large and small scales can be ascertained in screened modified gravity where environmental effects take place in the presence of matter. In this paper, we study the formation of structure in K-mouflage models \cite{Babichev:2009ee, Brax:2012jr}. The background cosmology has been analyzed in a companion paper \cite{Brax:2014aa}.

The K-mouflage mechanism is present   in scalar field theories with  noncanonical kinetic terms.  It is effective in regions of space where the gravitational acceleration is larger than a critical value \cite{Khoury:2013tda}. On large scales where matter is less dense, deviations from GR can be significant and affect the growth of structure. In particular, K-mouflage models do not converge towards GR in the large distance regime. As a result, K-mouflage models behave like a linear theory with a time dependent Newton constant up to quasilinear scales.

We study the perturbative regime of K-mouflage models before analyzing the nonlinear regime. Linear perturbation theory differs from $\Lambda$-CDM on large scales and is therefore amenable to clean comparisons with the measurements of the growth factor as forecast by the EUCLID mission \cite{Amendola:2012ys}. It turns out that the linear regime in the scalar sector gives a good description of the growth structure up to quasilinear scales owing to the relative irrelevance of nonlinear corrections in the scalar sector.  In the nonlinear regime, we use the spherical collapse to deduce the halo mass function and the deviation of the power spectrum from $\Lambda$-CDM. As expected the halo mass function is significantly affected for large masses while the power spectrum can see large deviations on nonlinear scales of order $1\ {\rm Mpc}$. This is a feature of  $f(R)$ models too \cite{Brax2013} which could be disentangled here inasmuch as it persists even for moderate redshifts in the K-mouflage case. In fact, the redshift dependence of the deviations from GR is also very different in models with the Vainshtein property such as Galileons \cite{Barreira:2013eea, Li:2013tda}, with fewer deviations at moderate redshifts than in the K-mouflage models. Hence we can expect that the three screening mechanisms could be disentangled by both analyzing the large-scale features, as chameleonlike models converge to GR contrary to K-mouflage and Vainshtein models, and the time evolution of their deviations from GR, as K-mouflage models show persistent ones up to $z=2$.

In section II, we introduce K-mouflage models, screening and the background cosmology. In section III, we analyze the perturbative regime of K-mouflage theories. In section IV, we focus on large scales and the ISW effect. In section V, we study the spherical collapse and apply these results to a halo model which allows us to tackle the cosmology of nonlinear scales. Finally in section VI, we calculate the power spectrum including  nonlinearities as defined by a halo model. In section VII, we compare our results with chameleonlike models and Galileons. We conclude in section VIII. Two appendices present the perturbation theory of K-mouflage models and a comparison of the power spectrum in K-mouflage models with the one where the only modification from $\Lambda$-CDM is due to the K-mouflage background cosmology.

\section{K-mouflage}
\label{K-mouflage}

We recall in this section the definition of the K-mouflage models that we consider and the
evolution of the cosmological background.

\subsection{Definition of the model}
\label{Definition of the model}

We consider scalar field models where the action in the
Einstein frame has the form
\beqa
S & = & \int \dd^4 x \; \sqrt{-g} \left[ \frac{\MPl^2}{2} R + {\cal L}_{\varphi}(\varphi)
\right] \nonumber \\
&& + \int \dd^4 x \; \sqrt{-\tg} \, {\cal L}_{\rm m}(\psi^{(i)}_{\rm m},\tg_{\mu\nu}) ,
\label{S-def}
\eeqa
where
 $g$ is the determinant of the metric tensor $g_{\mu\nu}$, and
$\psi^{(i)}_{\rm m}$ are various matter fields.
The scalar field $\varphi$ is explicitly coupled to matter via the
Jordan-frame metric $\tg_{\mu\nu}$, which is given by the conformal rescaling
\beq
\tg_{\mu\nu} = A^2(\varphi) \, g_{\mu\nu} ,
\label{g-Jordan-def}
\eeq
and $\tg$ is its determinant.
We have already considered various canonical scalar field models in previous
works \cite{BraxPV2012,Brax2013}, with
${\cal L}_{\varphi} = - (\pl\varphi)^2/2 - V(\varphi)$.
In this paper, we consider models with a nonstandard kinetic term
\beq
{\cal L}_{\varphi}(\varphi) = \cM^4 \, K(\chi) \;\;\; \mbox{with} \;\;\;
\chi = - \frac{1}{2\cM^4} \, \pl^{\mu}\varphi \, \pl_{\mu}\varphi .
\label{K-def}
\eeq
[We use the signature $(-,+,+,+)$ for the metric.]
The constant $\cM^4$ is an energy scale  of the order of the current energy density to retrieve the late-time accelerated expansion
of the Universe.
The canonical behavior
with a cosmological constant, $\rho_{\Lambda} = \cM^4$ [see Eq.(\ref{rho-phi-def})
below], is obtained in the weak-$\chi$ limit when:
\beq
\chi \rightarrow 0 : \;\;\; K(\chi) \simeq -1 + \chi + ... ,
\label{K-chi=0}
\eeq
where the dots stand for higher-order terms. We have chosen the minus sign of the constant $-1$
at $\chi=0$ using $\rho_{\Lambda} = - \cM^4 K(0)>0$ (for the models where the late-time
behavior corresponds to $\chi\rightarrow 0$).

The dynamics are determined by the  Klein-Gordon equation which reads now as
\beq
\frac{1}{\sqrt{-g}} \pl_{\mu} \left[ \sqrt{-g} \; \pl^{\mu} \varphi \; K' \right] -
\frac{\dd\ln A}{\dd\varphi} \; \rho_E = 0 ,
\label{KG-1}
\eeq
where $\rho_E=- g^{\mu\nu}T_{\mu\nu}$ is the Einstein-frame matter density,
and we note with a prime $K'=\dd K/\dd\chi$.

\subsection{Screening}

The suppression of the scalar field effect in dense environments due to K-mouflage is significant when the gradient of the scalar field satisfies
\be
\vert \nabla  \varphi\vert \gtrsim \cM^2
\ee
implying that a necessary condition on the Newtonian potential $\Psi_{\rm N}$ for screening in K-mouflage models is
\be
\vert \nabla \Psi_{\rm N} \vert  \gtrsim \frac{\cM^2}{2\beta M_{\rm Pl}} \label{curv}
\ee
where the coupling to matter $\beta= M_{\rm Pl} \frac{\dd\ln A}{\dd\varphi}$ is a slowly varying function of $\varphi$.
For the Newtonian potential around a dense object of mass $m$,  screening occurs inside the K-mouflage radius
\be
R_K= \left( \frac{\beta m}{4\pi  M_{\rm Pl} \cM^2} \right)^{1/2} .
\ee
For quasilinear scales in cosmology, screening appears when the wave number $k=1/x$ characterizing a given structure satisfies
\be
k\lesssim 3 \Omega_{\rm m0} A(\varphi_0) \beta(\varphi_0) \frac{H_0^2 M_{\rm Pl}}{\cM^2} \delta ,
\ee
where $\delta = (\rho-\bar\rho)/\bar\rho$ is the matter density contrast.
When $\cM^4 \sim 3 \Omega_{\Lambda 0} M_{\rm Pl}^2 H_0^2$  to reproduce the late-time cosmological constant,  we find that
\be
\frac{k}{H_0}\lesssim \sqrt{\frac{3}{\Omega_{\Lambda 0}}} \Omega_{\rm m0}  A(\varphi_0) \beta(\varphi_0) \delta
\ee
which is associated to superhorizon scales. As a result, all quasilinear objects in the Universe are unscreened in K-mouflage models.

Therefore, we expect that linear scales are maximally  affected in K-mouflage models. In particular the integrated Sachs-Wolfe effect becomes relevant,
where we have the approximation
\be
\frac{\Delta T_{\rm ISW}}{T_{\rm CMB}} \approx \frac{2}{c^2} \int_{\tau_{\rm LS}}^{\tau_0} \dd\tau \; \partial_\tau \Psi_{\rm N}
\ee
between the last scattering and now. Here $\tau=\int \dd t/a$ is the conformal time and
we have approximated the visibility function by unity after the last scattering and zero at earlier times.
Using the Poisson equation and defining by $D_+$ the matter density growing mode of linear perturbation theory, we find that on the largest scales the Integrated Sachs-Wolfe effect
behaves like
\beqa
\frac{\Delta T_{\rm ISW}}{T_{\rm CMB}} & \approx &
\frac{2}{c^2} \left[ \Psi_{\rm N}(\tau_0) - \Psi_{\rm N}(\tau_{\rm LS}) \right] \nonumber \\
& \propto & \left( \bar{A}_0 D_{+0} - \frac{\bar{A}_{\rm LS} D_{+\rm LS}}{a_{\rm LS}} \right) .
\eeqa
As a result, in K-mouflage models both the growth of structure (through $D_+$) and the background cosmology (through the variation of the function $A(\bar\varphi)$ as a function of the background field $\bar\varphi$) can lead to a significant change in the ISW effect from $\Lambda$-CDM.
We will analyze this effect in more details in Sec.~\ref{Sec-CMB}.

\subsection{Specific models}
\label{Specific-models}

We use the same three models as in the companion paper \cite{Brax:2014aa}
where we study the background
evolution for
\beqa
``\mbox{no-}\chi_* , \; K' \geq 1 " & : & \;\;   K(\chi) = -1 + \chi + K_0 \, \chi^m , \nonumber \\
&& K_0 > 0 ,  \;\; m \geq 2 ,
\label{K-power-1}
\eeqa
\beqa
``\mbox{with-}\chi_* , \; K' \leq 0 " & : & \;\;   K(\chi) = -1 + \chi + K_0 \, \chi^m , \nonumber \\
&& K_0 < 0 ,  \;\; m \geq 2 ,
\label{K-power-2}
\eeqa
and
\beq
``\mbox{with-}\chi_* , \; K' \geq 0 "  :  \;\;   K(\chi) = -1 + \chi - \chi^2 + \chi^3/4 .
\label{K-power-3}
\eeq
The first model (\ref{K-power-1}) corresponds to scenarios where $K'$ never comes across
a zero (``no-$\chi_*$'') during the background cosmological evolution and remains positive.
The second and third models correspond to scenarios where $K'$ comes across
a zero, $\chi_*>0$, at late times (in fact, at infinite time), from below [Eq.(\ref{K-power-2})]
or from above [Eq.(\ref{K-power-3})], as $\chi$ rolls down from $+\infty$.

More generally, the first two terms in Eq.(\ref{K-power-1}), $(-1+\chi)$, represent
the first-order expansion in $\chi$ of a generic function $K(\chi)$, as in
Eq.(\ref{K-chi=0}), so that we recover a canonical scalar field with a cosmological
constant for small time and spatial gradients.
The third term $K_0 \, \chi^m$ represents the large-$\chi$ behavior of the function
$K(\chi)$, or more precisely the relevant exponent at the time of interest.

As shown in the companion paper \cite{Brax:2014aa},
the models (\ref{K-power-2}) actually show ghost
instabilities. In particular, vacuum decay leads to a large production of photons
and observational constraints on the diffuse gamma ray background yield an upper bound
on the cutoff scale $\Lambda$ of the theory. This gives $\Lambda \leq 1$ keV for $m=2$
and $\Lambda \leq 4$ eV for $m=3$, hence these models are not very realistic.
Nevertheless, we include them in our study for illustration.

For the coupling function $A(\varphi)$, we again consider the simple power laws,
\beq
n \in {\mathbb N} , \;\; n \geq 1 : \;\;\;
A(\varphi) = \left( 1 + \frac{\beta\varphi}{n \MPl} \right)^n ,
\label{A-power-1}
\eeq
which include the linear case $n=1$,
and the exponential limit for $n \rightarrow +\infty$,
\beq
A(\varphi) = e^{\beta \varphi/\MPl} .
\label{A-exp-1}
\eeq
Without loss of generality, we normalized the field $\varphi$ (by the appropriate
additive constant) so that $A(0)=1$.

The action (\ref{S-def}) is invariant with respect to the transformation
$(\varphi,\beta) \rightarrow (-\varphi,-\beta)$; therefore, we can choose $\beta >0$.
Thus, in addition to the usual cosmological parameters, our system is defined by the
five parameters
\beq
\{\beta, n ; K_0 , m ; \cM^4 \} \;\; \mbox{with} \;\; \beta > 0, \; \cM^4 > 0 , \;
n \geq 1 , \; m \geq 2 ,
\label{param-def}
\eeq
except for the model (\ref{K-power-3}) where there are no parameters $\{K_0,m\}$ as the
kinetic function is fixed.
The scale $\cM$ is not an independent parameter. For a given value of the set
$\{ \beta, n ; K_0 , m \}$ and of $H_0$, it is fixed by the value of $\Omega_{\rm m0}$
today.
Thus, as in the companion paper \cite{Brax:2014aa},
in the numerical computations below, we choose
the same set of cosmological parameters today, given by the Planck results
\cite{Planck-Collaboration2013}.
Then, for any set $\{ \beta, n ; K_0 , m \}$, we tune $\cM$ to obtain the observed
dark energy density today. As noticed above, this means that
$\cM^4 \sim \bar\rho_{\rm de 0}$.

\subsection{Cosmological background}
\label{background}

We focus on the matter era and we only consider nonrelativistic pressureless matter
and the scalar field $\varphi$.
Then, as shown in the companion paper \cite{Brax:2014aa}, the Friedmann equations read
\beqa
3 \MPl^2 H^2 & = & \bar{\rho} + \bar{\rho}_{\varphi}^{\rm eff} ,
\label{Friedmann-3} \\
-2 \MPl^2 \dot{H} & = & \bar{\rho} + \bar{\rho}_{\varphi}^{\rm eff} + \bar{p}_{\varphi} ,
\label{Friedmann-4}
\eeqa
where $\rho$ is the conserved matter density, which is defined in terms of the Einstein-frame
matter density $\rho_E$ by
\beq
\rho = A^{-1} \rho_E , \;\;\; \mbox{and} \;\;\;
\dot{\bar{\rho}} = - 3 H \, \bar{\rho} \;\; \mbox{hence} \;\;
\bar\rho = \frac{\bar\rho_0}{a^3} ,
\label{rho-conserv}
\eeq
$\rho_{\varphi}$ and $p_{\varphi}$, are the scalar field energy density and pressure
(in the Einstein frame),
\beq
\bar{\rho}_{\varphi} = - \cM^4 \bar{K} + \dot{\bar\varphi}^2 \, \bar{K}'  , \;\;\;
\bar{p}_{\varphi} = \cM^4 \bar{K} ,
\label{rho-phi-def}
\eeq
and $\rho_{\varphi}^{\rm eff}$ is the effective scalar field density, defined by
\beq
\rho_{\varphi}^{\rm eff} = \rho_{\varphi} + [ A(\varphi)-1] \rho ,
\label{rho-phi-eff-def}
\eeq
which satisfies the standard conservation equation (the pressure $p_{\varphi}$
is not modified)
\beq
\dot{\bar{\rho}}_{\varphi}^{\rm eff} = - 3 H (\bar{\rho}_{\varphi}^{\rm eff}
+ \bar{p}_{\varphi} ) .
\label{conserv-1}
\eeq
Here and in the following, the overbar denotes background uniform quantities.

The Klein-Gordon equation (\ref{KG-1}) reads as
\beq
\pl_t \left( a^3 \dot{\bar\varphi} \bar{K}' \right) =
- \frac{\dd \bar{A}}{\dd\bar\varphi} \, \bar{\rho} \, a^3 ,
\label{KG-2}
\eeq
and the relevant cosmological solution is the solution of the integrated form
\beq
t \geq 0 : \;\; a^3 \dot{\bar\varphi} \bar{K}' = - \int_0^t \dd t' \; \bar\rho_0
\frac{\dd\bar{A}}{\dd\bar\varphi}(t') ,
\label{KG-int-1}
\eeq
with the boundary condition $\bar\varphi \rightarrow 0$ at $t \rightarrow 0$.
This gives the early-time power law behaviors
\beqa
t \rightarrow 0 & : & |\bar\varphi| \sim  t^{2(m-1)/(2m-1)} , \;\;
|\dot{\bar\varphi}| \sim t^{-1/(2m-1)} , \nonumber \\
&& \bar\rho_{\varphi}^{\rm eff} \sim \bar\rho_{\varphi} \sim \bar{p}_{\varphi}
\sim t^{-2m/(2m-1)} ,
\label{phi-t-0}
\eeqa
and the signs of $\bar\varphi$ and $\dot{\bar\varphi}$ are opposite to the sign of $K_0$,
\beq
t \rightarrow 0 : \;\; K_0 \dot{\bar\varphi} < 0 , \;\; K_0 \bar\varphi < 0 .
\label{K0-sign-phi}
\eeq
The solution (\ref{KG-int-1}) is an attractor tracker solution.
At early times, we recover the matter era, with
$\bar\rho_{\varphi}^{\rm eff} \ll \bar\rho$, and at late times we recover a cosmological-constant
behavior, with $\bar\rho_{\rm de} = \cM^4$ for the models (\ref{K-power-1}) and
$\bar\rho_{\rm de} = -\cM^4 K(\chi_*)$ for the models (\ref{K-power-2}) and
(\ref{K-power-3}). More precisely, for $t \rightarrow \infty$, far in the dark energy era, we obtain
the behaviors
\beq
``K' \geq 1 " : \;\; a(t) \sim e^{\cM^2 t/(\sqrt{3}M_{\rm Pl})} ,
\;\; \bar\varphi \rightarrow \mbox{constant} < 0 ,
\label{a-exp-L}
\eeq
\beq
``K' \leq 0 " : \;\; a(t) \sim e^{\sqrt{-K_*/3} \cM^2 t/M_{\rm Pl}} ,
\;\; \bar\varphi \simeq \sqrt{2 \chi_* {\cM^4}} t ,
\label{a-exp-L-neg1}
\eeq
and
\beq
``K' \geq 0 " : \;\; a(t) \sim e^{\sqrt{-K_*/3} \cM^2 t/M_{\rm Pl}} ,
\;\; \bar\varphi \simeq - \sqrt{2 \chi_* {\cM^4}} t ,
\label{a-exp-L-pos1}
\eeq
for the models (\ref{K-power-1}), (\ref{K-power-2}), and (\ref{K-power-3}).

In addition, from Eq.(\ref{KG-int-1}) and the fact that $\varphi/M_{\rm Pl}$ has not grown much
beyond unity until the current time $t_0$ (because of the BBN constraint, see the
companion paper), we have
\beq
0 \leq t \leq t_0  : \;\; \left| \frac{\beta\bar\varphi}{M_{\rm Pl}} \right| \lesssim 1
\;\;\; \mbox{and} \;\;\; \frac{\beta\bar\varphi}{M_{\rm Pl}} \sim
- \frac{\beta^2}{\bar{K}'} .
\label{beta-phi-1}
\eeq
This sets a condition on the parameters of the model. In particular, $\beta$ is typically less
than unity, except for models (\ref{K-power-1}) with a large value of $K_0$ that implies that
$\bar{K}'$ has remained large until today.

\section{Formation of large-scale structures}
\label{Formation-structures}

\subsection{Equations of motion}
\label{motion-large-scale}

To derive the continuity and Euler equations that govern the matter dynamics on large
scales, it is convenient to work in the Einstein frame, where the energy-momentum
tensor obeys
\beq
D_{\mu} T^{\mu\nu}_{(\rm m)} = - \rho_E \, \pl^{\nu} \ln A .
\label{conserv-T-1}
\eeq
[This follows from the conservation of the total energy-momentum tensor
$T^{\mu\nu}_{(\rm m)} + T^{\mu\nu}_{(\varphi)}$ of the sum of matter and
scalar field components,
using the Klein-Gordon equation (\ref{KG-1}) to simplify the part
$D_{\mu} T^{\mu\nu}_{(\varphi)}$, which gives rise to the right-hand side in
Eq.(\ref{conserv-T-1}).]
Then, considering perturbations in the conformal Newtonian gauge,
\beq
\dd s^2 = a^2 [ - (1+2 \Psi_{\rm N}) \dd\tau^2 + (1-2 \Psi_{\rm N}) \dd \vx^2 ] ,
\label{conform-Psi-1}
\eeq
where $\Psi_{\rm N}$ is Newton's gravitational potential and $\tau = \int \dd t/a$,
Eq.(\ref{conserv-T-1}) leads to
\beq
\frac{\pl \rho_E}{\pl\tau} + \nabla \cdot  ( \rho_E \vv ) + 3 {\cal H} \rho_E = \rho_E
\frac{\pl\ln A}{\pl\tau} ,
\label{contE-1}
\eeq
and
\beq
\frac{\pl \vv}{\pl\tau} + (\vv\cdot\nabla) \vv + \left( {\cal H}
+ \frac{\pl\ln A}{\pl\tau} \right) \vv = - \nabla \left( \Psi_{\rm N} + \ln A \right) ,
\label{Euler-1}
\eeq
where ${\cal H} = (\dd a/\dd\tau)/a = aH$ is the conformal expansion rate
and $\vv=\dd\vx/\dd\tau=a\dot{\vx}$ is the peculiar velocity.
Here and in the following, we work in the nonrelativistic ($v\ll c$) and weak field
($\Psi_{\rm N} \ll c^2$) limit.
The continuity equation (\ref{contE-1}) is obtained by contracting
Eq.(\ref{conserv-T-1}) with $u_{\nu}$ or from its $\nu=0$ component.
The Euler equation (\ref{Euler-1}) is obtained from the spatial $\nu=i$ components
of Eq.(\ref{conserv-T-1}), using Eq.(\ref{contE-1}) to simplify some terms.
In these calculations, as detailed in App.~\ref{appendix-Euler},
we encounter terms of the form $\vv\nabla\ln A$, which are
of order $(v^2)' \sim \nabla v^3 \sim v \nabla\Psi_{\rm N}$ as seen from the Euler
equation (\ref{Euler-1}). Therefore, they are negligible in the nonrelativistic and
weak field limit and must be dropped (as other terms such as $\nabla v^3$ or
$v \nabla\Psi_{\rm N}$). We omit them in Eqs.(\ref{contE-1})-(\ref{Euler-1}) and in
the following.
In terms of the density $\rho=A^{-1}\rho_E$ introduced in Eq.(\ref{rho-conserv}),
the continuity equation (\ref{contE-1}) reads, as in the usual $\Lambda$-CDM case,
as
\beq
\frac{\pl \rho}{\pl\tau} + \nabla \cdot  ( \rho \vv ) + 3 {\cal H} \rho = 0 ,
\label{cont-2}
\eeq
[where we used again the fact that $\vv\nabla\ln A$ is a higher-order correction
$\sim (v^2)'$].
Thus, the density $\rho$ is still conserved by the velocity flow, including peculiar
velocities, and it corresponds to the physical matter density

Next, from the Einstein equations we obtain the Poisson equation,
\beq
\frac{1}{a^2} \nabla^2 \Psi_{\rm N} = 4\pi {\cal G} (\delta\rho_E + \delta\rho_{\varphi})
= 4\pi {\cal G} (\delta\rho + \delta\rho_{\varphi}^{\rm eff}) ,
\label{Poisson-1}
\eeq
where $\delta\rho_i=\rho_i-\bar\rho_i$ are the density fluctuations. The
scalar field density $\rho_{\varphi}$ is still given as in Eq.(\ref{rho-phi-def}) by
\beq
\rho_{\varphi} = - \cM^4 K + \left( \frac{\pl\varphi}{\pl t} \right)^2 \, K'  ,
\label{rho-phi-def-2}
\eeq
and the effective scalar field density $\rho^{\rm eff}_{\varphi}$ by
Eq.(\ref{rho-phi-eff-def}).

The dynamics of the scalar field are given by the Klein-Gordon equation
(\ref{KG-1}), which reads as
\beq
\frac{1}{a^3} \frac{\pl}{\pl t} \left( a^3 \frac{\pl\varphi}{\pl t} \, K' \right)
- \frac{1}{a^2} \nabla \cdot \left( \nabla\varphi \, K' \right)
= - \frac{\dd A}{\dd\varphi} \, \rho ,
\label{KG-pert-1}
\eeq
while the argument $\chi$ of the kinetic function $K$ reads as
\beq
\chi = \frac{1}{2\cM^4} \left[ \left( \frac{\pl\varphi}{\pl t} \right)^2
- \frac{1}{a^2} ( \nabla\varphi)^2 \right] ,
\label{chi-phi-1}
\eeq
where we again used the weak field limit $\Psi_{\rm N} \ll c^2$.

\subsection{Small-scale (subhorizon) limit for the scalar field perturbations}
\label{Small-scale}

In contrast with most modified-gravity models, such as chameleon scenarios
\cite{Brax2013}, we cannot use the quasistatic approximation in its usual
form. In such models, the scalar field Lagrangian ${\cal L}_\varphi$ also contains
a potential $V(\varphi)$ and the Klein-Gordon equation (\ref{KG-pert-1})
contains an additional term, $- \dd V/\dd\varphi$, on the right-hand side.
Then, the quasistatic approximation assumes that the background field $\bar\varphi$
follows the minimum $\bar\varphi_*(t)$ of the effective potential
$V_{\rm eff}(\bar\varphi)= V(\bar\varphi)+\bar\rho(t) A(\bar\varphi)$.
In the same spirit, the Klein-Gordon equation of the inhomogeneous field
$\varphi=\bar\varphi+\delta\varphi$ is approximated as
$a^{-2} \nabla\cdot(\nabla\varphi K') = \dd V/\dd\varphi + \rho \, \dd A/\dd\varphi$.
This quasistatic approximation is valid provided $m c t \gg 1$, where $m^2 c^2 =
\pl^2V_{\rm eff}/\pl\varphi^2$, or one considers small scales $k c t \gg 1$,
and various functions such as the coupling function $A$ or the potential $V$
are smooth enough; see also the discussion in Sec.II.B.4 in \cite{Brax2013}.
These conditions are usually met by these models (e.g., because observational
constraints from the Solar System imply a large $m$).

In the models studied in this paper, there is no potential $V$ and the background
field $\bar\varphi$ cannot be given by a quasistatic approximation, where
time derivatives can be neglected.
In contrast, it is governed by the evolution equation (\ref{KG-int-1}). Nevertheless,
we can still consider a small-scale limit for the fluctuations of the scalar field.
Thus, within a perturbative approach, we write the field $\varphi$ and the
kinetic variable $\chi$ as
\beq
\varphi = \bar\varphi + \delta\varphi , \;\;\; \chi = \bar\chi+\delta\chi ,
\label{dphi-dchi-def}
\eeq
with
\beq
\bar\chi = \frac{\dot{\bar\varphi}^2}{2\cM^4}
\;\;\; \mbox{and} \;\;\; \delta\chi = - \frac{(\nabla\delta\varphi)^2}{2\cM^4 a^2} .
\label{dchi-def}
\eeq
This means that we neglect time derivatives of the scalar field fluctuations, in
the small-scale or subhorizon regime
\beq
c t k/a \gg 1.
\label{small-scale}
\eeq
In other words, as we expand the Klein-Gordon equation (\ref{KG-pert-1}) in
powers of $\delta\varphi$, for each order in $\delta\varphi$ we only keep the terms
with the highest power of $k$, that is, with the highest order of spatial derivatives.
Then, terms that would arise from the product $\dot{\bar\varphi}\pl\delta\varphi/\pl t$
in $\delta\chi$ will always be subdominant with respect to those that arise from
the product $(\nabla\delta\varphi)^2$. Therefore, we can simplify the analysis
by removing these subdominant terms from the start, by only keeping the
contribution $(\nabla\delta\varphi)^2$ in Eq.(\ref{dchi-def}).
In this small-scale approximation, the Klein-Gordon equation (\ref{KG-pert-1})
also simplifies as
\beq
\frac{1}{a^3} \frac{\pl}{\pl t} \left( a^3 \dot{\bar\varphi} \bar{K}' \right)
- \frac{1}{a^2} \nabla \cdot \left( \nabla\delta\varphi \, K' \right) =
- \frac{\dd \bar{A}}{\dd\bar\varphi} \, (\bar\rho +\delta\rho) ,
\label{KG-pert-2}
\eeq
where $\delta\rho$ is the matter density fluctuation.
Indeed, the time-derivative factor gives rise to terms of the form
$k^{2n}(\delta\varphi)^{2n}$, the spatial-derivative factor to
$k^{2n+2}(\delta\varphi)^{(2n+1)}$, and fluctuations of $A$ to $(\delta\varphi)^{n}$,
where powers of $k$ count the order of spatial derivatives.
Thus, terms with the highest power of $k$ per $\delta\varphi$ arise from the spatial
term. Therefore, we can remove subdominant terms by using the background
values $\bar{K}'$ in the first term in Eq.(\ref{KG-pert-2}) and
$\dd\bar{A}/\dd\bar\varphi$ on the right-hand side.
Subtracting the background solution (\ref{KG-2}) we obtain the small-scale
Klein-Gordon equation for the fluctuations of the scalar field
\beq
\frac{1}{a^2} \nabla \cdot \left( \nabla\delta\varphi \, K' \right) =
 \frac{\bar{A}\beta_1}{\MPl} \, \delta\rho .
\label{KG-pert-3}
\eeq
Here we have introduced the dimensionless coefficients
\beq
\beta_n(t) = \MPl^n \, \frac{\dd^n\ln A}{\dd\varphi^n}(\bar\varphi) .
\label{beta-bar-def}
\eeq
This equation can be inverted to give the functional $\delta\varphi[\delta\rho]$ by using
a perturbative approach.
Thus, we write the expansion in powers of the nonlinear density fluctuation,
\beq
\delta\varphi = \sum_{n=1}^{\infty} \delta\varphi^{(n)} \;\;\; \mbox{with}
\;\;\; \delta\varphi^{(n)} \propto (\delta\rho)^n ,
\label{phin-deltan-def}
\eeq
which reads in Fourier space (which we denote with a tilde),
\beqa
\delta\tilde{\varphi}(\vk) & = & \sum_{n=1}^{\infty} \int \dd\vk_1 \dots \dd\vk_n \;
\delta_D(\vk_1+\dots+\vk_n-\vk) \nonumber \\
&& \times \; h_n(\vk_1, \dots ,\vk_n) \, \delta\tilde{\rho}(\vk_1) \dots
\delta\tilde{\rho}(\vk_n) .
\label{phi-delta-hn-def}
\eeqa
Because $\delta\chi$ is quadratic in $\delta\varphi$, the left-hand side in
Eq.(\ref{KG-pert-3}) is odd in $\delta\varphi$ while the right-hand side is odd
over $\delta\rho$. Therefore, we have the symmetry
$(\delta\varphi,\delta\rho) \rightarrow (-\delta\varphi,-\delta\rho)$ and all even orders
in the expansion (\ref{phi-delta-hn-def}) vanish.

One can obtain the same results by keeping all terms in Eq.(\ref{KG-pert-1}),
including terms such as $\pl\delta\varphi/\pl t$, and looking for the leading terms in
the final expressions for each order $\delta\varphi^{(n)}$.
More precisely, from the scalings $k^{2n}(\delta\varphi)^{2n}$ and
$k^{2n+2}(\delta\varphi)^{(2n+1)}$ of the terms that arise from the time- and
space-derivatives of the Klein-Gordon equation (\ref{KG-pert-1}), one finds
$\delta\varphi^{(2n+1)} \sim k^{-2n-2}$ and
$\delta\varphi^{(2n)} \sim k^{-2n-2}$. Then, the even orders are subdominant
with respect to the scaling $\delta\varphi^{(n)} \sim k^{-n-1}$ satisfied by the odd
orders. This agrees with the result that even orders vanish in the
small-scale approximation (\ref{KG-pert-3}).

Introducing the coefficients
\beq
\kappa_n(t) = \frac{\dd^n K}{\dd\chi^n}(\bar\chi) , \;\;\;
\mbox{and hence} \;\;\; \bar{K}'=\kappa_1 ,
\label{kappan-def}
\eeq
we obtain at first order
\beq
\frac{\kappa_1}{a^2} \nabla^2 \delta\varphi^{(1)} = \frac{\bar{A}\beta_1}{\MPl} \,
\delta\rho , \;\;\;
h_1(k) =  \frac{-\bar{A}\beta_1 a^2}{\kappa_1 \MPl k^2} .
\label{h1-def}
\eeq
As explained above, the second order vanishes, $h_2=0$.
At third order, we obtain
\beqa
\frac{\kappa_1}{a^2} \nabla^2 \cdot \delta\varphi^{(3)} & = &
\frac{\kappa_2}{2\cM^4a^4} \biggl \lbrace \left[ \nabla \cdot \delta\varphi^{(1)} \right]^2 \,
\nabla^2 \cdot \delta\varphi^{(1)} \nonumber \\
&& + 2 \frac{\pl\delta\varphi^{(1)}}{\pl x_i}
\frac{\pl\delta\varphi^{(1)}}{\pl x_j} \frac{\pl^2\delta\varphi^{(1)}}{\pl x_i\pl x_j}
\biggl \rbrace ,
\label{phi3-def}
\eeqa
hence
\beqa
h_3(\vk_1,\vk_2,\vk_3) & = & \frac{\kappa_2 \bar{A}^3\beta_1^3 a^4}
{2 \kappa_1^4 \cM^4 \MPl^3} \nonumber \\
&& \hspace{-1cm} \times
\frac{(\vk_1\cdot\vk_2) k_3^2 + 2 (\vk_1\cdot\vk_3) (\vk_2\cdot\vk_3)}
{k^2 k_1^2 k_2^2 k_3^2} .
\label{h3-def}
\eeqa
Thus, even though the background field $\bar\varphi$ follows the dynamical
equation (\ref{KG-2}), the fluctuations $\delta\varphi$ are described by a quasistatic
approximation as time derivatives no longer appear in Eq.(\ref{KG-pert-3}).
This is possible because in the small-scale regime (\ref{small-scale}) fluctuations
have enough time to relax, on a local time scale $t_k \sim a/(ck)$ that is much shorter
than the Hubble time.

We can note that the quasistatic Klein-Gordon equation (\ref{KG-pert-3}) only defines
$\delta\varphi$ up to an additive constant $\delta\varphi_0$, because it only depends
on gradients of $\delta\varphi$.
We set this constant to zero as we require the scalar field fluctuations to vanish when
the density fluctuations vanish.
For a periodic density field, we can look for a periodic solution $\delta\varphi$
(thanks to the constraint $\overline{\delta\rho}=0$). This implies that at all orders
$n \geq 2$ the symmetrized kernels $h_n^s(\vk_1,..,\vk_n)$ vanish in the
limit where $\vk=\vk_1+..+\vk_n$ goes to zero while the individual wave numbers
$\{\vk_1,..,\vk_n\}$ remain finite.
[This is not obvious in Eq.(\ref{h3-def}) but one can check that the symmetrized
kernel $h_3^s=1/6 \sum_{\rm perm.} h_3(\vk_1,\vk_2,\vk_3)$ verifies
$h_3^s(\vk_1,\vk_2,\vk_3)=0$ when $\vk_1+\vk_2+\vk_3=0$.]

The results (\ref{h1-def}) and (\ref{h3-def}) only apply in the small-scale limit
(\ref{small-scale}). In particular, the divergences at $k_i\rightarrow 0$ are
not physical as these expressions are no longer valid for wave numbers below
$a/(ct)$.

\subsection{Perturbation theory}
\label{Perturbation-theory}

\subsubsection{Fifth force and Newtonian gravity}
\label{Fifth-force}

We have described in the previous section how the Klein-Gordon equation
(\ref{KG-pert-1}) can be solved for $\varphi[\rho]$, through the perturbative expansion
(\ref{phi-delta-hn-def}) in the density perturbation $\delta\rho$.
In a second step, this allows us to solve the dynamics of the system in a perturbative
manner, where we now expand in powers of the linear density field as in
the standard $\Lambda$-CDM perturbation theory.
This approach is identical to the method used in other modified-gravity models
with a canonical kinetic term, such as chameleon scenarios, or $f(R)$ models
\cite{Brax2013}.
To recover a framework that is similar to the standard GR case, we simply need
to express the Newtonian gravitational potential $\Psi_{\rm N}$ and the
coupling term $\ln A$ in Eq.(\ref{Euler-1}) in terms of the density field.
Then, the Euler and continuity equations (\ref{Euler-1})-(\ref{cont-2})
form a closed system in the fields $(\rho,\vv)$ that we can solve in a perturbative
manner. The difference with the standard GR case is that the nonlinearity is no longer
quadratic but includes vertices of all orders.

Using Eq.(\ref{h1-def}), we obtain from the linearized Klein-Gordon equation
\beq
\delta\varphi \sim \frac{\bar{A}\beta_1 a^2}{M_{\rm Pl} \kappa_1 k^2}
\, \delta\rho .
\label{dphi-drho-1}
\eeq
Then, from the background value (\ref{beta-phi-1}), we obtain in the small-scale
regime the relative field fluctuation
\beq
c t k/a \gg 1 : \;\;\; \frac{\delta\varphi}{\bar\varphi} \sim
\frac{a^2}{c^2t^2k^2} \, \frac{\delta\rho}{\bar\rho}
\ll \frac{\delta\rho}{\bar\rho} .
\label{dphi-drho-2}
\eeq
From the first property (\ref{beta-phi-1}) we also have $\bar{A} \simeq 1$,
$\beta_1 \simeq \beta$, $\delta A \sim \beta \delta\varphi/M_{\rm Pl}$,
and
\beq
ctk/a \gg 1: \;\;\; \delta\ln A \sim \frac{\delta A}{\bar{A}}
\sim \frac{\beta^2 a^2}{\bar{K}' c^2 t^2 k^2} \; \frac{\delta\rho}{\bar\rho}
\ll \frac{\delta\rho}{\bar\rho}  .
\label{dA-drho}
\eeq
In a similar fashion, the fluctuation $\delta\chi$ scales from Eq.(\ref{dchi-def})
as
\beq
ctk/a \gg 1: \;\;\; \frac{\delta\chi}{\bar\chi}
\sim \frac{a^2}{c^2t^2k^2} \left( \frac{\delta\rho}{\bar\rho} \right)^2
\ll \left( \frac{\delta\rho}{\bar\rho} \right)^2 .
\label{dchi-drho}
\eeq

The Newtonian gravitational potential $\Psi_{\rm N}$ is given by the modified
Poisson equation (\ref{Poisson-1}).
From Eqs.(\ref{rho-phi-def-2}) and (\ref{dphi-drho-2})-(\ref{dchi-drho}),
we have $\delta\rho_{\varphi} \ll \delta \rho$ in the small-scale regime, for moderate
density fluctuations, and the Poisson equation (\ref{Poisson-1}) simplifies as
\beq
ctk/a \gg 1: \;\;\;
\frac{1}{a^2} \nabla^2 \Psi_{\rm N} = 4\pi {\cal G} \bar{A} \delta\rho .
\label{Poisson-2}
\eeq
This is similar to the usual Poisson equation, with a linear dependence on the
density field fluctuations, but with a time dependent effective Newton constant
$\bar{A}(t) {\cal G}$.

From the coefficients (\ref{beta-bar-def}), the fifth-force gravitational potential
that enters the Euler equation (\ref{Euler-1}) reads as
\beq
\Psi_{\rm A} \equiv \ln A - \ln \bar{A} = \sum_{n=1}^{\infty} \frac{\beta_n}{\MPl^n n!}
\, (\delta\varphi)^n .
\label{PsiA-def}
\eeq
Substituting the expansion (\ref{phi-delta-hn-def}), we obtain the expansion of
$\Psi_{\rm A}$ in $\delta\rho$.
Together with Eq.(\ref{Poisson-2}), this provides the expression of the total
potential, $\Psi=\Psi_{\rm N}+\Psi_{\rm A}$, as a function of the matter density,
\beqa
\tilde{\Psi}(\vk) \!\! & = & \!\! \tilde{\Psi}_{\rm N} \! + \! \tilde{\Psi}_{\rm A} =
\sum_{n=1}^{\infty} \int \dd\vk_1 .. \dd\vk_n \;
\delta_D(\vk_1+..+\vk_n-\vk) \nonumber \\
&& \times \; H_n(\vk_1, .. ,\vk_n) \, \delta\tilde{\rho}(\vk_1) .. \delta\tilde{\rho}(\vk_n) .
\label{Psi-delta-Hn-def}
\eeqa
The first-order term reads as
\beq
H_1(k) = \frac{-a^2 \bar{A}}{2\MPl^2 k^2} \left( 1 + \frac{2\beta_1^2}{\kappa_1}
\right) .
\label{H1-def}
\eeq
Thus, at linear order the fifth force amplifies the Newtonian force by
a scale-independent factor, $\bar{A}(1+2\beta_1^2/\kappa_1)$.
The second-order term $H_2$ scales as $k^{-4}$ and as the even-order terms
of the expansion (\ref{phi-delta-hn-def}) it is subdominant, so that
$H_2=0$ in the small-scale regime.
The third-order term reads as
\beqa
H_3(\vk_1,\vk_2,\vk_3) & = & \frac{\kappa_2 \bar{A}^3\beta_1^4 a^4}
{2 \kappa_1^4 \cM^4 \MPl^4} \nonumber \\
&& \hspace{-1cm} \times
\frac{(\vk_1\cdot\vk_2) k_3^2 + 2 (\vk_1\cdot\vk_3) (\vk_2\cdot\vk_3)}
{k^2 k_1^2 k_2^2 k_3^2} ,
\label{H3-def}
\eeqa
which scales as $k^{-4}$.
The small-scale regime that we consider here, in the expressions
(\ref{H1-def})-(\ref{H3-def}) and $H_2=0$, applies to density fluctuations that are
not too large.
More precisely, at fixed nonlinear density contrast $\delta\rho/\rhob$, in the limit
$k\rightarrow \infty$ all higher-order terms are negligible and
the potential $\Psi$ becomes linear in $\delta\rho$.
Then, at a given small scale $ctk/a\gg 1$, as we let the density contrast go to
infinity the higher-order terms become relevant and as noticed above
odd-order terms first appear while even orders remain negligible, until
we further increase $\delta\rho$.
Thus, the expressions (\ref{H1-def})-(\ref{H3-def}) and $H_2=0$ apply in the
two regimes
\beq
\frac{ctk}{a} \gg 1, \;\; \frac{\delta\rho}{\bar\rho} \ll \frac{ctk}{a} : \;\;\;
\delta\varphi \simeq \delta\varphi^{(1)} , \;\;
\delta\Psi_{\rm A} \simeq \delta\Psi_{\rm A}^{(1)} ,
\label{psiA-1}
\eeq
and
\beqa
\frac{ctk}{a} \gg 1, \;\; \frac{ctk}{a} \lesssim \frac{\delta\rho}{\bar\rho} \ll
\left(\frac{ctk}{a}\right)^2 & : & \nonumber \\
&& \hspace{-3cm} \delta\varphi^{(3)} \gtrsim \delta\varphi^{(1)} , \;\;
\delta\varphi^{(2)} \ll \delta\varphi^{(1)} , \;\; \nonumber \\
&& \hspace{-3cm} \delta\Psi_{\rm A}^{(3)} \gtrsim \delta\Psi_{\rm A}^{(1)} , \;\;
\delta\Psi_{\rm A}^{(2)} \ll \delta\Psi_{\rm A}^{(1)} .
\label{psiA-2}
\eeqa
Thus, in the first regime (\ref{psiA-1}), associated with the small-scale limit
and moderate density fluctuations, the Klein-Gordon equation can be linearized
in $\varphi$ and the fifth force is actually linear in density fluctuations.
It is also scale independent in the sense that it multiplies the usual Newtonian
force by a scale-independent factor that only depends on time, as seen in
Eq.(\ref{H1-def}). This is the regime that applies to cosmological perturbation
theory, although for completeness we will include the higher-order correction
$\delta\Psi_{\rm A}^{(3)}$ below.
More precisely, by including the cubic term $\delta\Psi_{\rm A}^{(3)}$, we include
the first correction that appears as density fluctuations get large and reach
$\delta\rho/\bar\rho \sim ctk/a$. As density fluctuations further increase, we can
no longer truncate the expansions (\ref{phi-delta-hn-def}) and
(\ref{Psi-delta-Hn-def}), as odd terms $\delta\varphi^{(2n+1)}$ become of the same
order of magnitude or larger than $\delta\varphi^{(1)}$. Nevertheless, the
even-order terms remain negligible until $\delta\rho/\bar\rho \sim (ctk/a)^2$.

In this paper, we focus on cosmological perturbations with
$\delta\rho/\bar\rho \lesssim 200$. This corresponds to the perturbative regime for
density fluctuations and to collapsed halos down to the virial radius.
Then, we are in the regime (\ref{psiA-1}) where the small-scale limit and the
perturbative expansions (\ref{phi-delta-hn-def}) and (\ref{Psi-delta-Hn-def}) apply.
For studies of the Solar System (or inner regions of galaxies), we should go beyond
these expansions as density fluctuations become large enough to generate large
scalar field fluctuations. Then, the nonlinearities of the Klein-Gordon equation become
highly relevant and give rise to the K-mouflage mechanism, which ensures
a convergence back to General Relativity.

As noticed in Sec.III.B of the companion paper \cite{Brax:2014aa},
this high-density quasistatic regime corresponds to the
domain $\chi<0$ of the kinetic function $K(\chi)$, whereas cosmological perturbations
around the background only probe the domain around $\bar\chi>0$.
Therefore, the analysis of these two classes of phenomena can be treated in an independent
fashion, especially if we allow for general nonpolynomial functions $K(\chi)$ where
the behaviors at $\chi\rightarrow +\infty$ and $\chi\rightarrow -\infty$ are different
(e.g., with different power law exponents).
We leave the study of this nonlinear screening regime to a future paper.

\subsubsection{Closed system for density and velocity fields}
\label{Closed-system}

In the single-stream approximation, the formation of large-scale structures is
governed by the continuity and Euler equations (\ref{cont-2}) and
(\ref{Euler-1}). Using Eq.(\ref{dA-drho}), we can neglect the fluctuations
of $A$ in Eq.(\ref{Euler-1}), and the equations of motion read as
\beq
\frac{\pl \delta}{\pl\tau} + \nabla \cdot [ (1+\delta) \vv ) = 0 ,
\label{cont-3}
\eeq
\beq
\frac{\pl \vv}{\pl\tau} + (\vv\cdot\nabla) \vv + \left( {\cal H}
+ \frac{\dd\ln \bar{A}}{\dd\tau} \right) \vv = - \nabla \Psi ,
\label{Euler-3}
\eeq
where $\delta=\delta\rho/\bar\rho$ is the matter density contrast and $\Psi$ is
the total potential defined in Eq.(\ref{Psi-delta-Hn-def}).
Following the formalism described in \cite{Brax2013}, we can now solve the
dynamics in a perturbative approach in powers of the linear density contrast
$\delta_L$.
Introducing the time variable $\eta=\ln a$ and the two-component vector $\psi$,
\beq
\psi \equiv \left(\bea{c} \psi_1 \\ \psi_2 \ea \right) \equiv
\left( \bea{c} \delta \\ -(\nabla\cdot\vv)/\dot{a} \ea \right) ,
\label{psi-def}
\eeq
Eqs.(\ref{cont-3})-(\ref{Euler-3}) read in Fourier space as
\beqa
\frac{\pl\tpsi_1}{\pl\eta} - \tpsi_2 & = & \int \dd\vk_1\dd\vk_2 \;
\delta_D(\vk_1\!+\!\vk_2\!-\!\vk) \hat{\alpha}(\vk_1,\vk_2) \nonumber \\
&& \times \; \tpsi_2(\vk_1) \tpsi_1(\vk_2) ,
\label{continuity-2}
\eeqa
\beqa
\hspace{-0.5cm} \frac{\pl\tpsi_2}{\pl\eta} + \frac{k^2}{a^2H^2} \, \tilde{\Psi}
+ \left( \frac{1-3 w_{\varphi}^{\rm eff} \Omega_{\varphi}^{\rm eff}}{2}
+ \frac{\dd \ln \bar{A}}{\dd\eta} \right) \, \tpsi_2 & =  & \nonumber \\
&& \hspace{-7.5cm} \int\!\! \dd\vk_1\dd\vk_2 \; \delta_D(\vk_1\!+\!\vk_2\!-\!\vk)
\hat{\beta}(\vk_1,\vk_2) \tpsi_2(\vk_1) \tpsi_2(\vk_2) ,
\label{Euler-2}
\eeqa
with
\beq
\hat{\alpha}(\vk_1,\vk_2)= \frac{(\vk_1\!+\!\vk_2)\cdot\vk_1}{k_1^2} ,
\hat{\beta}(\vk_1,\vk_2)= \frac{|\vk_1\!+\!\vk_2|^2(\vk_1\!\cdot\!\vk_2)}{2k_1^2k_2^2} .
\label{alpha-beta-def}
\eeq
In the standard $\Lambda$-CDM cosmology, where the Newtonian gravitational
potential is linear in the density field, the continuity and Euler equations
(\ref{continuity-2})-(\ref{Euler-2}) are quadratic. In modified gravity models, such as
those studied in this paper, the potential $\Psi$ is nonlinear and contains terms of all
orders in $\delta\rho$. Therefore, we must introduce vertices of all orders and we write
Eqs.(\ref{continuity-2})-(\ref{Euler-2}) under the more concise form
\beq
\cO(x,x') \cdot \tpsi(x') = \sum_{n=2}^{\infty} K_n^s(x;x_1,..,x_n) \cdot
\tpsi(x_1) \dots \tpsi(x_n) ,
\label{O-Ks-def}
\eeq
where we have introduced the coordinates $x=(\vk,\eta,i)$, $i=1,2$ is the discrete
index of the two-component vector $\tpsi$, and repeated coordinates are integrated
over. The matrix $\cO$ reads as
\beqa
\hspace{-0.3cm} \cO(x,x') & \! = \! & \delta_D(\eta' \!-\! \eta) \, \delta_D(\vk' \!-\! \vk)
\nonumber \\
&& \hspace{-1.cm} \times \left( \bea{cc} \frac{\pl}{\pl\eta} & -1 \\  & \\
- \frac{3}{2} \Om (1\!+\!\epsilon_1) \;\; & \;\;  \frac{\pl}{\pl\eta} \!+\!
\frac{1\!-\! 3 w_{\varphi}^{\rm eff} \Omega_{\varphi}^{\rm eff}}{2} \!+\! \epsilon_2 \ea
\right) ,
\label{O-mod}
\eeqa
where $\epsilon_1(\eta)$ and $\epsilon_2(\eta)$ are given by
\beq
\epsilon_1(\eta) = \bar{A}-1+\frac{2\bar{A}\beta_1^2}{\kappa_1} , \;\;\;
\epsilon_2(\eta) =  \frac{\dd \ln \bar{A}}{\dd\eta} = \frac{\beta_1}{M_{\rm Pl}}
\frac{\dd\bar\varphi}{\dd\eta} .
\label{epsilon-def}
\eeq
These scale-independent functions of time measure the deviations of the equations
of motion at the linear level from the case of a $\Lambda$-CDM or uniform dark energy
scenario.
The function $\epsilon_1(\eta)$ is obtained from the first-order kernel of the
potential $\Psi$ as $1+\epsilon_1 = - 2 \MPl^2 a^{-2} k^2 \, H_1(k,\eta)$, using
Eq.(\ref{H1-def}).
The vertices $K_n^s$ are equal-time vertices of the form
\beqa
K_n^s(x;x_1,..,x_n) & = & \delta_D(\eta_1 \!-\! \eta) .. \delta_D(\eta_n \!-\! \eta)
\nonumber \\
&& \hspace{-2.5cm} \times \; \delta_D(\vk_1 \!+..+\! \vk_n \!-\! \vk) \;
\gamma_{i;i_1,..,i_n}^s(\vk_1,..,\vk_n;\eta) . \;\;\;
\label{Ks-def}
\eeqa
The nonzero vertices are the usual $\Lambda$-CDM ones,
\beq
\gamma_{1;1,2}^s(\vk_1,\vk_2) = \frac{\hat{\alpha}(\vk_2,\vk_1)}{2} , \;\;
\gamma_{1;2,1}^s(\vk_1,\vk_2) = \frac{\hat{\alpha}(\vk_1,\vk_2)}{2} , \nonumber
\eeq
\beq
\gamma_{2;2,2}^s(\vk_1,\vk_2) = \hat{\beta}(\vk_1,\vk_2) ,
\label{gamma-222}
\eeq
which are of order $n=2$ and do not depend on time, and the new vertices
associated with the modified gravitational potential (\ref{Psi-delta-Hn-def}),
\beqa
n \geq 2: \;\;\; \gamma_{2;1,..,1}^s(\vk_1,..,\vk_n;\eta) & = & -
\frac{k^2 \bar\rho^n}{a^2H^2 n!} \nonumber \\
&& \hspace{-1.5cm} \times \sum_{\rm perm.} H_n(\vk_1,..,\vk_n;\eta) , \;\;\;\;
\eeqa
where we sum over all permutations of $\{\vk_1,..,\vk_n\}$ to obtain symmetrized
kernels $\gamma^s$.
The first few vertices are given by
\beq
\gamma_{2;1,1}^s(\vk_1,\vk_2) = 0 ,
\label{gamma2}
\eeq
\beqa
\gamma_{2;1,1,1}^s(\vk_1,\vk_2,\vk_3) & = &
- \frac{9 \kappa_2 \bar{A}^3\beta_1^4 \Om^3 M_{\rm Pl}^2 H^4 a^2}
{2 \kappa_1^4 \cM^4 c^2} \nonumber \\
&& \hspace{-3cm} \times
\frac{(\vk_1\cdot\vk_2) k_3^2 + 2 (\vk_1\cdot\vk_3) (\vk_2\cdot\vk_3)}
{k_1^2 k_2^2 k_3^2} + 2 \mbox{perm.} , \;\;\;\;\;
\label{gamma3}
\eeqa
where ``2 perm.''  stands for two terms obtained by circular permutations
over $\{\vk_1,\vk_2,\vk_3\}$.

From the analysis in Sec.~\ref{Small-scale}, we can check that at all orders the
vertices decay as $k^2$ at low $k$,
\beq
n \geq 2 , \;\;\; k \rightarrow 0 : \;\;\;  \gamma^s_{2;1,..,1}(\vk_1,..,\vk_n) \sim k^2 ,
\label{gamma-n-k0}
\eeq
where the limit is taken by letting the sum $\vk=\vk_1+..+\vk_n$ go to zero while the
individual wave numbers $\{\vk_1,..,\vk_n\}$ remain finite
\footnote{As for the usual Newtonian gravity, this means that if the initial conditions
had very little power on large scales (i.e., the linear power spectrum $P_L(k)$ would
decay faster than $k^4$ at low $k$), nonlinearities would only generate a $k^4$ tail
at low $k$.}.

\subsubsection{Linear regime}
\label{linear}

On large scales, where density fluctuations are small, we can linearize the equation
of motion (\ref{O-Ks-def}). This gives ${\cal O}\cdot \tpsi_L=0$, where the subscript
``L'' denotes the linear solutions. Then, the density contrast linear modes
$D_{\pm}(\eta)$ are given by
\beq
\frac{\dd^2 D}{\dd\eta^2} + \left(
\frac{1-3w_{\varphi}^{\rm eff}\Omega_{\varphi}^{\rm eff}}{2} + \epsilon_2 \right)
\frac{\dd D}{\dd\eta} - \frac{3}{2} \Om (1+\epsilon_1) D = 0 .
\label{D-linear}
\eeq
In terms of the background quantities, we have
\beq
\left| \epsilon_1 \right| = \left| \bar{A}-1+\frac{2\bar{A}\beta_1^2}{\kappa_1} \right| \sim \left|
\frac{\beta^2}{\bar{K}'}  \right| ,
\label{eps1-1}
\eeq
where we used the property (\ref{beta-phi-1}),
and
\beq
\epsilon_2= \frac{\beta_1}{M_{\rm Pl}} \frac{\dd\bar\varphi}{\dd\eta}
\sim - \frac{\beta^2}{\bar{K}'} .
\label{eps2-1}
\eeq
In Eq.(\ref{eps1-1}) the sign of $\epsilon_1$ cannot be determined a priori (and it can
change with time as seen in Fig.~\ref{fig_eps_z} below) because the terms $(\bar{A} -1)
\simeq \beta\bar\varphi/M_{\rm Pl}$ and $2\bar{A}\beta^2/\bar{K}'$ are typically of opposite
signs, see Eq.(\ref{K0-sign-phi}), and of the same order.

As we have already seen, $\epsilon_1$ corresponds to a modification of Newton's
constant, while $\epsilon_2$ appears as a friction term in the Euler equation
(\ref{Euler-3}). They are both of order $\beta^2/\bar{K}'$.
As found in the companion paper \cite{Brax:2014aa}, this ratio also describes the deviation
of the background from the $\Lambda$-CDM scenario. In particular, it is constrained
by both the BBN upper bound on the variation of particle masses, see
Eq.(\ref{beta-phi-1}), and the requirement
$t_{\Lambda} \lesssim t_0$ that the dark energy behaves as a cosmological constant
in the recent Universe.

This common dependence on the ratio $\beta^2/\bar{K}'$ could be expected from
the form of the action (\ref{S-def}). Indeed, the coupling between the
scalar field $\varphi$ and  matter only occurs through the function
$A(\varphi)$ that relates the Einstein and Jordan metrics (\ref{g-Jordan-def})
(which appears as an effective modification of gravity from the matter point
of view).
A small $\beta$ means that this coupling function $A(\varphi)$ becomes
independent of $\varphi$ and almost equal to unity, see Eqs.(\ref{A-power-1})
and (\ref{A-exp-1}). Then, the Jordan metric becomes identical to the Einstein
metric and the matter no longer feels the scalar field, which remains at the origin.

On the other hand, a large $K'$ means that the scalar field is  sensitive to the nonlinearities
of the kinetic function $K(\chi)$. This gives rise to a large prefactor for time and
spatial derivatives of $\varphi$, so that the scalar field is frozen to zero,
plays no significant role and  General Relativity is recovered. This is also
the basis of the K-mouflage  mechanism.
Because the sign of $\beta$ is
irrelevant as it can be absorbed in a change of sign of $\varphi$, as noticed
in the text above Eq.(\ref{param-def}), the combination that sets the amplitude of the coupling
between matter and the scalar field is the ratio $\beta^2/K_0$, or more generally
$\beta^2/\bar{K}'$, in agreement with Eqs.(\ref{eps1-1})-(\ref{eps2-1}).

As described in the companion paper \cite{Brax:2014aa}, unless $|K_0| \gg 1$, we usually have
$|\bar{K}'| \simeq 1$ today, and cosmological large-scale structures are also
outside of their K-mouflage radius. Hence both the background and
cosmological structures feel the deviation from General Relativity. However,
at early times $\dot{\bar\varphi}$ and $\bar{K}'$ become large, so that
we recover the matter-dominated era as in the $\Lambda$-CDM scenario.

Thus, observations of the background quantities and of cosmological
density fluctuations are complementary, as they should be consistent with
each other. Density fluctuations may provide a more sensitive probe of the
deviations from General Relativity when we consider rare objects, such as clusters
of galaxies, which amplify the sensitivity to the details of the dynamics.
We will consider this in Sec.~\ref{Spherical} below.

Because the functions $\epsilon_i(\eta)$ do not depend on scale, the linear modes
$D_{\pm}(\eta)$ only depend on time and not on wave number,
as in the $\Lambda$-CDM cosmology.
As usual, we have both a linear growing mode $D_+(\eta)$ and a linear decaying mode
$D_-(\eta)$. Introducing the Wronskian of Eq.(\ref{D-linear}),
\beq
W(\eta) = e^{- \int_0^{\eta} \dd\eta' [ (1-3w_{\varphi}^{\rm eff}\Omega_{\varphi}^{\rm eff})/2
+\epsilon_2]} ,
\label{Wronskian}
\eeq
the decaying mode can be written as
\beq
D_-(\eta) = D_+(\eta) \int_{\eta}^{\infty} \dd\eta' \; \frac{W(\eta')}{D_+(\eta')^2} .
\label{D-}
\eeq
The growing mode $D_+(\eta)$ can be directly computed from Eq.(\ref{D-linear}).
As usual, we assume that the decaying mode has had time to decrease to a negligible
amplitude and we write the linear regime solution as
\beq
\tpsi_L(\vk,\eta) = \tdelta_{L0}(\vk) \; \left( \bea{c} D_+(\eta) \\
\frac{\dd D_+}{\dd\eta}(\eta) \ea \right) .
\label{psiL-def}
\eeq

\begin{figure}
\begin{center}
\epsfxsize=8.5 cm \epsfysize=5.8 cm {\epsfbox{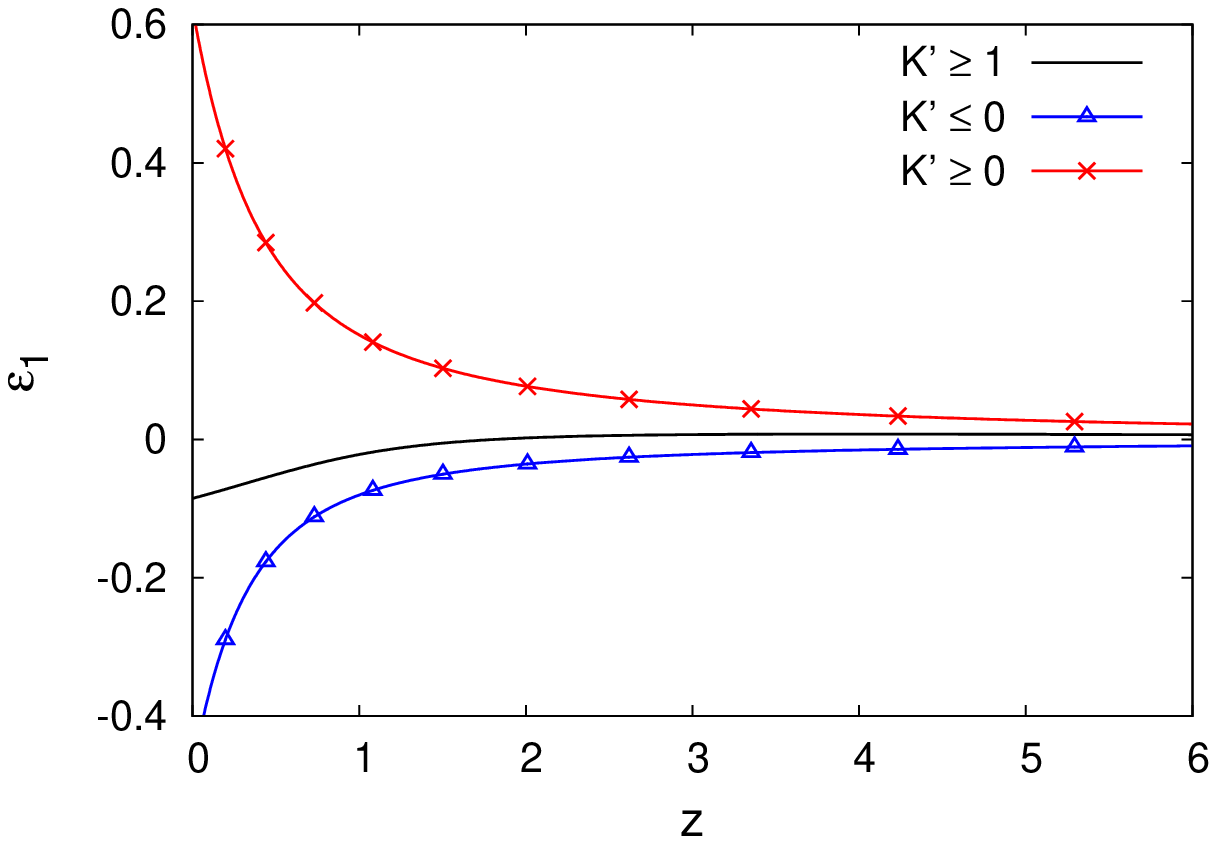}} \\
\epsfxsize=8.5 cm \epsfysize=5.8 cm {\epsfbox{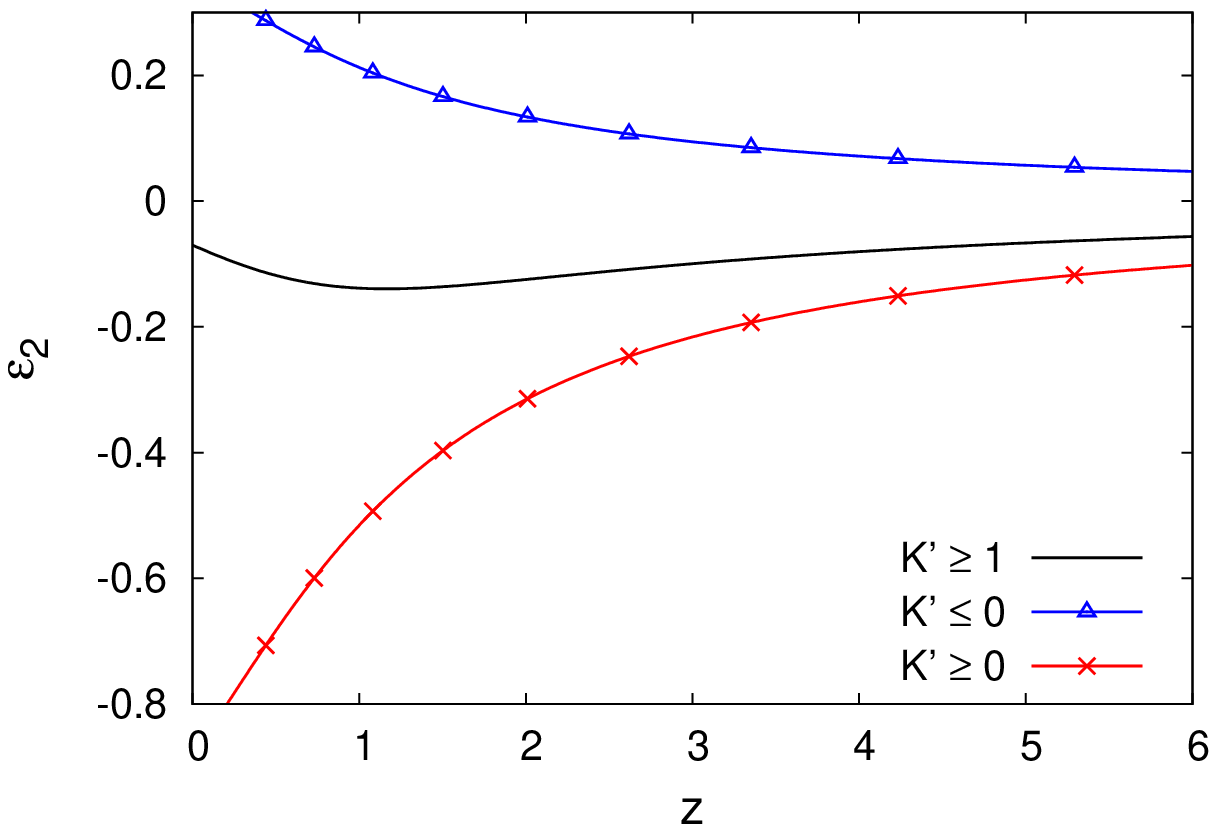}}
\end{center}
\caption{{\it Upper panel:} factor $\epsilon_1(z)$ for the models (\ref{K-power-1})
(with $K_0=1, m=3$, solid line), (\ref{K-power-2}) (with $K_0=-5, m=3$, line with triangles),
and (\ref{K-power-3}) (line with crosses). In all cases we use the exponential coupling function (\ref{A-exp-1}) with $\beta=0.3$.
{\it Lower panel:} factor $\epsilon_2(z)$ for the same models.}
\label{fig_eps_z}
\end{figure}

We show the factors $\epsilon_1(z)$ and $\epsilon_2(z)$ in Fig.~\ref{fig_eps_z},
for the models considered in Sec.~\ref{Specific-models}.
At high redshifts both $\epsilon_1$ and $\epsilon_2$ go to zero, as
$\bar\varphi\rightarrow 0$ and $|\bar{K}'|\rightarrow\infty$,
but it appears that the convergence to zero is faster for $\epsilon_1$.
For the models (\ref{K-power-1}) with $K_0 > 0$, at late times in the cosmological constant
regime, from Eq.(\ref{a-exp-L}) we find that $\epsilon_2$ shows an exponential decay to zero, as
$\epsilon_2 \sim e^{-\sqrt{3}\cM^2 t/M_{\rm Pl}}$, whereas $\epsilon_1$ goes to a finite
value along with $\bar\varphi$ and $\bar{A}$.
For the models (\ref{K-power-2}) and (\ref{K-power-3}), where $\bar\varphi$ keeps increasing linearly with time while $\bar{K}'$ goes to zero, $|\epsilon_1|$ goes to infinity at an exponential
rate while $\epsilon_2$ converges to a finite value.
The coefficient $\epsilon_2$ is positive, along with $\bar\varphi$, for the models (\ref{K-power-2})
where $\bar{K}' \leq 0$, and negative for the models (\ref{K-power-1}) and (\ref{K-power-3})
where  $\bar{K}' \geq 0$.

\begin{figure}
\begin{center}
\epsfxsize=8.5 cm \epsfysize=5.8 cm {\epsfbox{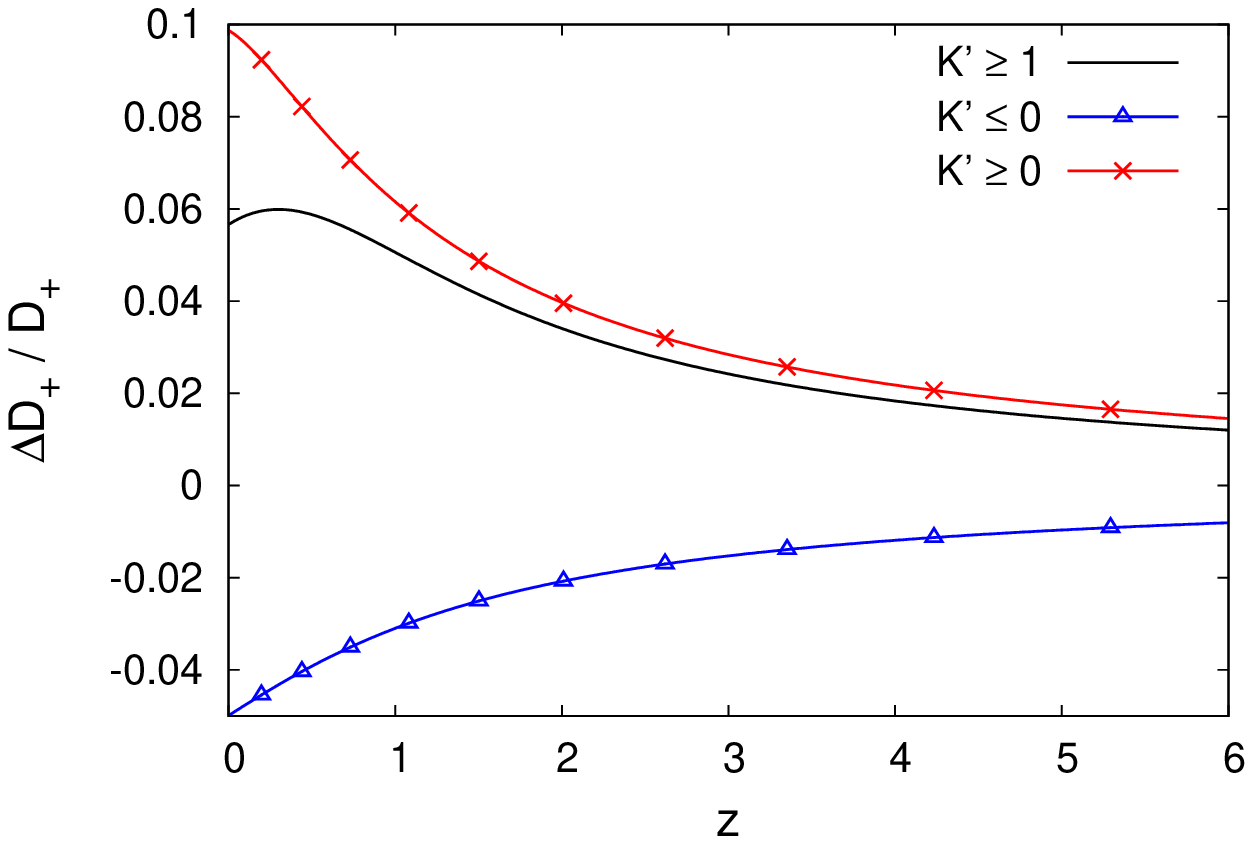}}\\
\epsfxsize=8.5 cm \epsfysize=5.8 cm {\epsfbox{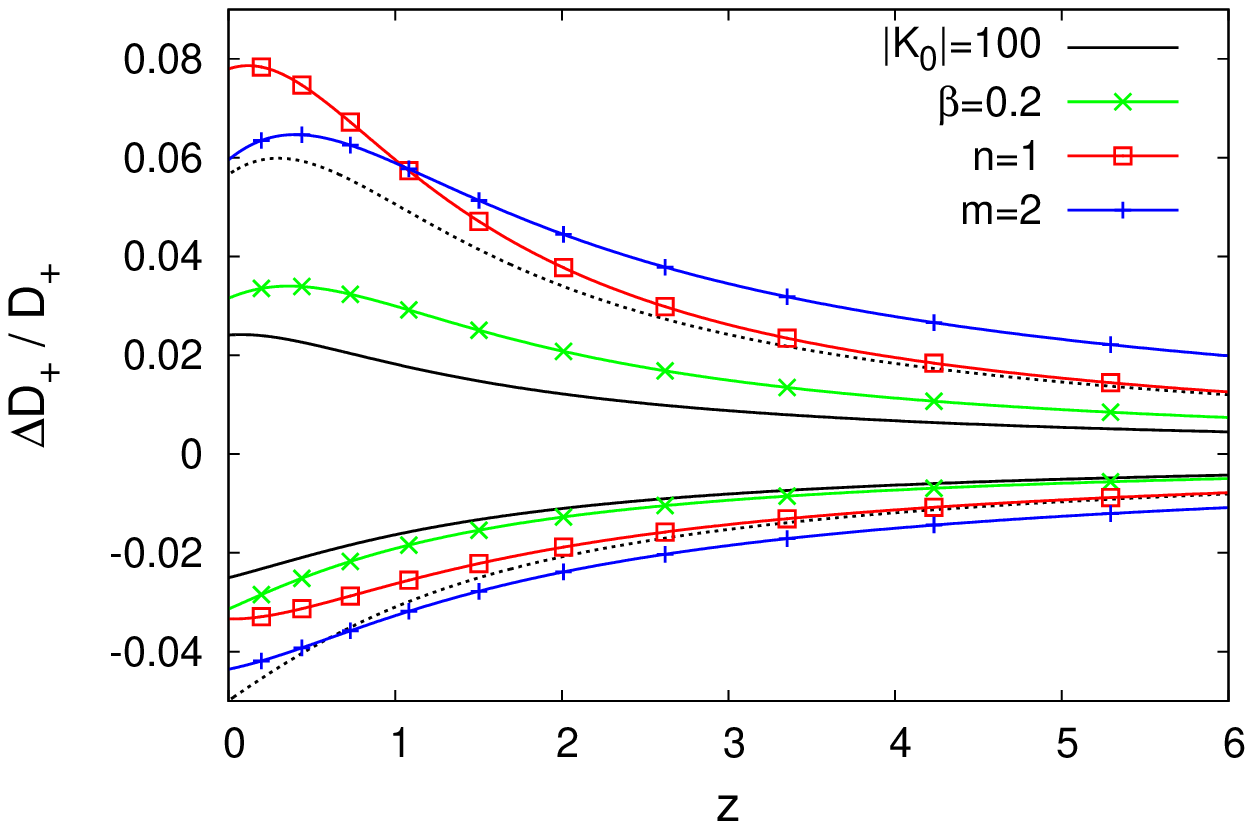}}
\end{center}
\caption
{Relative deviation
$[D_+(z)-D_{+\Lambda\rm CDM}(z)]/D_{+\Lambda\rm CDM}(z)$
of the linear growing mode from the $\Lambda-$CDM reference.
{\it Upper panel:} same models as in Fig.~\ref{fig_eps_z}.
{\it lower panel:} the upper dotted curve corresponds to the model (\ref{K-power-1}) with
$\{\beta=0.3,n=\infty;K_0=1,m=3\}$ (solid line in the upper panel) and the neighboring curves
correspond to the same model where we modify one among these four parameters, as indicated
by the legend (keeping the same sign of $K_0$). The lower dotted curve corresponds to the model (\ref{K-power-2}) with
$\{\beta=0.3,n=\infty;K_0=-5,m=3\}$ (triangles in the upper panel) and the neighboring curves
correspond to the same model where we again modify one among these four parameters, as indicated
by the legend.}
\label{fig_Dlin_z}
\end{figure}

In agreement with the discussion below Eq.(\ref{eps2-1}), for the models (\ref{K-power-1})
the deviations from zero of the coefficients $\epsilon_1$ and $\epsilon_2$ are of the same order
of magnitude as the deviations from $\Lambda$-CDM of the background quantities,
see also Figs.~1 and 2 of the companion paper \cite{Brax:2014aa}.
For the models (\ref{K-power-2}) and (\ref{K-power-3}), the deviations from $\Lambda$-CDM are somewhat amplified, as compared with the background quantities such as the Hubble expansion
rate. This is because $\epsilon_1$ diverges at late time, and $\epsilon_2$ involves
a time derivative, whereas $H(z)$ or $\Om(z)$ remain finite at late times and
are constrained to be equal to the $\Lambda$-CDM reference values today.

\begin{figure}
\begin{center}
\epsfxsize=8.5 cm \epsfysize=5.8 cm {\epsfbox{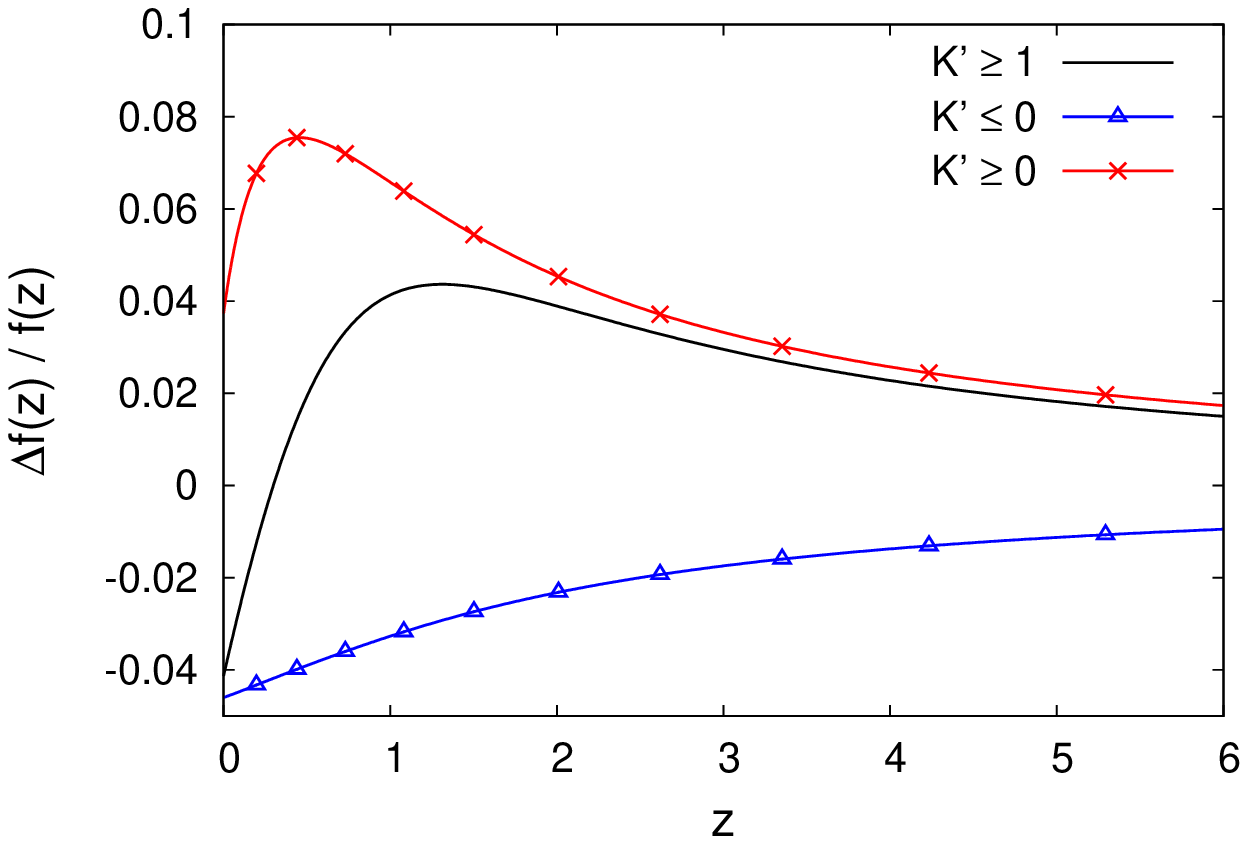}} \\
\epsfxsize=8.5 cm \epsfysize=5.8 cm {\epsfbox{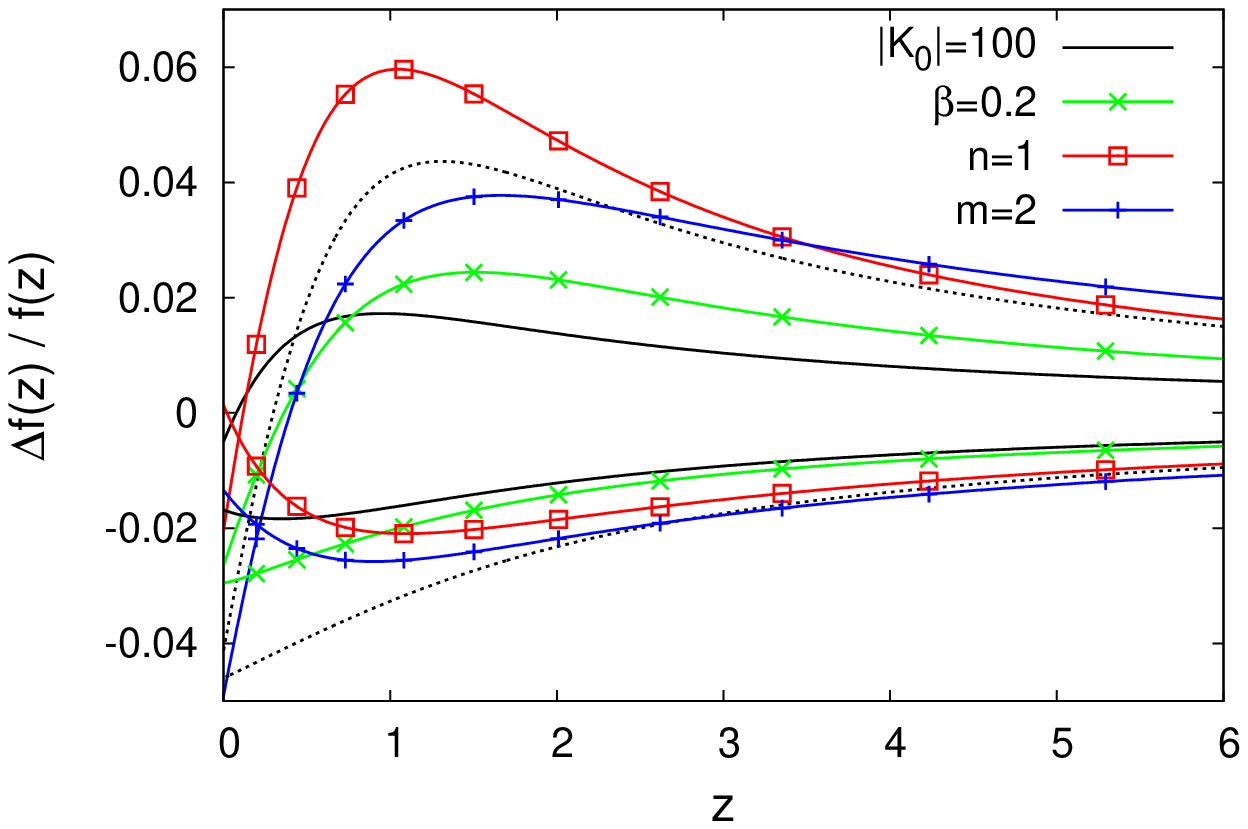}}
\end{center}
\caption{Relative deviation $[f(z)-f_{\Lambda \rm CDM}(z)]/f_{\Lambda \rm CDM}(z)$
of the linear growth rate $f(z) = \dd\ln D_+/\dd\ln a$.
{\it Upper panel:} same models as in the upper panel of Fig.~\ref{fig_Dlin_z}.
{\it Lower panel:} same models as in the lower panel of Fig.~\ref{fig_Dlin_z}.}
\label{fig_f_z}
\end{figure}

We show the linear growing mode $D_+(z)$ and the linear growth rate
$f(z) = \dd\ln D_+/\dd\ln a$ in the upper panels of Figs.~\ref{fig_Dlin_z} and \ref{fig_f_z},
for the same models as in Fig.~\ref{fig_eps_z}.
The relative deviations from the $\Lambda-$CDM reference are somewhat below those of the
factors $\epsilon_i$ at $z=0$ because the linear growing modes depend on the past history
of these factors.
As for the background expansion rate $H(z)$, the sign of $\bar{K}'$ sets the sign
of the deviation from the $\Lambda$-CDM reference.
Thus, the models (\ref{K-power-1}) with $K_0>0$, and (\ref{K-power-3}), which
both have $\bar{K}' \geq 0$, yield a smaller $H(z)$ (when we require a common normalization
today) and a larger linear growing mode $D_+(z)$, as well as a larger $f(z)$ at high redshift,
while opposite deviations are obtained for the model (\ref{K-power-2}) where $\bar{K}' \leq 0$.
Therefore, a positive $\bar{K}'$ (resp. a negative $\bar{K}'$) yields a faster (resp. slower)
growth of large-scale structures.

At early times, when the kinetic variable $\bar{\chi}$ is large and the kinetic function
$K(\chi)$ is governed by the highest power law $K_0 \chi^m$ in all three models
(\ref{K-power-1}), (\ref{K-power-2}), and  (\ref{K-power-3}), we obtain similar behaviors
for background and linear perturbation statistics, with only a change of sign along with the
coefficient $K_0$.
Thus, the high-redshift behavior is a direct probe of the sign of $\bar{K}'$ and of the large-$\chi$
power law asymptotic. At low redshift, where $\bar{\chi}$ becomes of order unity we can obtain
rather different behaviors, depending on the model parameters (e.g., sign of $\bar{K}'$ and
whether it goes to zero or to a nonzero value).

We consider the dependence of the linear growing modes on the model parameters in the
lower panels of Figs.~\ref{fig_Dlin_z} and \ref{fig_f_z}.
Taking as reference the models (\ref{K-power-1}) with $\{\beta=0.3,n=\infty;K_0=1,m=3\}$ and
(\ref{K-power-2}) with $\{\beta=0.3,n=\infty;K_0=-5,m=3\}$, we show our results when we
modify in turns either one of these four parameters.
In agreement with the discussion below Eq.(\ref{eps2-1}),
the relative deviations from the $\Lambda-$CDM reference
decrease for larger $|K_0|$ and smaller $\beta^2$.
As for background quantities, the main
dependence comes from the parameters $\beta$ and $K_0$, which set
the amplitude of the coupling between the scalar field and the matter
component [through the function $A(\varphi)$] and the normalization of
scalar field gradients [through the function $K(\chi)$]. More precisely,
the combination that sets the amplitude of the coupling
between matter and the scalar field is the ratio $\beta^2/K_0$, in agreement
with Eqs.(\ref{eps1-1})-(\ref{eps2-1}).
Changing the large-$\varphi$ and large-$\chi$ behaviors of the functions
$A(\varphi)$ (as $n$ goes from $+\infty$ to $1$) and $K(\chi)$ (as $m$ goes from
$3$ to $2$), only changes the predictions at the quantitative level and by a
modest amount.

\subsubsection{One-loop power spectrum}
\label{One-loop-power}

As in the standard perturbation theory, to go beyond linear order over $\delta_L$
we can look for a solution of the nonlinear equation of motion (\ref{O-Ks-def}) as a
perturbative expansion in powers of the linear density fluctuation $\delta_L$,
\beq
\psi(x) = \sum_{n=1}^{\infty} \psi^{(n)}(x) , \;\;\; \mbox{with} \;\;\;
\psi^{(n)} \propto \delta_L^n ,
\label{psi-n-def}
\eeq
and $\psi^{(1)}= \tpsi_L$.
Following the general approach described in \cite{Brax2013},
to compute the higher orders $\psi^{(n)}$ by recursion from Eq.(\ref{O-Ks-def}),
we introduce the retarded Green function $R_L$ of the linear operator $\cO$
(also called the linear propagator or response function), which obeys:
\beq
\cO(x,x') \cdot R_L(x',x'') = \delta_D(x-x'') ,
\label{RL-def}
\eeq
\beq
\eta_1 < \eta_2 : \;\;\; R_L(x_1,x_2)=0 ,
\eeq
and reads as
\beqa
\hspace{-0.5cm} R_L(x_1,x_2) & = & \frac{\Theta(\eta_1-\eta_2) \, \delta_D(\vk_1-\vk_2)}
{D_{+2}'D_{-2}-D_{+2}D_{-2}'} \; \times  \nonumber \\
&& \hspace{-1.8cm} \left( \! \bea{lr} D_{+2}'D_{-1}\!-\!D_{-2}'D_{+1} \;\;
& \;\;  D_{-2}D_{+1}\!-\!D_{+2}D_{-1} \\ & \\ D_{+2}'D_{-1}'\!-\!D_{-2}'D_{+1}'
&  D_{-2}D_{+1}'\!-\!D_{+2}D_{-1}' \ea \! \right) \;\;
\label{RL-1}
\eeqa
It involves both the linear growing and decaying modes, $D_+$ and $D_-$, and
$\Theta(\eta_1-\eta_2)$ is the Heaviside function, which ensures causality.
Then, from Eq.(\ref{O-Ks-def}) we obtain at second and third order
\beq
\tpsi^{(2)} = R_L \cdot K_2^s \cdot \tpsi^{(1)} \tpsi^{(1)} ,
\label{psi-2}
\eeq
\beq
\tpsi^{(3)} = 2 R_L \cdot K_2^s \cdot \tpsi^{(2)} \tpsi^{(1)} + R_L \cdot K_3^s \cdot
\tpsi^{(1)} \tpsi^{(1)} \tpsi^{(1)} .
\label{psi-3}
\eeq
The last term in Eq.(\ref{psi-3}) does not appear in the
standard $\Lambda$-CDM case. It is due to the vertex $\gamma^s_{2;1,1,1}$
associated with the term of order $(\delta\rho)^3$ of the nonlinear modified
gravitational potential $\Psi$.
Then, the two-point correlation $C_2$ of the field $\psi$ reads up to order $\delta_L^4$
as
\beqa
C_2(x_1,x_2) & \equiv & \lag \tpsi(x_1) \tpsi(x_2) \rag \nonumber \\
& = & \lag \tpsi^{(1)} \tpsi^{(1)} \rag + \lag \tpsi^{(2)}\tpsi^{(2)}\rag
+ \lag\tpsi^{(3)}\tpsi^{(1)}\rag \nonumber \\
&& + \lag\tpsi^{(1)}\tpsi^{(3)}\rag + ..
\label{C2-def}
\eeqa
Defining the equal-time matter density power spectrum as
\beq
\lag \tdelta(\vk_1,\eta) \tdelta(\vk_2,\eta) \rag = \delta_D(\vk_1+\vk_2) \; P(k_1,\eta) ,
\eeq
substituting the expressions (\ref{psiL-def}), (\ref{psi-2}), and (\ref{psi-3})
into Eq.(\ref{C2-def}) and using Wick's theorem, we obtain up to order $P_L^2$,
\beq
P(k) = P_{\rm tree}(k) + P_{\rm 1loop}(k) .
\label{Ptree+1loop}
\eeq
The ``tree'' contribution, associated with $\lag \tpsi^{(1)} \tpsi^{(1)} \rag$,
is simply the linear power spectrum,
\beq
P_{\rm tree} = P_L(k) ,
\label{P-tree}
\eeq
while the ``one-loop'' contribution corresponds to three diagrams,
\beq
P_{\rm 1loop} = P_{22} + P_{31} + P_{31}^{\Psi} .
\label{P-1loop}
\eeq
The contribution $P_{22}$ corresponds to $\lag \tpsi^{(2)}\tpsi^{(2)}\rag$ and
$P_{31}$ to $\lag\tpsi^{(3)}\tpsi^{(1)}\rag+\lag\tpsi^{(1)}\tpsi^{(3)}\rag$
using for $\tpsi^{(3)}$ the first standard term of Eq.(\ref{psi-3}).
The contribution $P_{31}^{\Psi}$ arises from the new second term of Eq.(\ref{psi-3}).
More details and the diagrams associated with these one-loop contributions can
be found in Ref.\cite{Brax2013}.
In particular, we obtain
\beqa
P_{31}^{\Psi} & = & 6 \int \dd\vk_1 \int_{-\infty}^{\eta} \dd\eta_1 \; R_{L,12}(\eta,\eta_1)
C_{L,11}(k;\eta,\eta_1) \;\;\; \nonumber \\
&& \times \; C_{L,11}(k_1;\eta_1,\eta_1) \; \gamma^s_{2;1,1,1}(\vk_1,-\vk_1,\vk;\eta_1) ,
\;\;\;
\label{P31-Psi}
\eeqa
and Eq.(\ref{gamma3}) gives
\beq
\int \dd {\bf \Omega}_{\vk_1} \; \gamma_{2;1,1,1}^s(\vk_1,-\vk_1,\vk) =
\frac{30 \pi \kappa_2 \bar{A}^3\beta_1^4 \Om^3 M_{\rm Pl}^2 H^4 a^2}
{\kappa_1^4 \cM^4 c^2 k_1^2} ,
\label{gamma3-P31}
\eeq
where ${\bf \Omega}_{\vk_1}$ is the unit vector of direction $\vk_1$.
Therefore, the angular average (\ref{gamma3-P31}) of the vertex
$\gamma_{2;1,1,1}^s$ no longer depends on the wave number $k$.
This implies that the one-loop contribution (\ref{P31-Psi}) is proportional to the
linear power spectrum $P_{L0}(k)$.
Thus, because the modification of gravity that arises from the models studied
in this paper is scale independent (in the small-scale regime $ctk/a\gg 1$ and
for density fluctuations that are not too large, as explained in
Secs.~\ref{Small-scale} and \ref{Fifth-force}), the vertices $h_n$, $H_n$, and
$\gamma_n$ do not show infrared cutoffs of the form $k^2/(m^2+k^2)$ with
$m \sim 1 h$/Mpc.
Rather, we obtain rational functions of wave numbers that can lead to
nonzero values at low $k$ as in Eq.(\ref{gamma3-P31}) [i.e., as compared with
$f(R)$ or other scalar field models, the infrared cutoff vanishes,
$m=0$].
[The angular average Eq.(\ref{gamma3-P31}) does not go to zero
at low $k$, in contrast with Eq.(\ref{gamma-n-k0}), because here
$k_3$ goes to zero along with the sum vector $k$.]
Then, higher-order contributions generated by the modification to gravity
can lead to a (small) time dependent renormalization of the linear power spectrum,
in the sense that $P_{31}^{\Psi}(k) \propto P_L(k)$ at low $k$.
From Eqs.(\ref{P31-Psi})-(\ref{gamma3-P31}) we obtain
\beq
P_{31}^{\Psi}(k) \sim P_L(k) \; \frac{\sigma_{\vs_L}^2 a^2}{c^2t^2} \;
\frac{\kappa_2 \beta^4}{\kappa_1^4} \ll P_L(k)
\label{P31-PL}
\eeq
where $\sigma^2_{\vs_L} = \lag {\vs}_L^2 \rag$ is the variance of the linear
displacement field [denoting $\vx(\vq,t) = \vq+\vs_L(\vq,t)$ as the trajectory of the
particle $\vq$ in Lagrangian perturbation theory at linear order].
This shows that in the small-scale regime (\ref{psiA-1}) the contribution
$P_{31}^{\Psi}$ is negligible, in agreement with the discussion below
Eq.(\ref{psiA-1}).

\begin{figure}
\begin{center}
\epsfxsize=8.5 cm \epsfysize=5.8 cm {\epsfbox{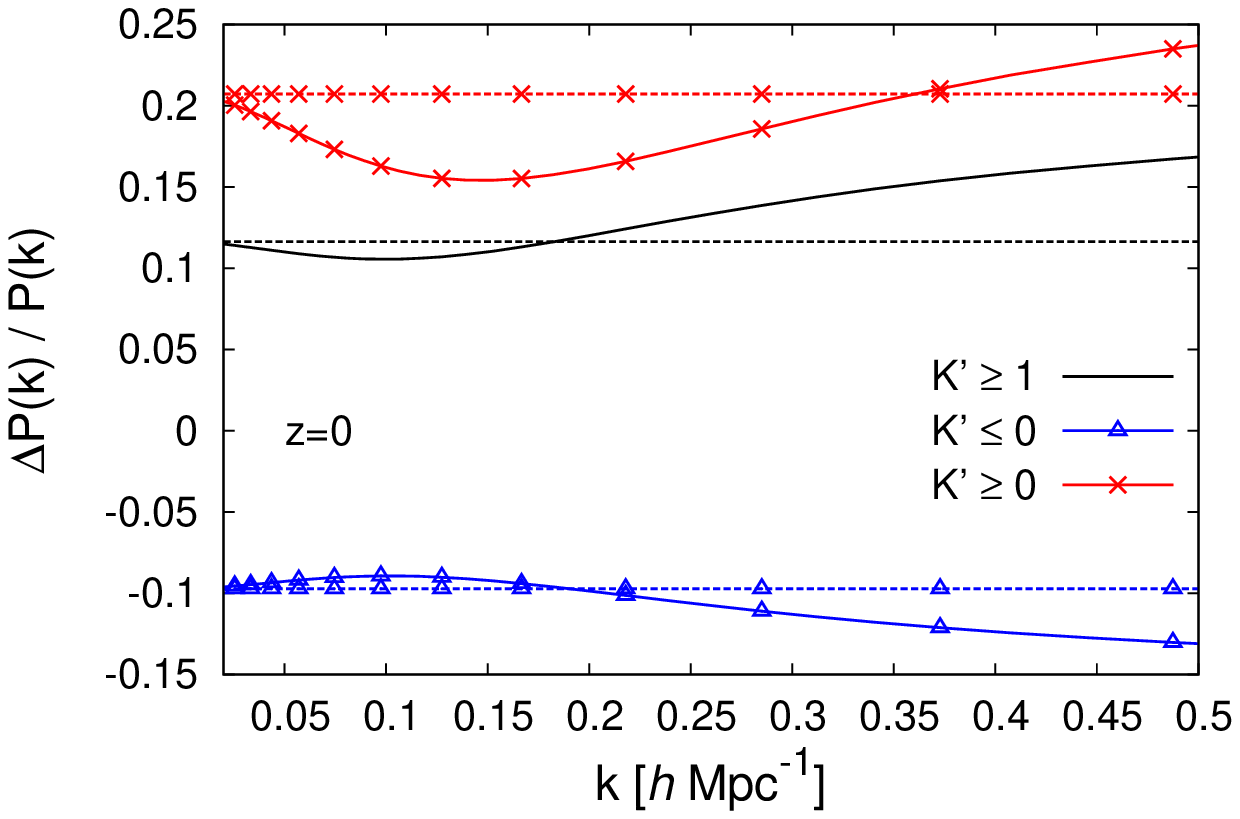}} \\
\epsfxsize=8.5 cm \epsfysize=5.8 cm {\epsfbox{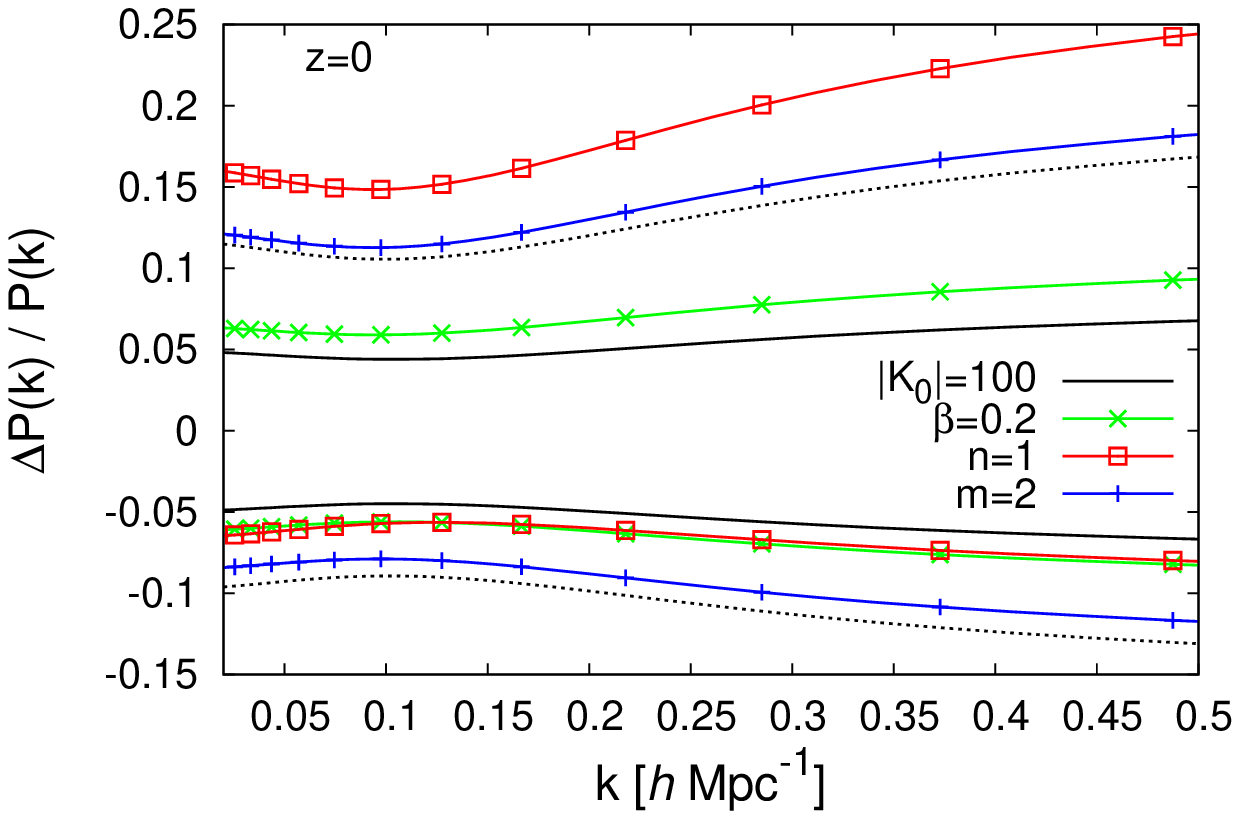}}
\end{center}
\caption{Relative deviation
$[P(k)-P_{\Lambda \rm CDM}(k)]/P_{\Lambda \rm CDM}(k)$
of the linear (dashed lines) and one-loop (solid lines) power spectra from the
$\Lambda$-CDM reference, at redshift $z=0$.
{\it Upper panel:} same models as in the upper panel of Fig.~\ref{fig_Dlin_z}.
{\it Lower panel:} same models as in the lower panel of Fig.~\ref{fig_Dlin_z}.
We only show the one-loop power spectra (\ref{Ptree+1loop}).}
\label{fig_dP1loop_z0}
\end{figure}

We show the linear and one-loop power spectra in Fig.~\ref{fig_dP1loop_z0}.
In agreement with the analysis in Sec.~\ref{linear} and the fact that the
coefficients $\epsilon_i(t)$ of Eq.(\ref{epsilon-def}) do not depend on scale,
the relative deviation of the linear power spectrum does not depend on wave number
(as long as $ctk/a \gg 1$).
In agreement with Fig.~\ref{fig_Dlin_z}, a positive $K_0$, or more generally $\bar{K}'$,
leads to a speeding-up of the matter clustering, and hence a greater matter density power spectrum,
while a negative $K_0$, or $\bar{K}'$, leads to a slower matter clustering.
This remains true at one-loop order. The one-loop correction first slightly decreases
the deviation from $\Lambda$-CDM, at $k\sim 0.1 h$Mpc$^{-1}$ for $z=0$, and
next amplifies the deviation at higher $k$ when density fluctuations become mildly
nonlinear (but one-loop perturbation theory does not extend beyond
$0.3 h/$Mpc at $z=0$).
We also checked that the one-loop contribution (\ref{P31-Psi})
is negligible, by comparing our results with those obtained when we set
$P_{31}^{\Psi}$ to zero.
Thus, in agreement with the discussion below Eq.(\ref{psiA-1}),
the nonlinearities are not due to the Klein-Gordon
equation, which can be kept at the linear level, but to the continuity and Euler
equations, more precisely to the usual vertices (\ref{gamma-222}), as in the
standard $\Lambda$-CDM scenario.

As for previous quantities, we can see in the lower panel that deviations from
the $\Lambda$-CDM predictions decrease for larger $K_0$ or smaller $\beta$.
The detailed shape of the coupling function $A(\varphi)$ appears to have a significant
impact on the power spectrum at the quantitative level, as we go from the exponential
form (\ref{A-exp-1}) to the linear form  (\ref{A-power-1}).

\section{Large-scale cosmic microwave anisotropies}
\label{Sec-CMB}

\subsection{Integrated Sachs-Wolfe effect}
\label{Sec-ISW}

\begin{figure}
\begin{center}
\epsfxsize=8.5 cm \epsfysize=5.8 cm {\epsfbox{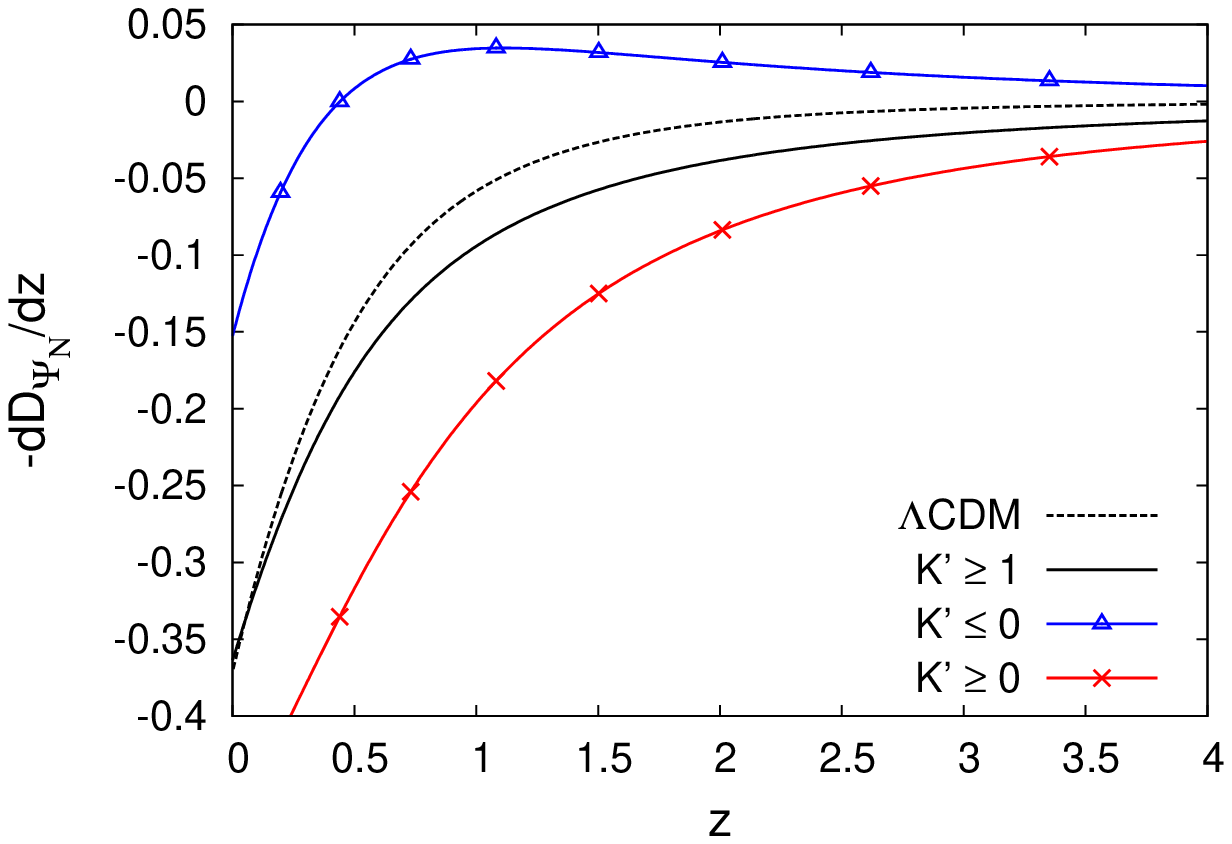}} \\
\epsfxsize=8.5 cm \epsfysize=5.8 cm {\epsfbox{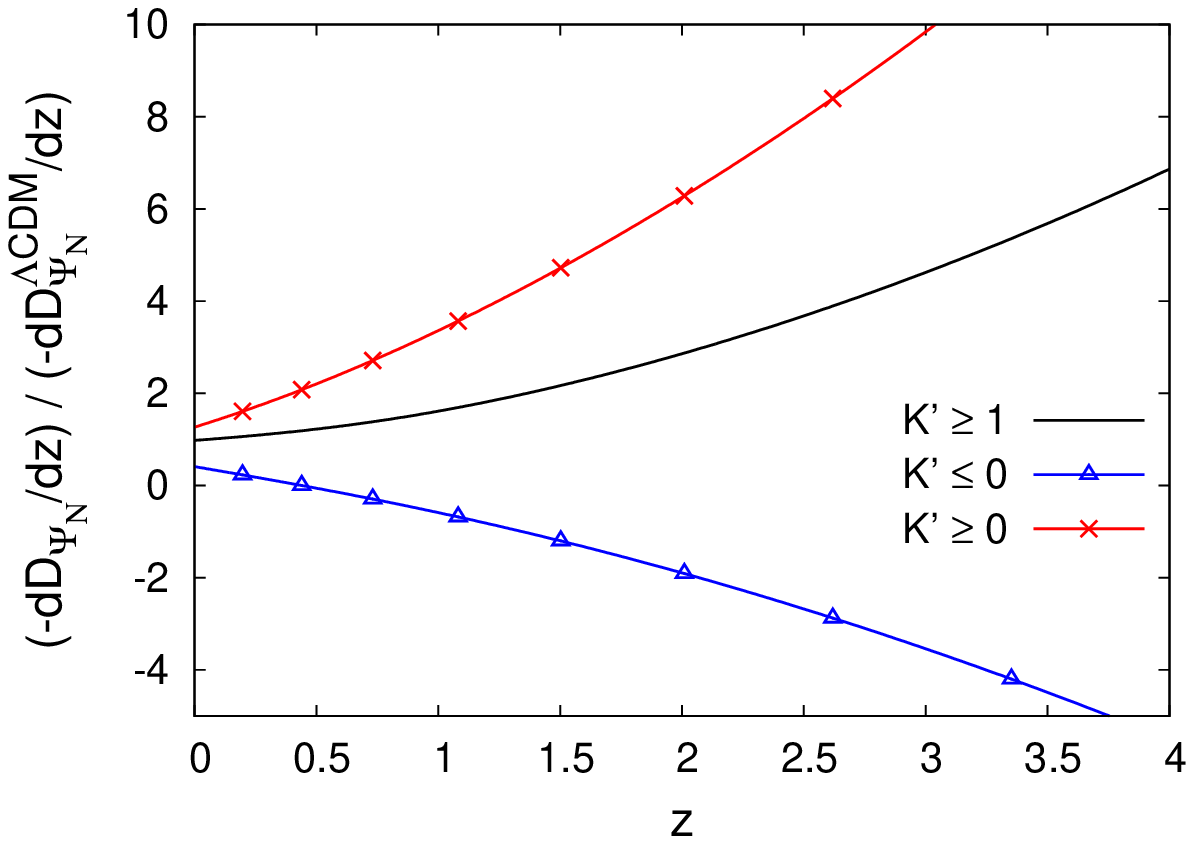}}
\end{center}
\caption{{\it Upper panel:} the factor $(-\dd D_{\Psi_{\rm N}}/\dd z)$ of
Eq.(\ref{dD-Psi}) for the reference $\Lambda$-CDM universe and the
scenarios of Fig.~\ref{fig_eps_z}.
{\it Lower panel:} ratio of these factors $(-\dd D_{\Psi_{\rm N}}/\dd z)$ to the
$\Lambda$-CDM reference.}
\label{fig_Dpsi_z}
\end{figure}

The Integrated Sachs-Wolfe effect (ISW) arises from the differential redshift effect
that is left as photons climb in and out of time dependent gravitational potentials
\cite{Sachs1967}.
This generates a large-scale fluctuation, $\Delta T_{\rm ISW}$, of the cosmic
microwave background (CMB), given by
\beq
\frac{\Delta T_{\rm ISW}}{\bar{T}_{\rm CMB}} = -\frac{2}{c^2} \int_0^{\infty} \dd z \;
e^{-\kappa(z)} \, \frac{\pl \Psi_{\rm N}}{\pl z} ,
\label{ISW-def}
\eeq
where $\kappa(z)$ is the optical depth to redshift $z$.
This effect can be constrained through the autocorrelation function of the
temperature fluctuations of the CMB and also through the cross-correlation
with the large-scale structures in the recent Universe, as galaxies and
clusters are correlated with the matter density and gravitational fields
\cite{Cooray2002a,Afshordi2004,Granett2009,Mainini2012}.
During the matter era, on large linear scales the gravitational potential
$\Psi_{\rm N}$ does not evolve with time. This means that the ISW effect
is dominated by low redshifts when dark energy modifies the linear growing mode
and makes $|\Psi_{\rm N}|$ decrease. This makes the ISW effect a probe of dark energy
properties and of modified-gravity theories.

From the modified Poisson equation (\ref{Poisson-2}), we obtain in the linear
regime $\nabla^2 \Psi_{\rm N} = 4\pi\cG \bar\rho_0 \, \bar{A}\delta_L/a$,
and we can write the linear mode $D_{\Psi_{\rm N}}$ of the
gravitational potential as
\beq
D_{\Psi_{\rm N}} = \bar{A} \, \frac{D_+}{a} ,
\label{D-Psi}
\eeq
where $D_+$ is the linear growing mode of the density contrast, given by
Eq.(\ref{D-linear}) and displayed in Fig.~\ref{fig_Dlin_z}.
The ISW effect (\ref{ISW-def}) involves the derivative of $D_{\Psi_{\rm N}}$ with
respect to time, or redshift, and we obtain
\beq
- \frac{\dd D_{\Psi_{\rm N}}}{\dd z} = a^2 \;  \frac{\dd D_{\Psi_{\rm N}}}{\dd a} =
\bar{A} D_+ ( \epsilon_2 + f -1 ) ,
\label{dD-Psi}
\eeq
where $\epsilon_2(z)=\dd\ln\bar{A}/\dd\ln a$ was already introduced in
Eq.(\ref{epsilon-def}) and shown in Fig.~\ref{fig_eps_z},
whereas $f(z)=\dd\ln D_+/\dd\ln a$ was shown in Fig.~\ref{fig_f_z}.
In the standard $\Lambda$-CDM cosmology we have $\bar{A}=1$ and
$\epsilon_2=0$.

Cross-correlations between the large-scale CMB temperature fluctuations (\ref{ISW-def})
and low-redshift galaxy surveys constrain the time derivative of the
gravitational potential at the redshift of the galaxy population, through the
correlation $\lag \Psi_{\rm N} \delta_{\rm g} \rag$.
Therefore, we show the factor $(-\dd D_{\Psi_{\rm N}}/\dd z)$
as a function of redshift in the upper panel of Fig.~\ref{fig_Dpsi_z}.

At high $z$, we recover the Einstein-de Sitter cosmology and the derivative
$\dd D_{\Psi_{\rm N}}/\dd z$ goes to zero for all models.
For the modified-gravity models, we still have $\bar{A} \simeq 1$ and
$f(z)$ remains close to the $\Lambda$-CDM reference, as seen in
Fig.~\ref{fig_f_z}. Then, the main source of deviation from the $\Lambda$-CDM
prediction is the new term $\epsilon_2$ in Eq.(\ref{dD-Psi}).
For scenarios with $\bar{K}'>0$, we have seen in Fig.~\ref{fig_eps_z} that
$\epsilon_2$ is negative, like $(f-1)$. Then, $(-\dd D_{\Psi_{\rm N}}/\dd z)$ is negative,
as in the $\Lambda$-CDM reference (i.e., linear gravitational potentials decay
with time in the dark energy era).
For the models (\ref{K-power-1}), where $\epsilon_2$ goes to zero at late times, the
ISW factor $(-\dd D_{\Psi_{\rm N}}/\dd z)$ always remains close to the $\Lambda$-CDM
reference, whereas for the models (\ref{K-power-3}), where $\epsilon_2$ converges to
a nonzero value, the deviation remains significant at low redshift.
For scenarios with $\bar{K}'<0$, we have seen in Fig.~\ref{fig_eps_z} that
$\epsilon_2$ is positive and decreases slowly at high redshift.
Then, the term $\epsilon_2$ can dominate over the factor $(f-1)$ and
we can see in Fig.~\ref{fig_Dpsi_z} that, for the model (\ref{K-power-2}) with $K_0=-5$,
$(-\dd D_{\Psi_{\rm N}}/\dd z)$ becomes positive at $z \gtrsim 0.4$.
Therefore, in such scenarios the ISW effect changes sign at high $z$ as the
linear gravitational potential first slowly grows with time when dark energy becomes
noticeable.
At low redshift the term $\epsilon_2$ is no longer dominant and
$(-\dd D_{\Psi_{\rm N}}/\dd z)$ is negative (i.e., linear gravitational potentials decay
with time) as in the standard $\Lambda$-CDM scenario.
Thus, cross-correlations between the ISW effect on the CMB and large-scale
structures at $z \sim 1$ would be a useful probe of such models, as the
sign of the correlation itself would discriminate between the two categories
$\bar{K}'>0$ and $\bar{K}'<0$.

The lower panel in Fig.~\ref{fig_Dpsi_z} shows the ratio of $(-\dd D_{\Psi_{\rm N}}/\dd z)$
to the $\Lambda$-CDM reference. In agreement with the upper panel and the
discussion above, this ratio remains positive for scenarios with $\bar{K}' > 0$ and becomes negative
at high redshift for scenarios with $\bar{K}'<0$.
In both cases, because of the factor $\epsilon_2$, which decays rather slowly with
redshift, $(-\dd D_{\Psi_{\rm N}}/\dd z)$ decays more slowly than the $\Lambda$-CDM
prediction at high $z$. This yields a ratio to the $\Lambda$-CDM reference
that grows at high $z$.
However, in practice it is difficult to probe these high-redshift behaviors, because
most of the signal comes from the low-redshift range where linear gravitational
potentials show a significant time dependence, as seen in the upper panel.

\subsection{Low-$\ell$ CMB anisotropies}
\label{Sec-Cl}

\begin{figure}
\begin{center}
\epsfxsize=8.5 cm \epsfysize=5.8 cm {\epsfbox{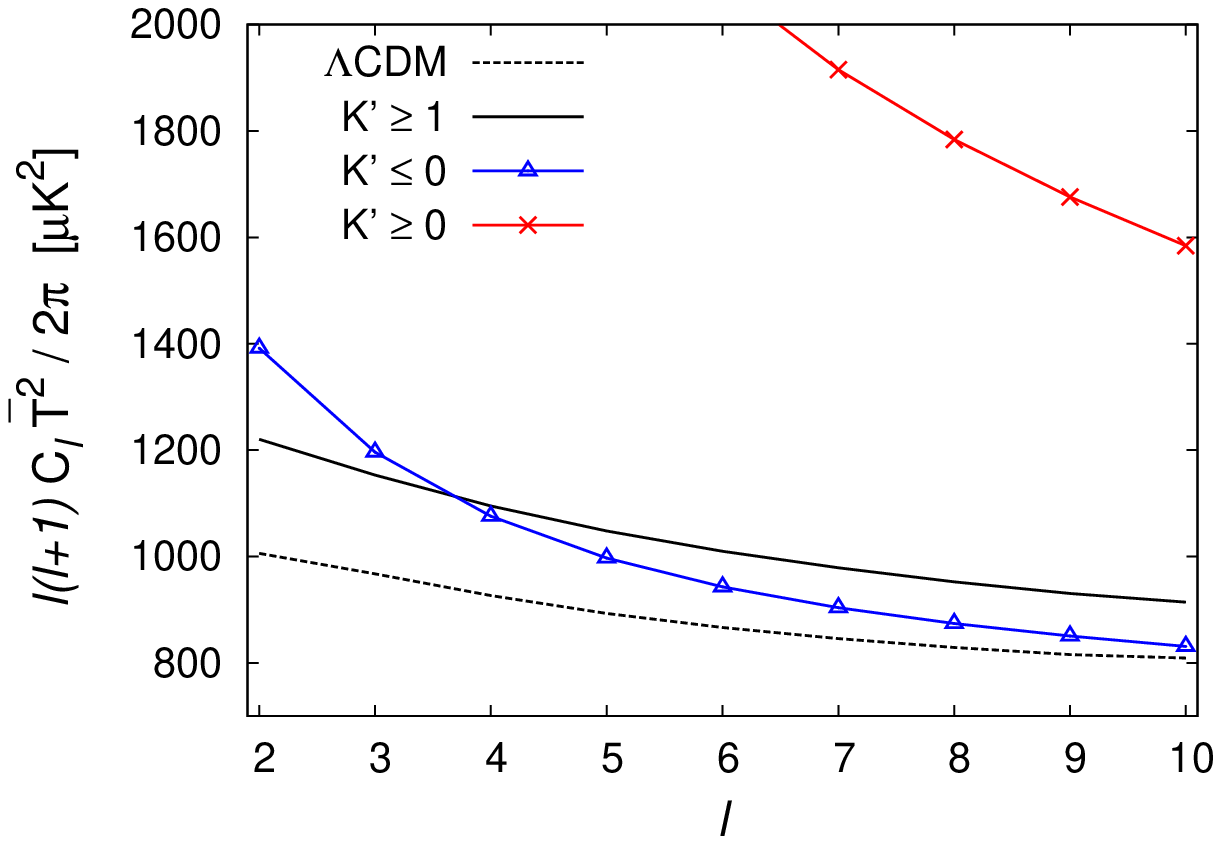}}\\
\epsfxsize=8.5 cm \epsfysize=5.8 cm {\epsfbox{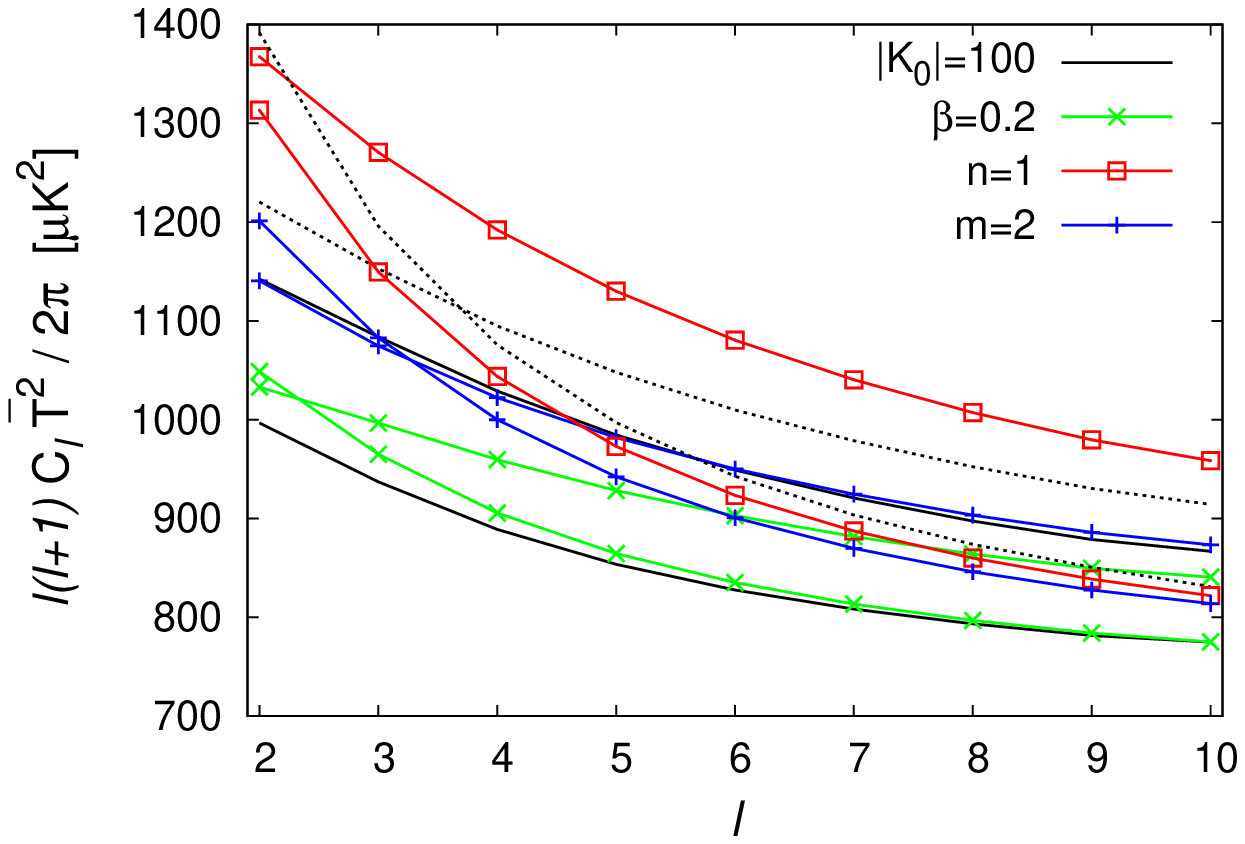}}
\end{center}
\caption{CMB angular power spectrum $\ell(\ell+1) C_{\ell} \bar{T}^2/(2\pi)$ at low multipoles.
{\it Upper panel:} same models as in the upper panel of Fig.~\ref{fig_Dlin_z}.
{\it Lower panel:} same models as in the lower panel of Fig.~\ref{fig_Dlin_z}.}
\label{fig_Cl}
\end{figure}

At low-$\ell$ multipoles, the CMB anisotropies are governed by the
Sachs-Wolfe (SW) and integrated Sachs-Wolfe (ISW) effects, and the temperature
fluctuation $\Delta T_{\rm CMB}/\bar{T}_{\rm CMB}$ in a direction ${\vec \gamma}$
on the sky reads as
\beq
\frac{\Delta T_{\rm CMB}}{\bar{T}_{\rm CMB}} \simeq \frac{1}{3c^2} \Psi_{\rm N}(\tau_{\rm LS})
+ \frac{2}{c^2} \int_{\tau_{\rm LS}}^{\tau_0} \dd\tau \, \frac{\pl\Psi_{\rm N}}{\pl\tau} ,
\label{dT-1}
\eeq
where $\tau = \int \dd t/a$ is the conformal time, $\tau_{\rm LS}$ and $\tau_0$ its value
at the last-scattering surface and today. Here we have used
$\Delta T/\bar{T} \simeq -2 \Psi_{\rm N}/(3c^2)$ at $\tau_{\rm LS}$ and we have
approximated the opacity as $\kappa=0$ after the last-scattering surface and
$\kappa=+\infty$ before.
The first term in Eq.(\ref{dT-1}) is the Sachs-Wolfe effect (due to the initial temperature
fluctuation and gravitational potential) and the second term is the integrated
Sachs-Wolfe effect (\ref{ISW-def}).
Expanding as usual the temperature fluctuations on the sky in spherical harmonics,
$\Delta T ({\vec \gamma})/\bar{T} = \sum_{\ell,m} a_{\ell,m} Y^{\ell}_m({\vec \gamma})$,
and defining the multipole power spectrum as
$\lag a_{\ell,m} a_{\ell',m'}^*\rag = \delta_{\ell,\ell'} \delta_{m,m'} C_{\ell}$,
we obtain
\beqa
C_{\ell} & \!\! = \!\! & (4\pi)^2 \int_0^{\infty} \!\! \dd k k^2 \, P_{\Psi_{\rm N},0}(k) \biggl \lbrace
\frac{D_{\Psi_{\rm N}}(\tau_{\rm LS})}{3} j_{\ell}[kc(\tau_0 \! - \! \tau_{\rm LS})] \nonumber \\
&& + 2 \int_{\tau_{\rm LS}}^{\tau_0}
\dd \tau \frac{\dd D_{\Psi_{\rm N}}}{\dd\tau} j_{\ell}[kc(\tau_0-\tau)] \biggl \rbrace^2 ,
\eeqa
where we have defined the gravitational potential linear power spectrum as
$\lag \tpsi_{\rm N}(\vk_1,\tau_1) \tpsi_{\rm N}(\vk_2,\tau_2) \rag = \delta_D(\vk_1+\vk_2)
\, D_{\Psi_{\rm N}}(\tau_1) D_{\Psi_{\rm N}}(\tau_2) \, c^4 P_{\Psi_{\rm N},0}(k_1)$,
and the linear mode $D_{\Psi_{\rm N}}$ was given in Eq.(\ref{D-Psi}).
Because we only consider low multipoles $\ell$, and hence large scales and low
wave numbers $k$, we write the linear power spectrum as
\beq
P_{\Psi_{\rm N},0}(k) = {\cal N} \; k^{n_s-4} ,
\label{Pk-Psi-ns}
\eeq
and the integration over $k$ yields
\beqa
\!\! C_{\ell} & \! = \! & \frac{2^{n_s} \pi^3 {\cal N} \, \Gamma[\ell \!+\! \frac{n_s-1}{2}]}
{c^{n_s-1} \Gamma[2-\frac{n_s}{2}]}
\biggl \lbrace \frac{D_{\Psi_{\rm N}}(\tau_{\rm LS})^2 (\tau_0 \!-\! \tau_{\rm LS})^{1-n_s}
\Gamma[3 \!-\! n_s]}{9 \, \Gamma[2-\frac{n_s}{2}] \, \Gamma[\ell+\frac{5-n_s}{2}]}
\nonumber \\
&& + 4 \int_{\tau_{\rm LS}}^{\tau_0} \dd\tau_1\dd\tau_2 \,
\frac{\dd\ D_{\Psi_{\rm N}}}{\dd\tau}(\tau_1) \frac{\dd\ D_{\Psi_{\rm N}}}{\dd\tau}(\tau_2)
\nonumber \\
&& \times
\frac{(\tau_0 \!-\! \tau_1)^{\ell} (\tau_0 \!-\! \tau_2)^{\ell} \;
_2F_1(\ell+\frac{n_s-1}{2},\ell+1;2\ell+2;y_{12})}
{(2\tau_0 \!-\! \tau_1 \!-\! \tau_2)^{2\ell+n_s-1} \, \Gamma[\ell+\frac{3}{2}]}
\nonumber \\
&& + \frac{4 D_{\Psi_{\rm N}}(\tau_{\rm LS})}{3} \int_{\tau_{\rm LS}}^{\tau_0} \dd\tau \,
\frac{\dd\ D_{\Psi_{\rm N}}}{\dd\tau}(\tau) \,
\frac{(\tau_0 \!-\! \tau)^{\ell} (\tau_0 \!-\! \tau_{\rm LS})^{\ell}}
{(2\tau_0 \!-\! \tau \!-\! \tau_{\rm LS})^{2\ell+n_s-1}}
\nonumber \\
&& \times
\frac{_2F_1(\ell+\frac{n_s-1}{2},\ell+1;2\ell+2;y)}
{\Gamma[\ell+\frac{3}{2}]} \biggl \rbrace ,
\label{Cl-SW-ISW}
\eeqa
where the arguments of the hypergeometric functions are
$y_{12}=4(\tau_0-\tau_1)(\tau_0-\tau_2)/(2\tau_0-\tau_1-\tau_2)^2$
and
$y=4(\tau_0-\tau)(\tau_0-\tau_{\rm LS})/(2\tau_0-\tau-\tau_{\rm LS})^2$.
The first term in Eq.(\ref{Cl-SW-ISW}) comes from the Sachs-Wolfe effect,
$({\rm SW})^2$, the second term from the integrated Sachs-Wolfe effect,
$({\rm ISW})^2$, and the third term from the cross-correlation $({\rm SW}) \times
({\rm ISW})$.

We show the CMB power spectrum on large angular scales, from Eq.(\ref{Cl-SW-ISW}),
in Fig.~\ref{fig_Cl}.
We compare the modified gravity-scenarios to the $\Lambda$-CDM reference in the
upper panel, and we consider the sensitivity to other model parameters in the
lower panel.
The (SW)$^2$ contribution is the same in all scenarios, because we recover the
$\Lambda$-CDM cosmology and the same initial conditions at early times.

In agreement with Fig.~\ref{fig_Dpsi_z}, for the models with $\bar{K}'>0$ the ISW
effect is greater than the $\Lambda$-CDM prediction and of opposite sign to the
SW effect ($\dd D_{\Psi_{\rm N}} / \dd\tau$ is negative).
It happens that the greater value of (ISW)$^2$ more than compensates the
lower value of (SW)$\times$(ISW) (which has a greater amplitude but is negative)
and the full CMB power spectrum $C_{\ell}$ is larger than the $\Lambda$-CDM
prediction.

For the models with $\bar{K}'<0$, the ISW
effect is smaller than in the $\Lambda$-CDM case and its has a different sign,
because as seen in Fig.~\ref{fig_Dpsi_z} the time derivative
$\dd D_{\Psi_{\rm N}} / \dd\tau$ is now positive at high redshifts, $z \gtrsim 0.4$.
Then, the (ISW)$^2$ contribution is smaller than the $\Lambda$-CDM prediction
but the cross-term (SW)$\times$(ISW) is now positive and this more than compensates
the decrease of (ISW)$^2$, and the full CMB power spectrum $C_{\ell}$ is again
larger than the $\Lambda$-CDM result.

Although low-$\ell$ measurements have large error bars due to the cosmic variance,
the model (\ref{K-power-3}) shown in the upper panel of Fig.~\ref{fig_Cl} is
already excluded by data (e.g., WMAP9 \cite{Hinshaw2013}).
However, as shown in the upper and lower panels, the exact values of $C_{\ell}$
depend on the parameters of the model and low-$\ell$ measurements of $C_{\ell}$
cannot rule out the full class of models.
In particular, it might be possible to choose parameters, or functions $K(\chi)$
and $A(\varphi)$, so that the net effect is to decrease the CMB power spectrum
at the lowest $\ell$, which would provide a better match to data than the
$\Lambda$-CDM reference.
However, we have not made a detailed search of the parameter space to find out
whether such a result can be achieved.
On the other hand, the results shown in Fig.~\ref{fig_Cl} suggest that ``generic''
models typically yield a stronger growth of the CMB power at low $\ell$ than
for the $\Lambda$-CDM scenario.

\section{Spherical collapse}
\label{Spherical}

\subsection{Spherical dynamics}
\label{Spherical-dynamics}

From Eq.(\ref{Euler-3}) the particle trajectories read in physical coordinates $(\vr,t)$
as
\beq
\ddot{\vr} + \frac{\dd\ln \bar{A}}{\dd t} \dot{\vr} - \left( \frac{\ddot{a}}{a} + \frac{\dot{a}}{a}
\frac{\dd\ln \bar{A}}{\dd t} \right) \vr = - \nabla_r ( \Psi_{\rm N} + \ln A ) ,
\label{traject-1}
\eeq
where $\nabla_r = \nabla/a$ is the gradient operator in physical coordinates.
To study the spherical collapse before shell crossing, it is convenient to label each
shell by its Lagrangian radius $q$ or enclosed mass $M$, and to introduce its
normalized radius $y(t)$ by
\beq
y(t) = \frac{r(t)}{a(t) q} \;\;\; \mbox{with} \;\;\; q = \left(\frac{3M}{4\pi\bar\rho_0}\right)^{1/3} ,
\;\;\; y(t=0) = 1 .
\label{y-def}
\eeq
In particular, the matter density contrast within radius $r(t)$ reads as
\beq
1+ \delta(<r) = y(t)^{-3} ,
\label{deltaR-def}
\eeq
where $\delta = (\rho - \bar\rho)/\bar\rho$ is the matter density contrast.
In terms of $y(t)$, Eq.(\ref{traject-1}) also reads as
\beqa
\frac{\dd^2 y}{\dd\eta^2} + \left(
\frac{1-3 w_{\varphi}^{\rm eff} \Omega_{\varphi}^{\rm eff}}{2}
\!+\! \frac{\dd\ln \bar{A}}{\dd\eta} \right) \frac{\dd y}{\dd\eta} & = & \nonumber \\
&& \hspace{-4cm} - \frac{3 \Omega_{\rm m} \, y}{8\pi {\cal G} \bar\rho \, r}
\frac{\pl}{\pl r} ( \Psi_{\rm N} + \ln A ) ,
\label{yt-2}
\eeqa
where as in Sec.~\ref{Perturbation-theory} we use $\eta=\ln a$ as the time coordinate.

The Newtonian potential is given by the modified Poisson equation
(\ref{Poisson-2}), which gives in spherically symmetric configurations the
Newtonian force
\beq
F_{\rm N} = - \frac{\pl\Psi_{\rm N}}{\pl r} = - \bar{A} \frac{\cG \delta M(<r)}{r^2} .
\label{FN-def}
\eeq
The Klein-Gordon equation (\ref{KG-pert-3}) reads in spherically symmetric
configurations as
\beq
\frac{1}{r^2} \frac{\pl}{\pl r} \left( r^2 \frac{\pl\varphi}{\pl r} \bar{K}' \right) =
\frac{\bar{A}\beta_1}{M_{\rm Pl}} \delta\rho ,
\label{KG-spher}
\eeq
hence
\beq
\frac{\pl\varphi}{\pl r} = \frac{\bar{A}\beta_1}{4\pi M_{\rm Pl}\kappa_1}
\frac{\delta M(<r)}{r^2} .
\label{phi-spher}
\eeq
In Eqs.(\ref{traject-1})-(\ref{phi-spher}) we have assumed the small-scale regime,
$ctk/a\gg 1$, where relative fluctuations of $A$ are negligible as compared
with relative density fluctuations, see Eq.(\ref {dA-drho}).
Then, the fifth force reads from Eq.(\ref{phi-spher}) as
\beq
F_{\rm A} = - \frac{\pl\ln A}{\pl r} \simeq - \frac{\beta_1}{M_{\rm Pl}}
\frac{\pl\varphi}{\pl r} = \frac{2\beta_1^2}{\kappa_1} F_{\rm N} .
\label{FA-def}
\eeq
Thus, it is proportional to the Newtonian force (\ref{FN-def}) with a time dependent
prefactor. Moreover, this prefactor is again of the form $\beta^2/\bar{K}'$,
as for the deviations from $\Lambda$-CDM of the background and of particle
masses, see the companion paper and Eq.(\ref{beta-phi-1}) here,
and of matter clustering in the perturbative regime, see Sec.~\ref{Closed-system}
and Eqs.(\ref{eps1-1})-(\ref{eps2-1}).
This is not surprising because these are different probes of the same underlying model.
Then, the equation of motion (\ref{yt-2}) of the mass shell $M$ reads as
\beqa
\frac{\dd^2 y}{\dd\eta^2} + \left(
\frac{1-3 w_{\varphi}^{\rm eff} \Omega_{\varphi}^{\rm eff}}{2}
+ \epsilon_2 \right) \frac{\dd y}{\dd\eta} && \nonumber \\
&& \hspace{-3cm}  + \frac{\Om}{2} (y^{-3}-1)
y ( 1 + \epsilon_1 ) = 0 , \;
\label{yt-3}
\eeqa
where the factors $\epsilon_1$ and $\epsilon_2$ were defined in
Eq.(\ref{epsilon-def}).
The usual $\Lambda$-CDM dynamics are recovered when the factors
$\epsilon_i$ are set to zero, and the background terms
$w_{\varphi}^{\rm eff} \Omega_{\varphi}^{\rm eff}$ and $\Om$ follow the
$\Lambda$-CDM evolution.
Moreover, we can see that the nonlinear spherical dynamics (\ref{yt-3})
involve the same factors as the evolution equation (\ref{D-linear}) of the
linear modes.
This is made possible by the small-scale and moderate-density
regime (\ref{psiA-1}), which we used to derive Eq.(\ref{yt-3}).
In particular, this has  allowed us to write the linearized Klein-Gordon equation
(\ref{KG-spher}).
It is clear that if higher orders in Eq.(\ref{KG-spher}) had been
relevant, which is the case in the regime (\ref{psiA-2}) and at higher densities,
they would have given rise to new factors beyond the terms $\epsilon_1$
and $\epsilon_2$.

Thanks to the scale independence of the modification of gravity
brought by the model studied here, in the small-scale regime (\ref{psiA-1}),
we preserve a key property of the spherical $\Lambda$-CDM dynamics:
the motions of different mass shells are decoupled before shell crossing.
This greatly simplifies the analysis of the spherical collapse.
This property is not satisfied by other models of modified gravity, such as
$f(R)$ theories or dilaton models, where the fifth force shows an explicit
scale dependence that couples the motions of different shells.
This feature only applies to the regime (\ref{psiA-1}), which is sufficient
for our purposes as we consider density contrast $\delta \lesssim 200$.
If we considered higher-density regions, such as inner galaxy cores or the
Solar System, there would be a departure from the expression (\ref{FA-def}). In that case, nonlinearities
of the Klein-Gordon equation become important and give rise to the
K-mouflage mechanism that eventually leads to a recovery of General Relativity.
We do not consider this regime in this paper.

\subsection{Linear density contrast threshold}
\label{linear-threshold}

\begin{figure}
\begin{center}
\epsfxsize=8.5 cm \epsfysize=6. cm {\epsfbox{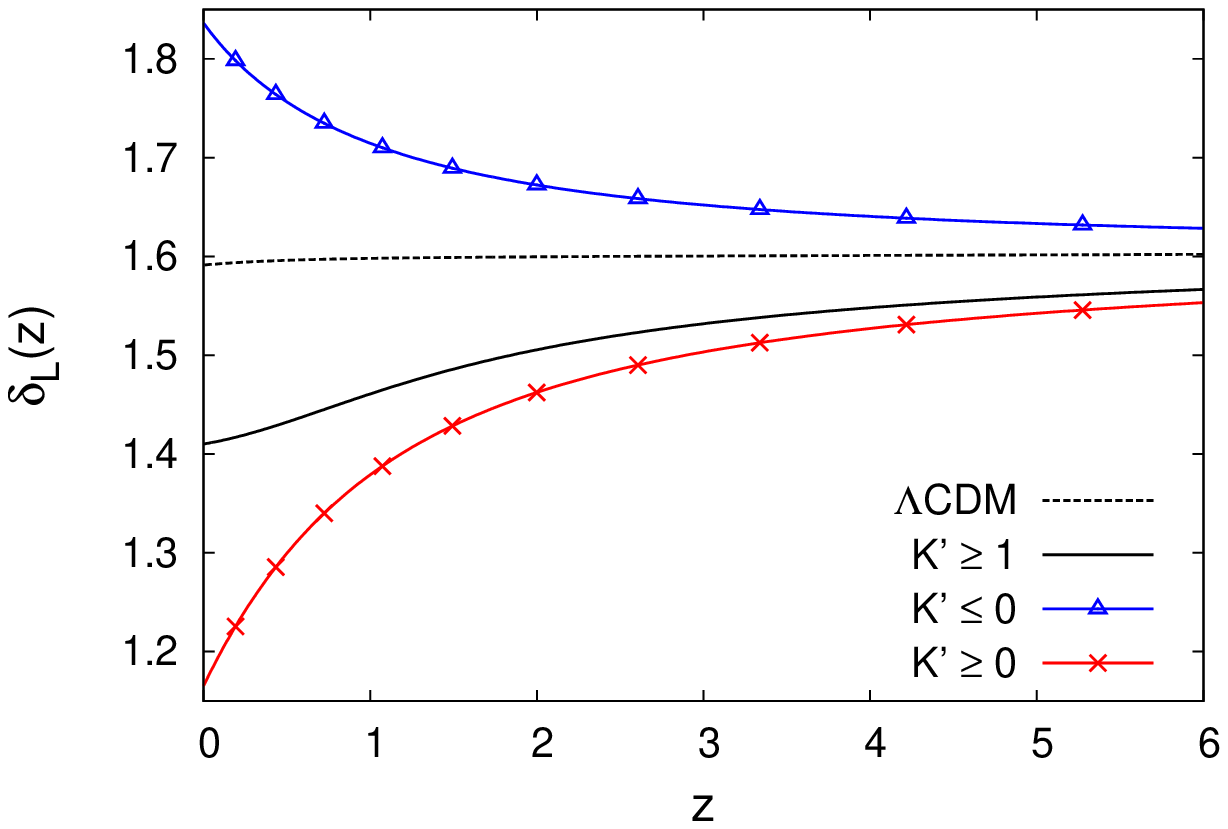}} \\
\epsfxsize=8.5 cm \epsfysize=6. cm {\epsfbox{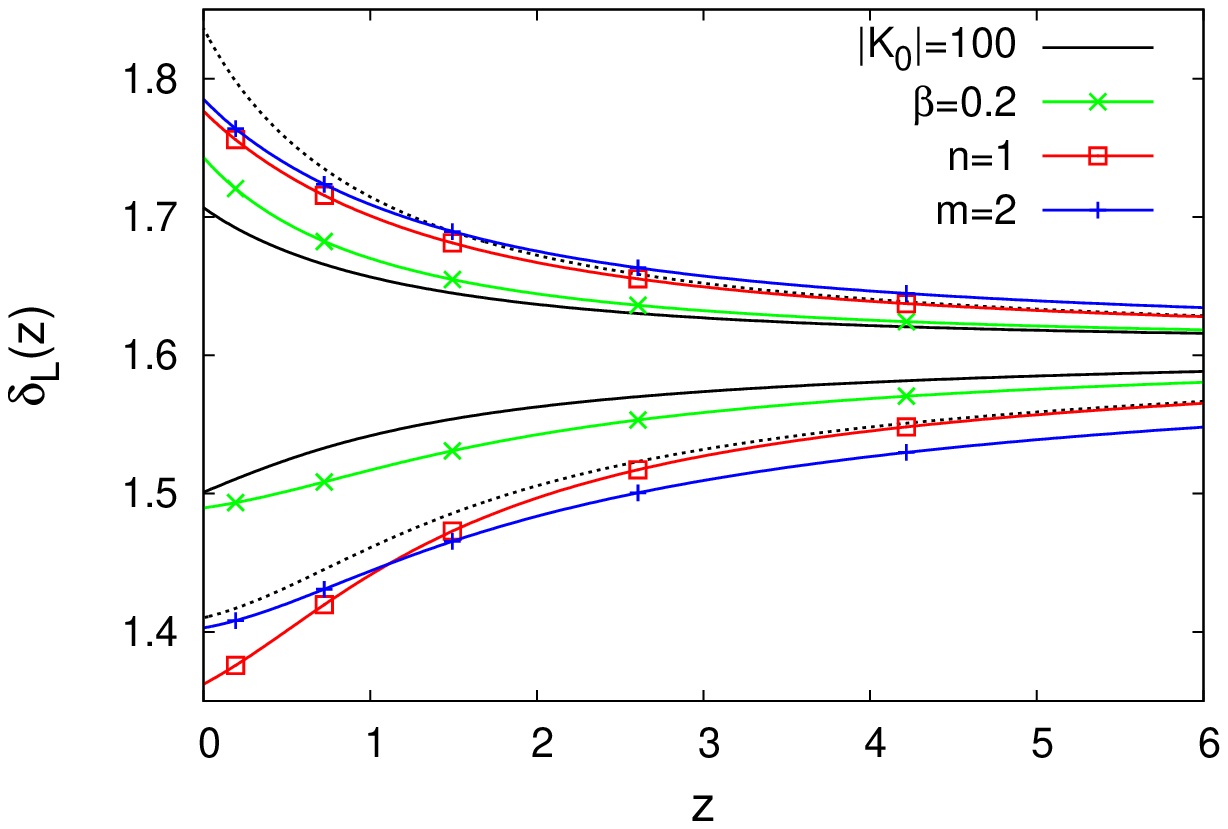}}
\end{center}
\caption{Linear density contrast threshold $\delta_{L(\Lambda)}(z)$.
{\it Upper panel:} same models as in the upper panel of Fig.~\ref{fig_Dlin_z}.
{\it Lower panel:} same models as in the lower panel of Fig.~\ref{fig_Dlin_z}.}
\label{fig_deltaLM_z0}
\end{figure}

By solving the equation of motion (\ref{yt-3}) we can numerically compute
the linear density contrast threshold $\delta_L(M,z)$ that corresponds to
a nonlinear density contrast of 200. (We choose a nonlinear overdensity of
200 to define virialized halos. This allows us to compare with previous works
and to use the same rescaled halo mass function.)
Because the modification of gravity is scale independent in the regime
(\ref{psiA-1}), the mass $M$ no longer appears in the equation of motion
(\ref{yt-3}). Therefore, the linear threshold $\delta_L(M,z)$ is actually independent
of the halo mass $M$, as in the $\Lambda$-CDM scenario.

In practice, rather than the linear threshold $\delta_L$ we consider
the linear threshold $\delta_{L(\Lambda)}$ associated with the initial conditions.
Indeed, if we wish to estimate the impact of the modification of gravity on
nonlinear matter clustering and on the halo mass function, we are not really interested
in the linear density contrast today, $\delta_L$, associated with a nonlinear
density contrast $\delta=200$, as $\delta_L$ cannot be directly observed.
Rather, we are interested in the initial (or early-time) linear threshold $\delta_{Li}$,
at a high redshift $z_i$, which is required to produce at a later time (e..g, today)
a nonlinear density contrast $\delta$. Indeed, from $\delta_{Li}$ we can estimate
from the initial Gaussian density field $\delta_{Li}(\vx)$ whether this threshold
corresponds to a rare or common density fluctuation.
In the usual $\Lambda$-CDM scenario, one usually ``translates'' both the
initial linear threshold $\delta_{Li}$ and the initial density field $\delta_{Li}(\vx)$,
or the root mean square density fluctuation $\sigma_{Li}$, to the present
time by multiplying them by the common linear growth factor $D_+(z_0)/D_+(z_i)$.
This avoids introducing the ``initial'' redshift $z_i$ as initial conditions are
expressed in terms of the current linear density field.

However, because we compare different cosmological scenarios, with slightly
different linear growing modes but with the same high-redshift linear
power spectrum, we must go back to the high redshift $z_i$. More precisely,
to compare the efficiency of the matter clustering process between these
cosmological scenarios, we wish to compare the probabilities associated with
a given nonlinear threshold $\delta=200$ today. This means that we wish to
compare the initial linear thresholds $\delta_{Li}$ required in each scenario
to reach the same $\delta$ today (because the initial Gaussian conditions are taken
to be the same, far in the matter era).
Nevertheless, to avoid introducing an explicit arbitrary high redshift $z_i$, and to
follow the usual practice, we translate all initial thresholds $\delta_{Li}$ to the
current time (or to the redshift $z$ of interest), by multiplying all of them by the
same $\Lambda$-CDM linear growth factor
$D_{+\Lambda\rm CDM}(z_0)/D_{+\Lambda\rm CDM}(z_i)$.
[In contrast, the ``true'' linear threshold $\delta_L$ in each cosmology is obtained
by multiplying by its own linear growth factor $D_+(z_0)/D_+(z_i)$].

In this fashion, the comparison between the various $\delta_{L(\Lambda)}$
gives a direct hint of the various probabilities to reach $\delta=200$ and of how
far the nonlinear matter clustering is advanced between the various models,
starting from the same linear power spectra at high $z$.
In contrast, if we consider the ``true'' linear thresholds $\delta_L$ the comparison
is biased by the fact that in different cosmologies the same $\delta_L$ actually
corresponds to different initial conditions at a given high $z$.
(Going back to the initial redshift $z_i$, as we implicitly do here, is also
more convenient in more general modified-gravity models where the
linear growing modes become scale dependent at late times.)

We show our results in Fig.~\ref{fig_deltaLM_z0}. In agreement with the
results of Sec.~\ref{Perturbation-theory}, where we have found that for models with $\bar{K}'>0$
the scalar field accelerates the clustering of matter as it leads to greater linear
growing modes and one-loop power spectra, we find that a smaller linear
density contrast  $\delta_{L(\Lambda)}(z)$ is required to reach the same nonlinear
overdensity of $200$ than in the $\Lambda$-CDM scenario.
Conversely, models with $\bar{K}'<0$ lead to a greater linear threshold $\delta_{L(\Lambda)}$.
In all cases, we recover the $\Lambda$-CDM reference value at high redshift.
The departure from the $\Lambda$-CDM reference grows faster at low $z$ for the
models (\ref{K-power-2}) and (\ref{K-power-3}) where $\bar{K}'\rightarrow 0$ at late times,
in agreement with Figs.~\ref{fig_eps_z} and \ref{fig_Dlin_z}, and the greater amplitude of the
characteristic ratio $\beta^2/\bar{K}'$.

Again, the lower panel shows that a higher value of $|K_0|$, or more generally $|\bar{K}'|$,
and a smaller $\beta$, lead to a smaller deviation from the $\Lambda$-CDM reference.
The characteristic exponents $n$ and $M$ of the coupling function $A(\varphi)$ and
of the kinetic function $K(\chi)$ do not have a great quantitative impact.

\subsection{Halo mass function}
\label{Halo mass function}

\begin{figure}
\begin{center}
\epsfxsize=8.5 cm \epsfysize=6. cm {\epsfbox{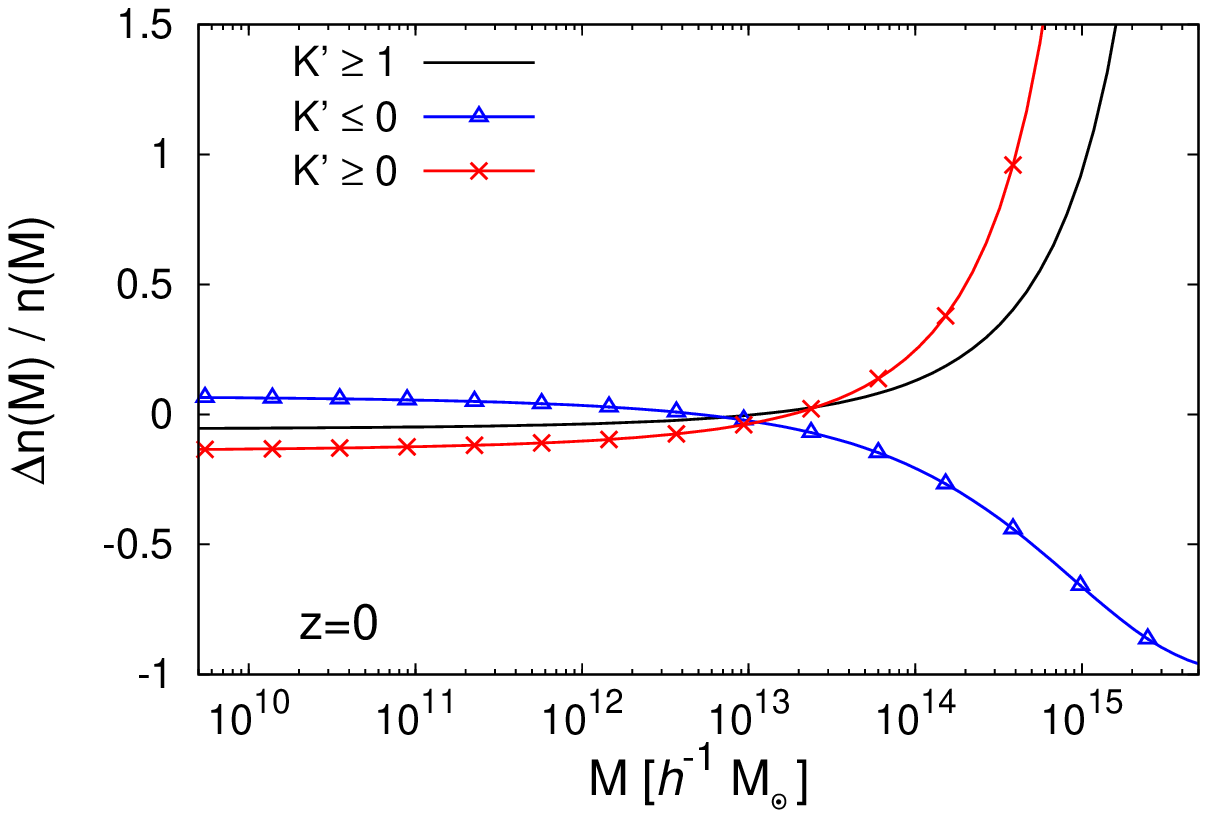}} \\
\epsfxsize=8.5 cm \epsfysize=6. cm {\epsfbox{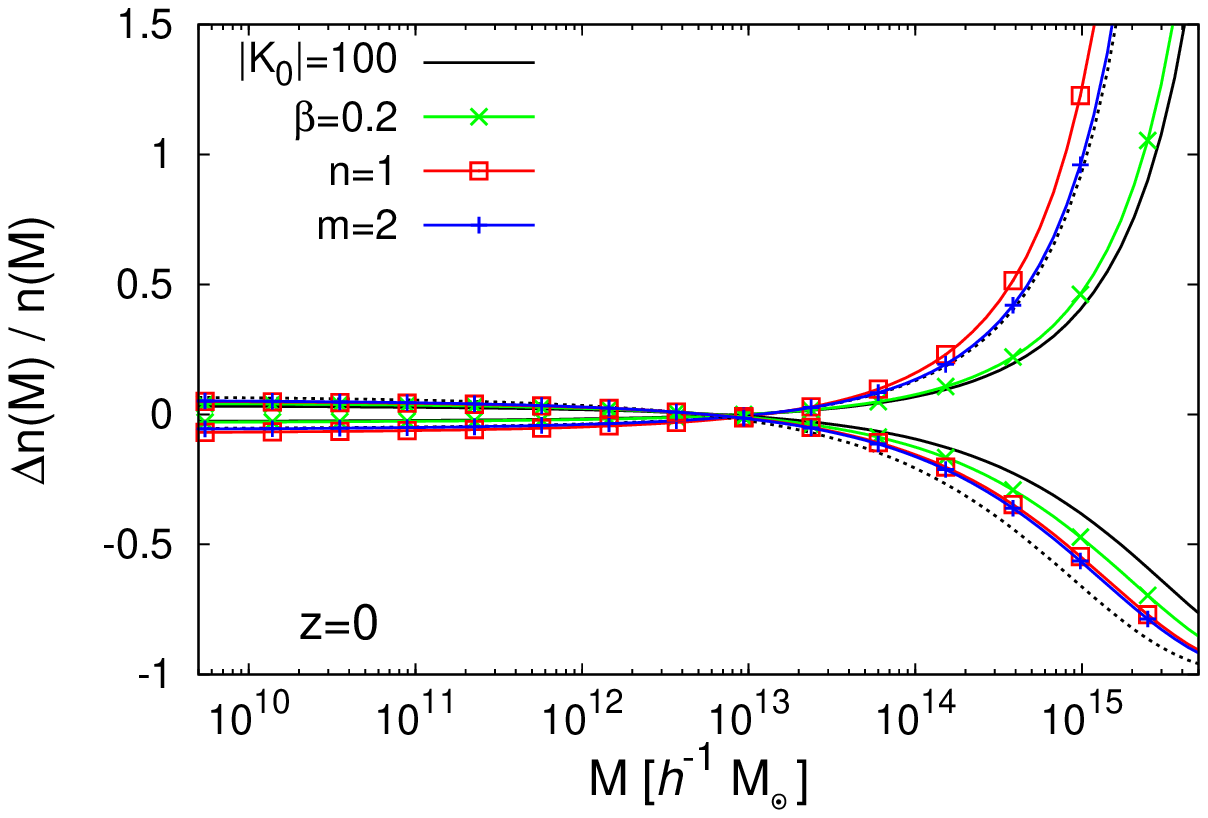}}
\end{center}
\caption{Relative deviation
$[n(M)-n_{\Lambda\rm CDM}(M)]/n_{\Lambda\rm CDM}(M)$
of the halo mass function from the $\Lambda$-CDM reference, at $z=0$.
{\it Upper panel:} same models as in the upper panel of Fig.~\ref{fig_Dlin_z}.
{\it Lower panel:} same models as in the lower panel of Fig.~\ref{fig_Dlin_z}.}
\label{fig_dnM_z0}
\end{figure}

As usual, we write the comoving halo mass function $n(M) \dd M/M$ as
\beq
n(M) \frac{\dd M}{M} = \frac{\bar\rho_0}{M} f(\nu) \frac{\dd\nu}{\nu} ,
\;\;\; \mbox{with} \;\;\; \nu = \frac{\delta_{L(\Lambda)}}{\sigma_{(\Lambda)}(M)} .
\label{nM-def}
\eeq
Here $\sigma_{(\Lambda)}(M)$ is the root mean square of the linear density contrast
at scale $M$ and $\delta_{L(\Lambda)}$ is the linear density contrast associated with
the nonlinear density threshold of $200$ that defines the virialized halos,
both being translated from the initial conditions by the $\Lambda$-CDM
growth factor $D_{+\Lambda\rm CDM}(z_0)/D_{+\Lambda\rm CDM}(z_i)$
as explained in Sec.~\ref{linear-threshold}.
Thus, the scaling variable $\nu$ directly measures the probability of
density fluctuations in the Gaussian initial conditions.
Then, we take for the scaling function $f(\nu)$ the fit to $\Lambda$-CDM simulations
obtained in \cite{Valageas2009}, which obeys the exponential tail
$f(\nu) \sim e^{-\nu^2/2}$ at large $\nu$.
This means that the mass function (\ref{nM-def}) shows the correct large-mass
tail, which is set by the Gaussian initial conditions and the relationship between
the current nonlinear density contrast $\delta$ and the associated initial linear
density contrast $\delta_{Li}$ [or equivalently $\delta_{L(\Lambda)}$], which
was obtained in Sec.~\ref{linear-threshold}.
The deviation from the $\Lambda$-CDM reference at low mass is not meant
to be accurately reproduced by this model (e.g., we neglect any dependence
on modified gravity of the exponent of the low-mass power law tail).
However, the low-mass range is not very important for our purposes and it
is constrained by the normalization condition $\int (M/\rhob_0) n(M) \dd M/M =1$,
which is automatically satisfied by our simple approximation.

We show our results in Fig.~\ref{fig_dnM_z0}.
As usual, the deviation from the $\Lambda$-CDM reference is most important
at high mass because the Gaussian cutoff $e^{-\nu^2/2}$ amplifies the
sensitivity of rare events to the collapse dynamics.
In agreement with the results obtained in the previous sections,
models with $K_0>0$, or more generally $\bar{K}'>0$, lead to a faster matter clustering
and to a greater number of rare massive halos, while $\bar{K}'<0$ leads to fewer massive halos.
(The normalization constraint to unity implies that the deviation from the
$\Lambda$-CDM mass function changes sign between the high-mass and low-mass
tails.) Again, a higher $|K_0|$ or a lower $|\beta|$ implies smaller deviations from the
$\Lambda$-CDM reference and the results are mostly sensitive to the parameters
$K_0$ and $\beta$.

\section{Nonlinear matter power spectrum}
\label{nonlinear-Pk}

\begin{figure}
\begin{center}
\epsfxsize=8.5 cm \epsfysize=6. cm {\epsfbox{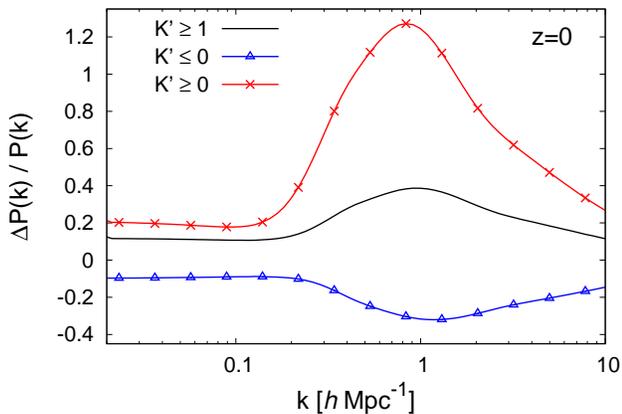}} \\
\epsfxsize=8.5 cm \epsfysize=6. cm {\epsfbox{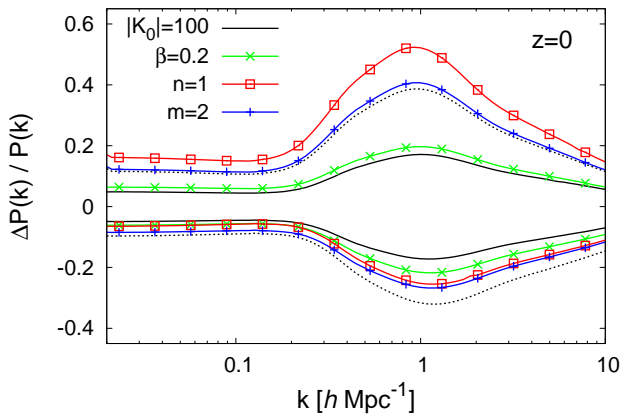}}
\end{center}
\caption{Relative deviation
$[P(k)-P_{\Lambda \rm CDM}(k)]/P_{\Lambda \rm CDM}(k)$
of the nonlinear matter density power spectrum from the $\Lambda$-CDM reference,
at redshift $z=0$.
{\it Upper panel:} same models as in the upper panel of Fig.~\ref{fig_Dlin_z}.
{\it Lower panel:} same models as in the lower panel of Fig.~\ref{fig_Dlin_z}.}
\label{fig_dPk_z0}
\end{figure}

\begin{figure}
\begin{center}
\epsfxsize=8.5 cm \epsfysize=6. cm {\epsfbox{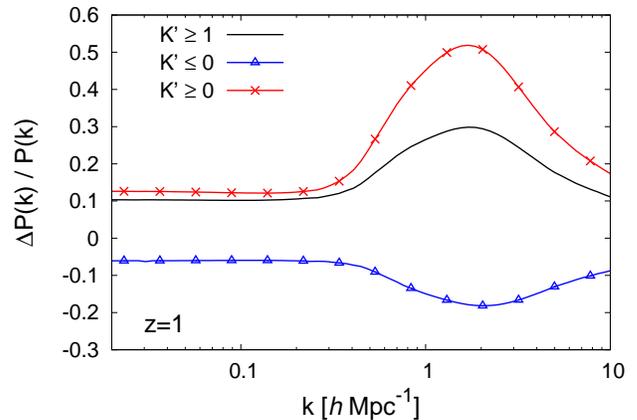}} \\
\epsfxsize=8.5 cm \epsfysize=6. cm {\epsfbox{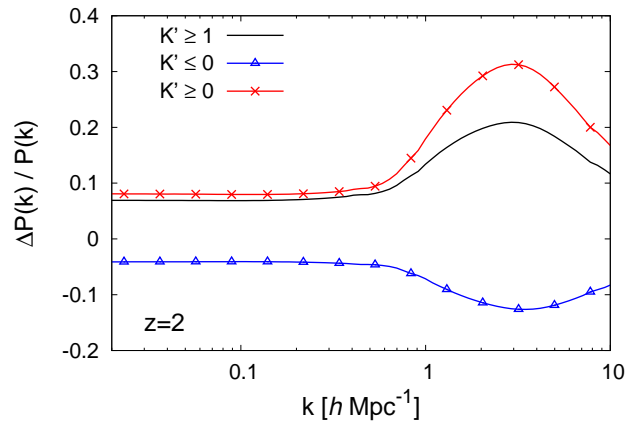}}
\end{center}
\caption{Same as in the upper panel of Fig.~\ref{fig_dPk_z0}, but at
redshifts $z=1$ (upper panel) and $z=2$ (lower panel).}
\label{fig_dPk_z1z2}
\end{figure}

Following the method built in \cite{Valageas2013} for the $\Lambda$-CDM
scenario, and applied to several modified-gravity models in \cite{Brax2013},
we combine the one-loop perturbation theory obtained in
Sec.~\ref{Perturbation-theory} with a halo model to obtain the nonlinear matter density
power spectrum from small to large wave numbers. By construction, this power spectrum
agrees with Eq.(\ref{Ptree+1loop}) when it is expanding up to order $P_{L0}^2$.
In the halo model that governs its high-$k$ limit, we take into account the
impact of the modified gravity on the nonlinear dynamics through the
halo mass function (\ref{nM-def}) (i.e., through the acceleration or slowing down
of the spherical collapse), but we neglect the impact of the modified gravity
on the halo profiles (i.e., we keep the NFW profile from \cite{Navarro1997}
and the mass-concentration relation from \cite{Valageas2013}).

We show our results in Figs.~\ref{fig_dPk_z0} and \ref{fig_dPk_z1z2}.
At low $k$ we recover the one-loop power spectra shown in
Fig.~\ref{fig_dP1loop_z0}, with an almost $k$-independent relative deviation from the
$\Lambda$-CDM reference, because of the scale independence of our
modified-gravity models in the regime (\ref{psiA-1}).
The deviations are amplified on mildly nonlinear scales, $k \sim 1 h$Mpc$^{-1}$
at $z=0$, as they become sensitive to later stages of the nonlinear dynamics
and to the large-mass tail of the halo mass function (see for instance
\cite{Valageas2013a}).
At higher $k$ the relative deviations decrease because the power spectrum is
governed by the low-mass tail of the halo mass function and the inner halo
density profiles. However, we may underestimate the signal at
$k \gtrsim 10 h$Mpc$^{-1}$ because we neglected the impact of the modification
of gravity on these halo profiles.

Again, the sign of the deviation from the $\Lambda$-CDM reference depends on
the sign of $K_0$, or $\bar{K}'$, and the results are mostly sensitive to $K_0$ and $\beta^2$.

The deviation from the $\Lambda$-CDM prediction decreases at high redshift,
because we normalize the linear power spectra to the same initial value at early
times, far in the matter era.
However, we can see in Fig.~\ref{fig_dPk_z1z2} that this decrease is rather slow
and that significant deviations are already present at $z=2$ in the matter power
spectrum.

The comparison with the background results obtained in the companion paper \cite{Brax:2014aa}
shows that the relative deviations
are significantly greater, by about a factor $10$, for $P(k)$ than for background
quantities such as the Hubble expansion rate $H(z)$.
Therefore, large-scale structures provide a useful probe of such modified-gravity
scenarios.
In particular, it is possible to keep a background evolution that is very close the
the $\Lambda$-CDM cosmology, at the percent level, while obtaining significant
departures in terms of the matter clustering, at the $10\%$ level
[in terms of $P(k)$, higher-order statistics such as the bispectrum, or the high-mass
tail of the halo mass function, can show even greater deviations].

\section{Comparison with other models}

In this section, we compare our results with the ones of the chameleon-$f(R)$ models \cite{Brax:2013mua,Li:2012by}, the dilatons \cite{Brax:2011ja}, the symmetrons \cite{Brax:2012nk}, the DGP model \cite{Li:2013nua} and the Galileon theories \cite{Barreira:2013eea,Li:2013tda} when the N-body simulations are available.

Let us start with the chameleons, dilatons and symmetrons. For all these models, the background follows the one of $\Lambda$-CDM. At the perturbative level, and first in the linear theories, deviations from GR occur on scales lower than the Compton wavelength of the scalar field \cite{Brax:2004px}. As Solar System tests and the screening of the Milky Way imply that the cosmological range of the scalar must be less than 1 Mpc \cite{Brax:2011aw}, the effects of these models on linear scales are suppressed and only in the quasilinear to mildly nonlinear regimes one can expect to see significant deviations. Symmetrons and dilatons screen gravity in a stronger way in the local environment implying that constraints on these models are less severe than on chameleon-$f(R)$ theories. This implies that the effects of the symmetron and to a lesser extent of the dilaton on large-scale structures are enhanced compared to chameleon-$f(R)$ models. Typically, one expects to see a peak in the deviations from GR on the scales corresponding to the range of the scalar field, especially in the power spectrum of density fluctuations \cite{Brax2013}. On small and large scales, the models converge towards GR. On small scales, this is due to the screening effect and on large scales this is also the screening property outside the Compton radius.

For the DGP model on the self-accelerated branch \cite{Li:2013nua}, the background is modified compared to $\Lambda$-CDM with an increase of $H$ compared to $\Lambda$-CDM. At the perturbative level, the DGP model leads to
a scale-independent decrease of Newton's constant on large and linear scales. On nonlinear scales, the Vainshtein mechanism reduces the negative deviations from GR and applies on scale as large as $0.1 h {\rm Mpc}^{-1}$. Moreover, the overall deviation from GR is significantly reduced, even on large scales, at redshifts $z=1$ and $z=2$. This comes from the fact that the deviation of the background from $\Lambda$-CDM is a late-time effect.

The cubic and quartic Galileon models \cite{Barreira:2013eea,Li:2013tda} have been simulated with differences between the two models. In both cases, the background evolution follows a tracker solution where $\dot \varphi$ goes like $1/H$, i.e. the field varies more with time in the recent past implying a stronger effect on the growth of structure on large scales in the linear regime. In both the cubic and quartic scales, the effective Newton constant is larger than in GR in sparse regions of the Universe. The Vainshtein mechanism operates on very mildly quasilinear scales as soon as $0.03 h {\rm Mpc}^{-1}$ where nonlinear effects cannot be neglected. Moreover, in the densest regions the effective Newton constant changes sign in the quartic case and becomes smaller than in GR. As in the DGP case, the deviations from GR decrease significantly from $z=0$ to $z=1$ and beyond $z=1$ the nonlinear effects of the Galileon models can be neglected and the models behave like linear theories with a time dependent but scale-independent Newton constant.

The K-mouflage models also have a modified background evolution which becomes prominent in the recent past of the Universe and deviations from $\Lambda$-CDM are significant from a redshift $z\sim 5$ with maximal extension between $z=1$ and $z=2$. At the perturbative level, linear effects depend on a time dependent Newton constant and a new friction term implying that the growing mode and the growth rate are enhanced for $K_0>0$ and depleted when $K_0<0$ compared to $\Lambda$-CDM. As the K-mouflage models do not screen the quasilinear structures of the Universe, we find that linear features persist beyond the linear regime. We find that the one loop contribution to perturbation theory is negligible and that, in fact, the K-mouflage models behave like a linear model in the scalar sector. Nonlinearities are only present as usual in the matter sector. These nonlinearities imply that the deviation from $\Lambda$-CDM of the power spectrum have a peak at the onset of the nonlinear regime around $1\ h {\rm Mpc}^{-1}$ for $z=0$. In the nonlinear regime, the critical linear density contrast is smaller than the $\Lambda$-CDM one for $K_0>0$ in a scale-independent way, leading to an increase in the number of large-mass clusters for $K_0>0$. We have also studied the evolution of the deviations from $\Lambda$-CDM with the redshift and shown that they increase in the recent past corresponding to the maximal deviations of the background from $\Lambda$-CDM between $z=1$ and $z=2$. Moreover, when subtracting the power spectrum calculated with the modified background (QCDM, see App.~\ref{Modified-background}), we find that the influence of the new friction term and the modified Newton constant on linear scales and their extension to the nonlinear regime are not negligible and vary very little from $z=0$ to $z=2$, apart from a shift in the peak of the power spectrum due to the change of the onset of nonlinearities with the redshift.

Hence we can draw general conclusions about the differences between the three types of screening mechanisms at the cosmological level. Models with the Vainshtein and K-mouflage properties have a modified background cosmology when
chameleonlike models follow $\Lambda$-CDM. The deviations from $\Lambda$-CDM are more pronounced for K-mouflage models at redshifts of a few than for Vainshtein-like models. Structures grow differently too. On linear scales, chameleons converge to GR while models of the K-mouflage and Vainshtein types show deviations there. K-mouflage models do not screen quasilinear structures whereas chameleonlike and Vainshtein models do. Moreover, K-mouflage models have deviations from $\Lambda$-CDM which persist up to redshifts of a few when for the other types of screening the effects become less significant.
These features are sufficiently different to hope to distinguish these models if modified gravity effects were to be detected by future surveys.

\section{Summary and conclusion}

\subsection{Summary}

Before we conclude this paper, let us briefly summarize the main properties of the
K-mouflage models studied here and our results:

- K-mouflage models involve an additional scalar field, $\varphi$, with a nonstandard
nonlinear kinetic term, ${\cal M}^4 K[-(\pl\varphi)^2/2 {\cal M}^4]$, where ${\cal M}^4 $ is of the order of the critical density now. Here we consider models where the field $\varphi$
is also conformally coupled to matter fields through the Jordan metric
$\tilde{g}_{\mu\nu} = A^2(\varphi) g_{\mu\nu}$.

- The nonlinearity of the Lagrangian, which gives rise to terms $\bar{K}' (\pl\delta\varphi)^2$ for the fluctuations
with respect to the cosmological  background, provides a ``screening mechanism'' as the  large prefactor
$\bar{K}'$ freezes the fluctuations and suppresses the fifth force in the high-density and small-scale
regimes. This provides a convergence to GR on small astrophysical scales (studied in an upcoming
paper) and at high redshift \cite{Brax:2014aa}. On the other hand, in contrast with some other
modified-gravity scenarios (e.g., chameleon and Galileon models), linear cosmological structures
are unscreened and show deviations from $\Lambda$-CDM up to the Hubble scale.
Moreover, the dark energy background evolution of K-mouflage models only behaves as a cosmological constant at low redshifts. 

- The equations of motion obtained for matter on cosmological scales are the usual continuity
equation, a modified Euler equation (with an additional friction term and an additional fifth-force
potential term), and a modified Poisson equation (with a time dependent effective Newton constant).
The scalar field fluctuations obey a time dependent Klein-Gordon equation, as the background
field evolves with time.

- Even though this background does not follow a quasistatic evolution, on small scales
(far below the horizon) the scalar field fluctuations obey a quasistatic regime (because
spatial gradients dominate over time derivatives) and are ``slaved'' to the same-time density
fluctuations (i.e., the Klein-Gordon equation takes the form of a nonlinear Poisson equation).

- For cosmological structures, from the cosmic web down to clusters of galaxies, which show
moderate density contrasts ($\delta \lesssim 200$), this quasistatic Klein-Gordon equation
can be linearized. Therefore, these models provide an explicit nonlinear example
where the scalar field sector can be linearized on cosmological scales (while the nonlinearity
appears on smaller astrophysical scale and ensures the convergence back to GR).

- Quasilinear scales can be studied using cosmological perturbation theory as in the standard
$\Lambda$-CDM scenario, taking into account the new linear friction and fifth-force terms
in the Euler equation. In particular, the nonlinearities are due to the usual transport terms
that are identical to those found in the $\Lambda$-CDM case.

- The linear regime growth factors differ from the $\Lambda$-CDM predictions through
time dependent terms, but in contrast with some other modified-gravity models, they
do not show an additional scale dependence (because large scales remain unscreened
up to the horizon). The sign of the deviation from $\Lambda$-CDM is set by the sign of
the derivative $\bar{K}'$, as for background quantities \cite{Brax:2014aa}.
For instance, models with $K'>0$ yield a smaller Hubble expansion rate $H(z)$
(with a common normalization today) and larger linear growth rates $D_+(z)$ and $f(z)$.
More precisely, the quantity that governs the deviations from the $\Lambda$-CDM
predictions, both for the background and the perturbations, is the ratio $\beta^2/\bar{K}'$, where
$\beta$ is the coupling constant to the matter. As for small-scale screening, these deviations
are suppressed in models with a large nonlinear factor $\bar{K}'$.

- Because large linear scales deviate from the $\Lambda$-CDM predictions, large-scale
CMB anisotropies also show a significant deviation through the ISW effect.
In particular, the cross-correlation between the large-scale CMB temperature fluctuations
and low-redshift galaxy surveys can change sign for models with $K'<0$, and the amplitude
of the relative deviation from $\Lambda$-CDM grows with redshift.
This also gives rise to a deviation for the low-$\ell$ CMB multipoles $C_{\ell}$,
but this generically yields more power than the $\Lambda$-CDM prediction over $\ell \leq 10$
whatever the sign of $\bar{K}'$.

- To go beyond the perturbative regime, we have also studied the spherical collapse dynamics.
For the same reason as the absence of scale dependence in the linear regime, the
spherical collapse is only modified by time dependent but scale-independent factors.
This also means that, as in GR or Newtonian gravity, different mass shells remain uncoupled
until shell crossing. This simplifies the analysis and it leads to a time dependent linear
density contrast threshold $\delta_{L}(z)$ for the collapse of virialized halos.
This yields in turn a deviation for the large-mass tail of the halo mass function, that again depends
on the sign of $\bar{K}'$.

- Combining perturbation theory and the spherical collapse dynamics, we have estimated the
matter power spectrum up to mildly nonlinear scales $k \lesssim 10 h$Mpc$^{-1}$.
We recover a constant relative deviation from $\Lambda$-CDM on linear scales and a peak
on weakly nonlinear scales, $k \sim 1 h$Mpc$^{-1}$, due to the amplification associated with
the nonlinear matter dynamics and the large-mass tail of the halo mass function.
Therefore, large-scale structures provide a useful probe of such models as the deviations
from $\Lambda$-CDM for $P(k)$, or the halo mass function, can be greater by a factor of 10
than those of background quantities, such as $H(z)$.

- These deviations decrease rather slowly at higher redshift and
remain non-negligible at $z=2$ (as compared to $z=0$).
This is due to the fact that the dark energy component only slowly becomes subdominant at
high $z$, because its energy density actually grows (but at a smaller rate than the matter
density).
This feature is rather different from the behavior obtained in some other modified-gravity models
[e.g., $f(R)$ theories or dilaton models] where the background is almost identical to
$\Lambda$-CDM and the deviations for matter perturbations are only significant at low $z$.

\subsection{Conclusions}

In conclusion, K-mouflage is an alternative to the screening by the chameleon or the Vainshtein mechanisms with striking features on the growth of large-scale structures. The most significant one is certainly the
absence of screening of large astrophysical  objects on cosmological scales, such as galaxy clusters. In this regime, the scalar theory behaves like a linear field theory leading to a time dependent modification of Newton's constant and an increase/decrease of the growth of structure compared to $\Lambda$-CDM depending on the ratio $\beta^2/\bar K'$ corresponding to the square of the effective coupling to matter when the bare coupling $\beta$ is rescaled by the wave function normalization of the field $|\bar K'|^{1/2}$. For models where deviations from the $\Lambda$-CDM behavior at the background level are at the percent level, the deviations of the power spectrum of the density contrast on mildly nonlinear scales is enhanced compared to the linear part of the spectrum and can reach ten percent. Moreover, the convergence to the Einstein- de Sitter behavior of perturbations in the past is rather slow due to the properties of the background cosmology. Indeed, at the background level, the Hubble rate converges to
the Einstein- de Sitter case in the distant past due to the screening of the scalar field in the high-density environment of the early Universe while it converges to a $\Lambda$-CDM behavior in the very recent past. In the intermediate regime around redshifts of $1\lesssim z \lesssim 2$, the Hubble rate can differ significantly from its $\Lambda$-CDM counterpart. This translates into a relative persistence of the deviations from $\Lambda$-CDM which differs from other screening mechanisms, up to redshifts of a few.
We leave a more detailed analysis of cosmological observational constraints to future works.
K-mouflage could also have different features on smaller scales where the density contrast is larger than in galaxy clusters. In this regime, the nonlinearities of the models reappear and cannot be neglected, especially
on scales of the order of the K-mouflage radius. This is left for future work.

\begin{acknowledgments}

This work is supported in part by the French Agence Nationale de la Recherche under Grant ANR-12-BS05-0002. Ph. B. acknowledges partial support from the  European Union FP7  ITN
INVISIBLES (Marie Curie Actions, PITN- GA-2011- 289442) and from the Agence Nationale de la Recherche under contract ANR 2010 BLANC 0413 01.

\end{acknowledgments}

\appendix

\section{Perturbations}
\label{appendix-Euler}

In this appendix we derive the perturbation equations used in the text.
First of all, the energy-momentum tensor of a CDM fluid is
\be
T^{\mu\nu}=\rho_E u^\mu u^\nu
\ee
where $\rho_E$ is the energy density in the Einstein frame and $u^\mu$ is the velocity 4-vector normalized such that
$
u^\mu u_\mu=-1 ,
$
where indices are raised or lowered using the Einstein metric $g_{\mu\nu}$.
Notice that we have the identification
$
u^\mu= \frac{\dd x^\mu}{\dd{\cal T}}
$
where ${\cal T}$ is the proper time and
$
\dd{\cal T}^2= -g_{\mu\nu} \dd x^\mu \dd x^\nu .
$
In terms of perturbations in the conformal Newton gauge, where
\be
\dd s^2=a^2 [ -(1+2\Psi_{\rm N}) \dd\tau^2 +(1-2\Psi_{\rm N}) \dd \vx^2 ] ,
\ee
we have
$
u^\mu= a^{-1}(1-\Psi_{\rm N} + v_j v^j/2, v^i)
$
where $v^i$ is the velocity vector of CDM particles identified to leading order as
$
v^i= \frac{\dd x^i}{\dd\tau}
$
and we also have to leading order
$
\dd{\cal T}= a \dd\tau \sqrt{1+\Psi_{\rm N}-v^iv_i/2}.
$
Here and throughout this paper, the particles are nonrelativistic and Newton's potential is small, and we only keep terms up to first order over $\Psi_{\rm N}$ and $v^2$
(within virialized halos we have $v^2 \sim \Psi_{\rm N}$).
The Bianchi identity and the Klein-Gordon equations imply that matter is not conserved but
\be
D_\mu T^{\mu\nu}_{(\rm m)} = - \rho_E \, \partial^\nu (\ln A) .
\label{con}
\ee
This follows directly from
$
G^{\mu\nu}= 8\pi \cG [ T^{\mu\nu}_{(\rm m)} +T^{\mu\nu}_{(\varphi)} ]
$
and
$
D_\mu G^{\mu\nu}=0 ,
$
leading to
$
D_\mu T^{\mu\nu}_{(\rm m)} =- D_\mu T^{\mu\nu}_{(\varphi)}
$
where
$
T^{\mu\nu}_{(\varphi)}= K' \partial^\mu\varphi \partial^\nu\varphi + g^{\mu\nu} \cM^4 K
$.
The Klein-Gordon equation is
\be
D_\mu( \partial^\mu \varphi K')= -\beta\frac{ T}{M_{\rm Pl}} ,
\ee
where
\be
\beta= M_{\rm Pl} \frac{\dd\ln A}{\dd\varphi}.
\ee
Using this we easily get (\ref{con}) with $T=-\rho_E$.
This gives
explicitly
\be
\dot \rho_E u^\nu + 3h\rho_E  u^\nu +\rho_E u^\mu D_\mu u^\nu= - \rho_E \,  \partial^\nu(\ln A)
\label{big}
\ee
where we have introduced
$
\dot\rho_E=u^\mu D_\mu \rho_E
$
 and
 the local Hubble rate
$
 3h=D_\mu u^\mu .
$

Contracting  with $u_\nu$ and using $u^2=-1$, we get
\be
\dot \rho_E + 3h \rho_E =\frac{\dot  A}{A} \rho_E.
\label{cr}
\ee
It is easy to see that in perturbations
$
\dot \rho_E= a^{-1} \left( \rho_E' + v^i\partial_i \rho_E\right) ,
$
whereas
$
3h= 3H + \frac{\theta}{a} ,
$
where $'=\pl/\pl\tau$ and $H=a'/a^2$, and we have defined $\theta= \partial_i v^i$.
Here and in the following, we neglect terms of order $\Psi_{\rm N}$ or $v^2$ as compared
with unity, as well as their time derivatives $\Psi_{\rm N}'$ or $v^j v_j'$, and we only
keep the first-order spatial gradients such as $\pl^i \Psi_{\rm N}$ or $v^j \pl_j v^i$
(because we consider large-scale structures that evolve on the Hubble time scale
but are much smaller than the horizon).
Therefore this is explicitly
\be
\rho_E' +v^i\partial_i \rho_E + (3{\cal H} +\theta)\rho_E = (\ln A)' \rho_E + v^i\partial_i (\ln A) \rho_E ,
\label{gory}
\ee
using ${\cal H}=a'/a$.
Let us use (\ref{cr}), and defining
\be
\rho_E = A \rho ,
\ee
we get
\be
\dot \rho +3 h \rho=0 ,
\ee
which is nothing but the usual conservation equation for $\rho$.
This can be done also explicitly in (\ref{gory})
which becomes
after a few trivial steps
\be
\rho' +\partial_i(\rho v^i) + 3{\cal H} \rho =0 ,
\ee
as one expects.

The nonconservation equation (\ref{big}) can now be drastically simplified and leads to
\be
u^\nu D_\nu u^\mu= -\frac{\dot A}{A} u^\mu -\partial^\mu (\ln A) .
\ee
This is the generalized geodesic equation. Specializing to $\mu=i$, we get
\be
\partial_\tau v^i +{\cal H} v^i +v^j\partial_j v^i= - \left(\frac{A'}{A}  + v^j\partial_j \ln A \right)v^i
-\partial^i (\Psi_{\rm N} + \ln A)
\ee
as expected for Newton's law and its relativistic corrections. Notice that there is a term in $v^i v^j \partial_j \ln A$ which is small or of order $v^2\ll 1 $ in the nonrelativistic limit compared to
$\partial^i \ln A$, and can be safely dropped.

\section{Modified background and dynamics}
\label{Modified-background}

\subsection{Effect of the factors $\epsilon_1$ and $\epsilon_2$}

\begin{figure}
\begin{center}
\epsfxsize=8.5 cm \epsfysize=6. cm {\epsfbox{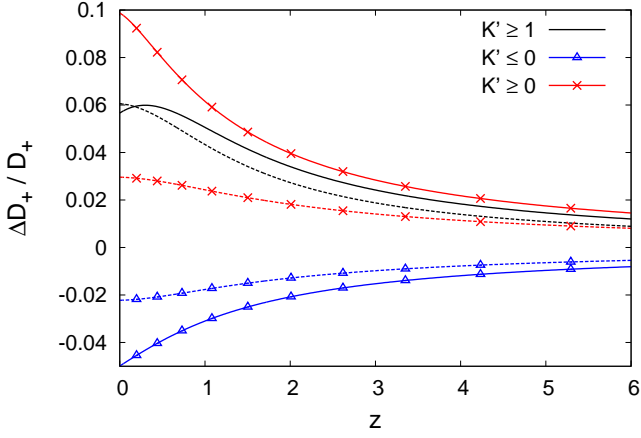}}
\end{center}
\caption{Relative deviation
$[D_+(z)-D_{+\Lambda\rm CDM}(z)]/D_{+\Lambda\rm CDM}(z)$
of the linear growing mode from the $\Lambda-$CDM reference.
Solid lines: same models as in the upper panel of Fig.~\ref{fig_Dlin_z}.
Dashed lines: same models but without the factors
$\epsilon_i$ in Eq.(\ref{D-linear}).}
\label{fig_Dlin_z_noeps}
\end{figure}

\begin{figure}
\begin{center}
\epsfxsize=8.5 cm \epsfysize=6. cm {\epsfbox{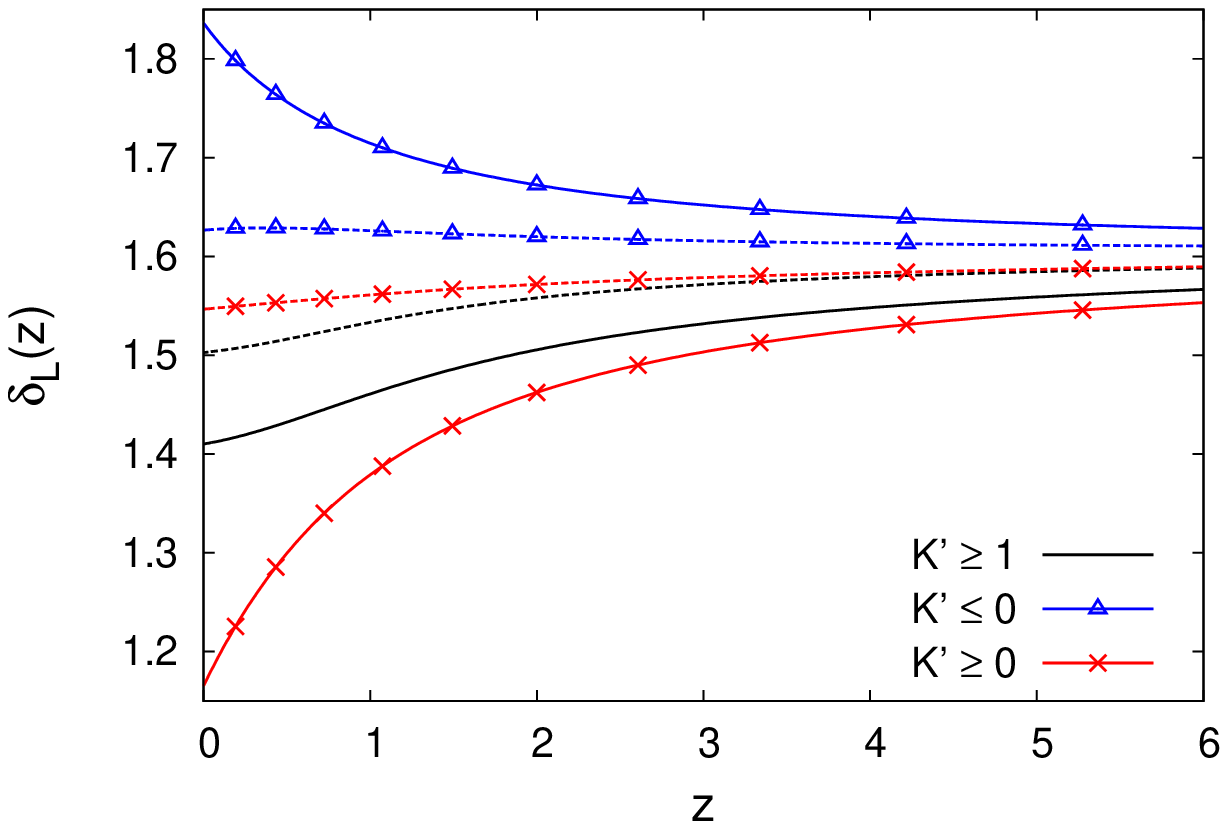}} \\
\epsfxsize=8.5 cm \epsfysize=6. cm {\epsfbox{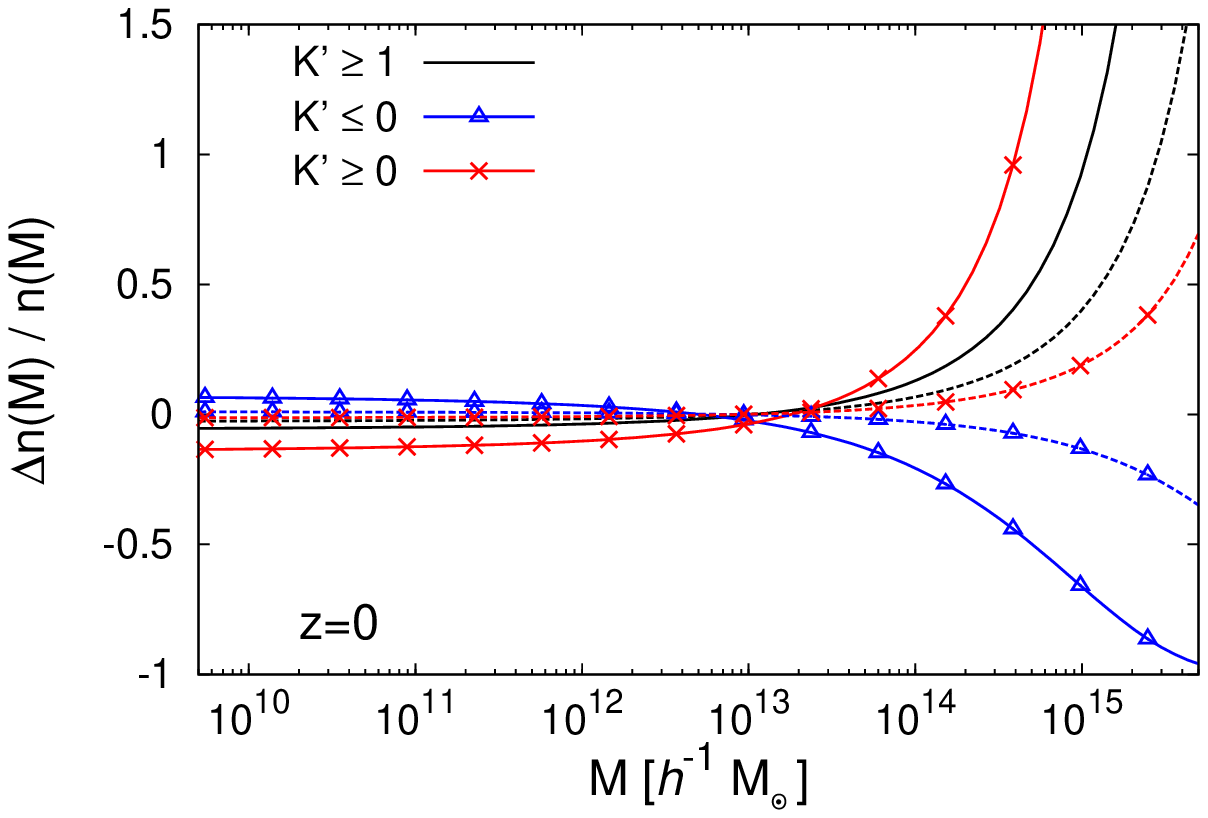}}
\end{center}
\caption{{\it Upper panel:} linear density contrast threshold $\delta_{L(\Lambda)}(z)$,
for the same models as in the upper panel of Fig.~\ref{fig_deltaLM_z0},
taking into account the factors $\epsilon_i$ in Eq.(\ref{yt-3}) (solid lines)
or setting them to zero (dashed lines).
{\it Lower panel:} relative deviation
$[n(M)-n_{\Lambda\rm CDM}(M)]/n_{\Lambda\rm CDM}(M)$
of the halo mass function from the $\Lambda$-CDM reference, for the
same cases.}
\label{fig_deltaLM_z0_noeps}
\end{figure}

\begin{figure}
\begin{center}
\epsfxsize=8.5 cm \epsfysize=6. cm {\epsfbox{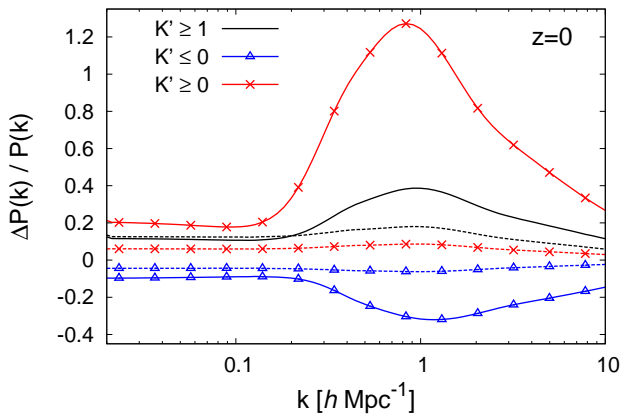}}
\end{center}
\caption{Relative deviation
$[P(k)-P_{\Lambda \rm CDM}(k)]/P_{\Lambda \rm CDM}(k)$
of the nonlinear matter density power spectrum from the $\Lambda$-CDM reference,
at redshift $z=0$, for the same models as in Fig.~\ref{fig_dPk_z0}, where
we keep the factors $\epsilon_i$ (solid lines) or set them to zero (dashed lines).}
\label{fig_dPk_z0_noeps}
\end{figure}

The linear growing modes $D_+(\eta)$ obtained from Eq.(\ref{D-linear})
deviate from the $\Lambda$-CDM reference because of two effects:
(a) the background evolution is different from $\Lambda$-CDM (the factors
$w_{\varphi}^{\rm eff}\Omega_{\varphi}^{\rm eff}$ and $\Om$ show a different
redshift dependence), and
(b) there are two new terms $\epsilon_1$ and $\epsilon_2$ in the evolution equation,
due to the coupling between the matter density and velocity fluctuations and the
scalar field $\varphi$.

To disentangle these two effects, we compare in Fig.~\ref{fig_Dlin_z_noeps}
the results we obtain from the full Eq.(\ref{D-linear})), also shown in
Fig.~\ref{fig_Dlin_z}, with those we obtain when we set the factors
$\epsilon_1$ and $\epsilon_2$ to zero in Eq.(\ref{D-linear})) (while keeping
the modified-gravity background).
We can see that a significant part of the deviation from the $\Lambda$-CDM reference
is merely due to the change of background evolution, especially at $z \geq 1$.
However, the importance of the factors $\epsilon_i$ depends somewhat on the
values of the parameters of the model and they cannot be
discarded. In particular, for the model (\ref{K-power-3}) the effect of the factors
$\epsilon_i$ is quite large.

In a similar fashion, the nonlinear spherical dynamics involve both the
modified background factors $w_{\varphi}^{\rm eff}\Omega_{\varphi}^{\rm eff}$
and $\Om$, and the new terms $\epsilon_1$ and $\epsilon_2$, see
Eq.(\ref{yt-3}). We compare in Fig.~\ref{fig_deltaLM_z0_noeps} the linear density
threshold $\delta_{L(\Lambda)}$ and the halo mass function $n(M)$ obtained
when we include the factors $\epsilon_i$ or not.
The deviation from the $\Lambda$-CDM reference keeps the same
sign whether we include these factors or not, as for the linear modes shown
in Fig.~\ref{fig_Dlin_z_noeps}, but the quantitative impact of the factors $\epsilon_i$
is greater and they cannot be neglected.

We also compare the matter density power spectra obtained by keeping
the factors $\epsilon_i$ or setting them to zero in Fig.~\ref{fig_dPk_z0_noeps}.
The difference between these cases agrees with the results found in
Figs.~\ref{fig_Dlin_z_noeps} and \ref{fig_deltaLM_z0_noeps}.
Neglecting the factors $\epsilon_i$ would significantly distort the shape of
the deviation from the $\Lambda$-CDM power spectrum and underestimate
this deviation, especially on mildly nonlinear scales.

\subsection{Comparison with the QCDM reference}

\begin{figure}
\begin{center}
\epsfxsize=8.5 cm \epsfysize=6. cm {\epsfbox{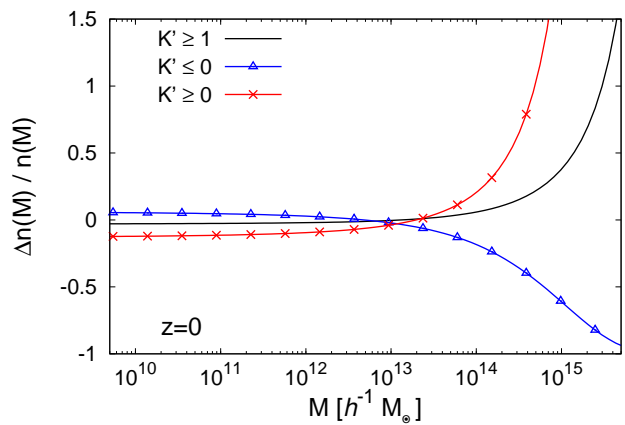}}
\end{center}
\caption{Relative deviation
$[n(M)-n_{\rm QCDM}(M)]/n_{\rm QCDM}(M)$
of the halo mass function from the QCDM reference, for the
same cases as in the upper panel of Fig.~\ref{fig_dnM_z0}.}
\label{fig_deltaLM_z0_eps_noeps}
\end{figure}

\begin{figure}
\begin{center}
\epsfxsize=8.5 cm \epsfysize=6. cm {\epsfbox{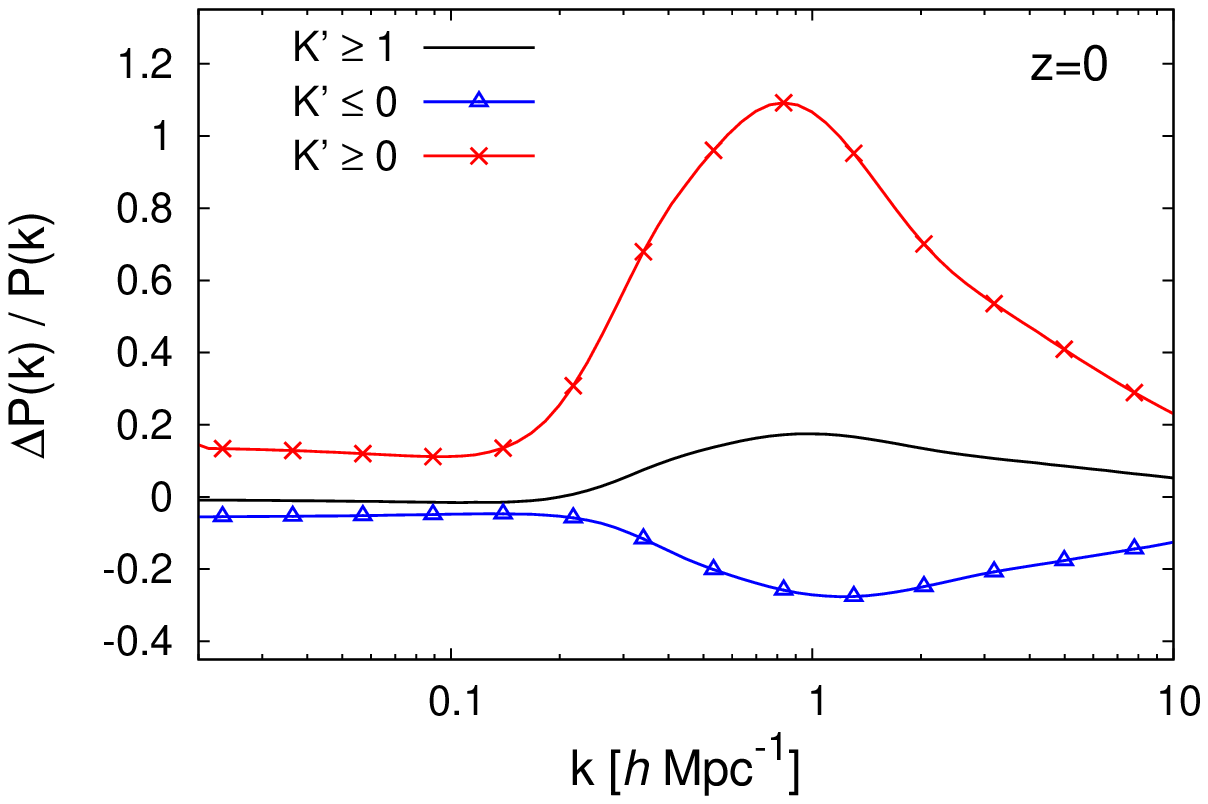}}\\
\epsfxsize=8.5 cm \epsfysize=6. cm {\epsfbox{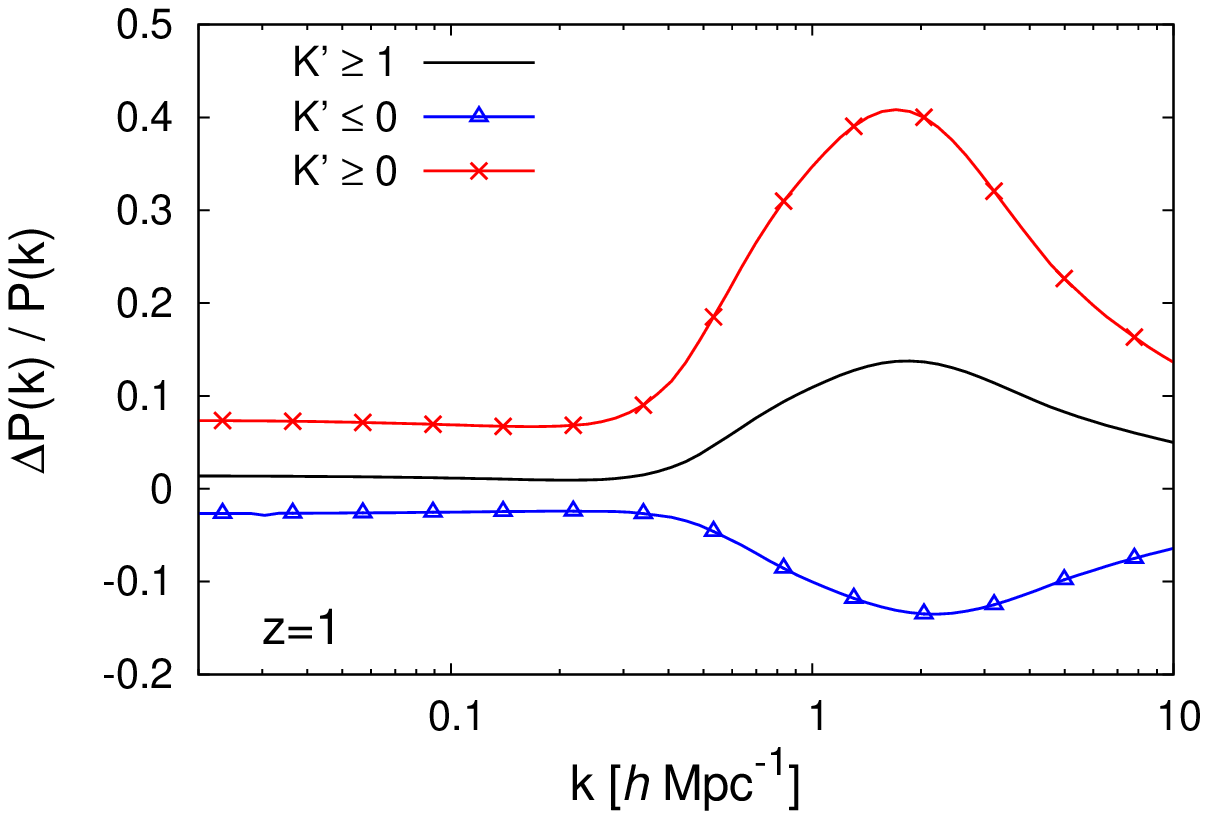}}\\
\epsfxsize=8.5 cm \epsfysize=6. cm {\epsfbox{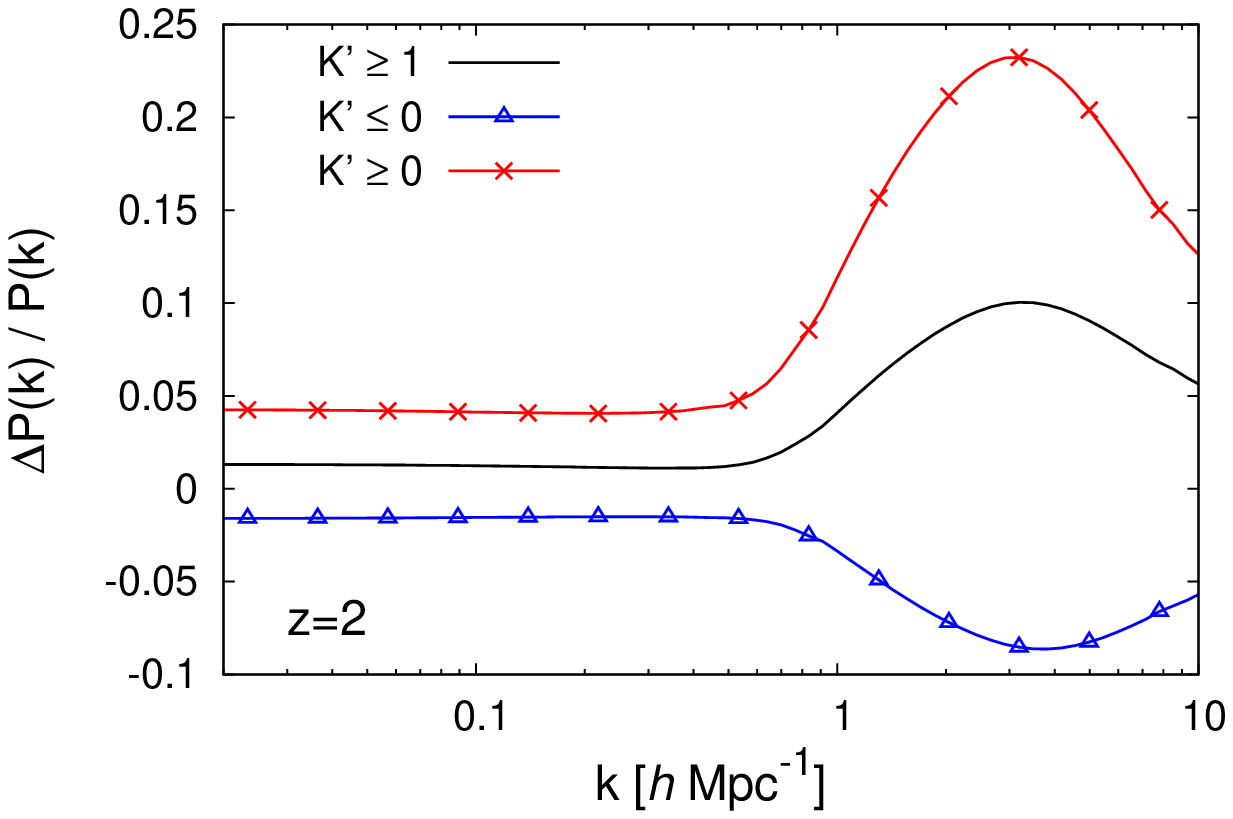}}
\end{center}
\caption{Relative deviation
$[P(k)-P_{\rm QCDM}(k)]/P_{\rm QCDM}(k)$
of the nonlinear matter density power spectrum from the QCDM reference,
at redshifts $z=0, 1$, and $2$, for the same models as in the upper panel of Fig.~\ref{fig_dPk_z0}.}
\label{fig_dPk_eps_noeps}
\end{figure}

Finally, we have studied the effect of the parameters $\epsilon_{1,2}$ by subtracting the power spectrum and halo mass function calculated with the
modified background (QCDM).
In agreement with the results above, the deviations from GR are not entirely due to the modified background and one can clearly see in Figs.~\ref{fig_deltaLM_z0_eps_noeps}
and \ref{fig_dPk_eps_noeps} that the modified
perturbation dynamics play a significant role. The order of magnitude of the deviation does not
change much from $z=0$ to $z=2$, a key feature of the K-mouflage models, except for the model
(\ref{K-power-3}) where there is a strong evolution at $z \lesssim 2$ and a slower evolution at
higher redshifts than for models (\ref{K-power-1}) and (\ref{K-power-2}).

\bibliography{refK}   

\end{document}